\documentclass[a4paper,fleqn,usenatbib]{mnras}
\pdfoutput=1
\usepackage{newtxtext,newtxmath}

\usepackage[T1]{fontenc}
\usepackage{ae,aecompl}

\usepackage{graphicx}	
\usepackage{amsmath}	
\usepackage{amssymb}	

\title[The KR in AS1063 and M1149]{The Kormendy Relation of Galaxies in the Frontier Fields Clusters: Abell\,S1063 and MACS\,J1149.5+2223}

\author[L. Tortorelli et al.]{Luca Tortorelli$^{1,2}$\thanks{E-mail: torluca@phys.ethz.ch},
Amata Mercurio$^{2}$, Maurizio Paolillo$^{3,4}$, Piero Rosati$^{5}$,\newauthor Adriana Gargiulo$^{6}$, Raphael Gobat$^{7}$, Italo Balestra$^{8}$, G. B. Caminha$^{5,9}$,\newauthor Marianna Annunziatella$^{10}$, Claudio Grillo$^{11,12}$, Marco Lombardi$^{12}$, Mario Nonino$^{13}$,\newauthor Alessandro Rettura$^{14}$, Barbara Sartoris$^{10,13}$ and Veronica Strazzullo$^{15}$
\\
$^{1}$Institute for Particle Physics and Astrophysics, ETH Z\"urich, Wolfgang-Pauli-Str. 27, 8093 Z\"urich, Switzerland\\
$^{2}$INAF-Osservatorio Astronomico di Capodimonte, Salita Moiariello 16, I-80131 Napoli, Italy\\
$^{3}$Dipartimento di Fisica "Ettore Pancini", Universit\`a degli Studi di Napoli "Federico II", Via Cinthia 21, 80126 Napoli, Italy\\
$^{4}$INFN - Unit\`a di Napoli, via Cintia, 80126 Napoli, Italy\\
$^{5}$Dipartimento di Fisica e Scienze della Terra, Universit\`a di Ferrara, Via Saragat 1, I-44122, Ferrara, Italy\\
$^{6}$INAF - Istituto di Astrofisica Spaziale e Fisica Cosmica Milano, via Bassini 15, 20133 Milano, Italy\\
$^{7}$School of Physics, Korea Institute for Advanced Study, Hoegiro 85, Dongdaemun-gu, 02455 Seoul, Republic of Korea\\
$^{8}$University Observatory Munich, Scheinerstrasse 1, D-81679 Munich, Germany\\
$^{9}$Kapteyn Astronomical Institute, University of Groningen, P.O. Box 800,9700 AV, Groningen, The Netherlands\\
$^{10}$Dipartimento di Fisica - Sezione di Astronomia, Universit\`a di Trieste, via Tiepolo 11, I-34131 Trieste, Italy\\
$^{11}$Dark Cosmology Centre, Niels Bohr Institute, University of Copenhagen, Juliane Maries Vej 30, DK-2100 Copenhagen, Denmark\\
$^{12}$Dipartimento di Fisica, Universit\`a di Milano, via Celoria 16, I-20133 Milan, Italy\\
$^{13}$INAF - Osservatorio Astronomico di Trieste, via G. B. Tiepolo 11, I-34143 Trieste, Italy\\
$^{14}$IPAC, Caltech, KS 314-6, 1200 E. California Blvd, Pasadena, CA 91125, USA\\
$^{15}$Department of Physics, Ludwig-Maximilians-Universitat, Scheinerstr. 1, D-81679 Munchen, Germany
}

\date{Accepted 2018 March 5. Received 2018 February 19; in original form 2017 November 7}

\pubyear{2018}

\begin{document}
\label{firstpage}
\pagerange{\pageref{firstpage}--\pageref{lastpage}}
\maketitle

\begin{abstract}
We analyse the Kormendy relations (KRs) of the two Frontier Fields clusters, Abell\,S1063, at z = 0.348, and MACS\,J1149.5+2223, at z = 0.542, exploiting very deep Hubble Space Telescope photometry and VLT/MUSE integral field spectroscopy. With this novel dataset, we are able to investigate how the KR parameters depend on the cluster galaxy sample selection and how this affects studies of galaxy evolution based on the KR. We define and compare four different galaxy samples according to: (a) S\'ersic indices: early-type ('ETG'), (b) visual inspection: 'ellipticals', (c) colours: 'red', (d) spectral properties: 'passive'. The classification is performed for a complete sample of galaxies with m$_{\textit{F814W}} \le$ 22.5 ABmag (M$_{*}$ $\gtrsim 10^{10.0}$ M$_{\odot}$). To derive robust galaxy structural parameters, we use two methods: (1) an iterative estimate of structural parameters using images of increasing size, in order to deal with closely separated galaxies and (2) different background estimations, to deal with the Intracluster light contamination. The comparison between the KRs obtained from the different samples suggests that the sample selection could affect the estimate of the best-fitting KR parameters. The KR built with ETGs is fully consistent with the one obtained for ellipticals and passive. On the other hand, the KR slope built on the red sample is only marginally consistent with those obtained with the other samples. We also release the photometric catalogue with structural parameters for the galaxies included in the present analysis.
\end{abstract}





\section{Introduction}

Galaxies are characterized by a wide range of masses, sizes and morphologies, but their physical parameters, such as total luminosity, effective radius and central stellar velocity dispersion are correlated. The study of the relation among such properties, and their variation as a function of redshift, leads to a better understanding of the physical processes that underlie the formation and evolution of galaxies themselves.

Observations show the existence of a correlation between the effective radius R$_{\mathrm{e}}$ of an early-type galaxy (ETG) and the mean surface brightness within that radius $\left \langle \mu \right \rangle_{\mathrm{e}}$, $\left \langle \mu \right \rangle_{\mathrm{e}} = \alpha + \beta \log{\mathrm{R}_{\mathrm{e}}}$, known as the Kormendy relation (hereafter KR, \citealt{Kormendy1977}), stating that more luminous ellipticals are larger and have a lower characteristic surface brightness. This relation provides information on the distribution of the light profiles and the sizes of ETGs, thus it can be used as a proxy to investigate their present and past evolution, in particular their size-evolution over cosmic time. 

It is still a matter of debate whether the size of ETGs increases with the cosmic time. Several authors found that high-z (z $\gtrsim$ 1.4) massive galaxies are more compact, i.e. they have a smaller effective radius, than local ETGs \citep{Daddi2005, Trujillo2006, Longhetti2007}. This size evolution is also present at moderate redshifts: the effective radius of ETGs should increase at least by a factor of $\sim$ 1.5--2 since z $ \sim 1.2$ to explain the discrepancy of the evolution of $\left \langle \mu \right \rangle_{\mathrm{e}}$ as a function of redshifts in the KR, assuming a pure luminosity evolution (e.g., \citealt{Longhetti2007, Rettura2010, Trujillo2011}). However, other studies provide contrasting results when samples of ETGs are strictly selected on the basis of their morphology (e.g., \citealt{Mancini2010,Saracco2010,Stott2011,Jorgensen2013}) and when the number density and stellar population properties of compact ETGs at high and low redshifts are taken into account (e.g., \citealt{Saglia2010, Poggianti2013, Belli2014,Fagioli2016}). This seems to imply that the sample selection plays an important role in disantangling between the two scenarios. Since the KR can be used to study the size-evolution over cosmic time, understanding whether there is a bias in the determination of its parameters due to the sample selection becomes crucial.

The KR applies to ETGs defined according to their light profiles. However, since ETGs are expected to be also red in photometric colours and passive in terms of occurring star formation, often, in literature, this relation has been obtained using a sample of galaxies selected according to their colours and/or their spectra. This is especially true at high-z or with low resolution data, where a proper morphological classification is hard to determine. In order to investigate possible biases due to the different definitions of the galaxies on which the KR is measured, it is useful to select and compare four different samples: early-type, ellipticals, red and passive galaxies. Indeed, one wouldn't expect perferct overlap between these 4 classes even in principle. For example, it is well-known that dusty star-forming galaxies can have red-sequence-like colours, and passive discs can be expected to stay disky for quite some time after quenching.

This work aims at investigating how the KR changes as a function of the sample selection using two Frontier Fields clusters: Abell S1063 (hereafter AS1063) and MACSJ1149.5+2223 (hereafter M1149), spanning the redshift range 0.35--0.54. These redshifts witness significant evolution in the cluster galaxy population \citep{Poggianti2006, Desai2007}. In fact, z $ \sim 0.4$ represents the peak of the infall rate of field galaxies into a cluster \citep{Kauffmann1995} and of the transformation of these infalling field galaxies from star-forming disc-dominated galaxies into passively evolving spheroids, through, e.g., ram-pressure stripping when infalling through a cluster core. Therefore, they constitute the perfect environments in which one can conduct this study, since one deals with galaxies with mixed colours and/or morphologies that may bias the classification (e.g., a spiral entering the cluster and loosing gas through ram-pressure stripping appears passive, but still it has a disc-like morphology). In order to do that, exquisite photometric and spectroscopic data have to be used, such as the Frontier Fields clusters, which are those with the deepest Hubble Space Telescope (\textit{HST}) photometry currently available, and MUSE integral field spectroscopic data.

AS1063 is a massive cluster (total mass M $ \sim 2.42 \times 10^{15}\ \mathrm{M}_{\odot}$, \citealt{Caminha2016}) at z $= 0.348 \pm 0.001$ \citep{Karman2015} with a large velocity dispersion ($1380 \pm 32$) km s$^{-1}$ \citep{Caminha2016} and it is the southern-most Frontier Fields (Sect. 2) cluster, while M1149 at z $= 0.542 \pm 0.001$ \citep{Grillo2016}, with an estimated total mass of $\sim 2.5 \times 10^{15}\ \mathrm{M}_{\odot}$ \citep{Zheng2012}, was discovered as part of the MACS survey \citep{Ebeling2000}.

In Sect. 2, we describe the dataset and the membership of cluster galaxies whose surface photometry analysis is detailed in Sect. 3. Sect. 4 is devoted to the definition of the samples of early-type, elliptical, red and passive galaxies and to their comparison. In Sect. 5 we discuss the results on the analysis of the KR and we compare them with other results in literature in Sect. 6. Finally, in Sect. 7, we summarize our results and present our conclusions.

Unless otherwise stated, we give errors at the 68 per cent confidence level (hereafter c.l.) and we report circularized effective radii. Throughout this paper, we use H$_0$ = 70 km s$^{-1}$ Mpc$^{-1}$ in a flat cosmology with $\Omega_{\mathrm{M}}$ = 0.3 and $\Omega_{\Lambda}$ = 0.7. In the adopted cosmology, 1\arcmin\ corresponds to 0.295 Mpc at z = 0.348 and 0.373 Mpc at z = 0.542. 

\section{Dataset and Membership}

\begin{figure*}
\centering
\includegraphics[scale=0.10]{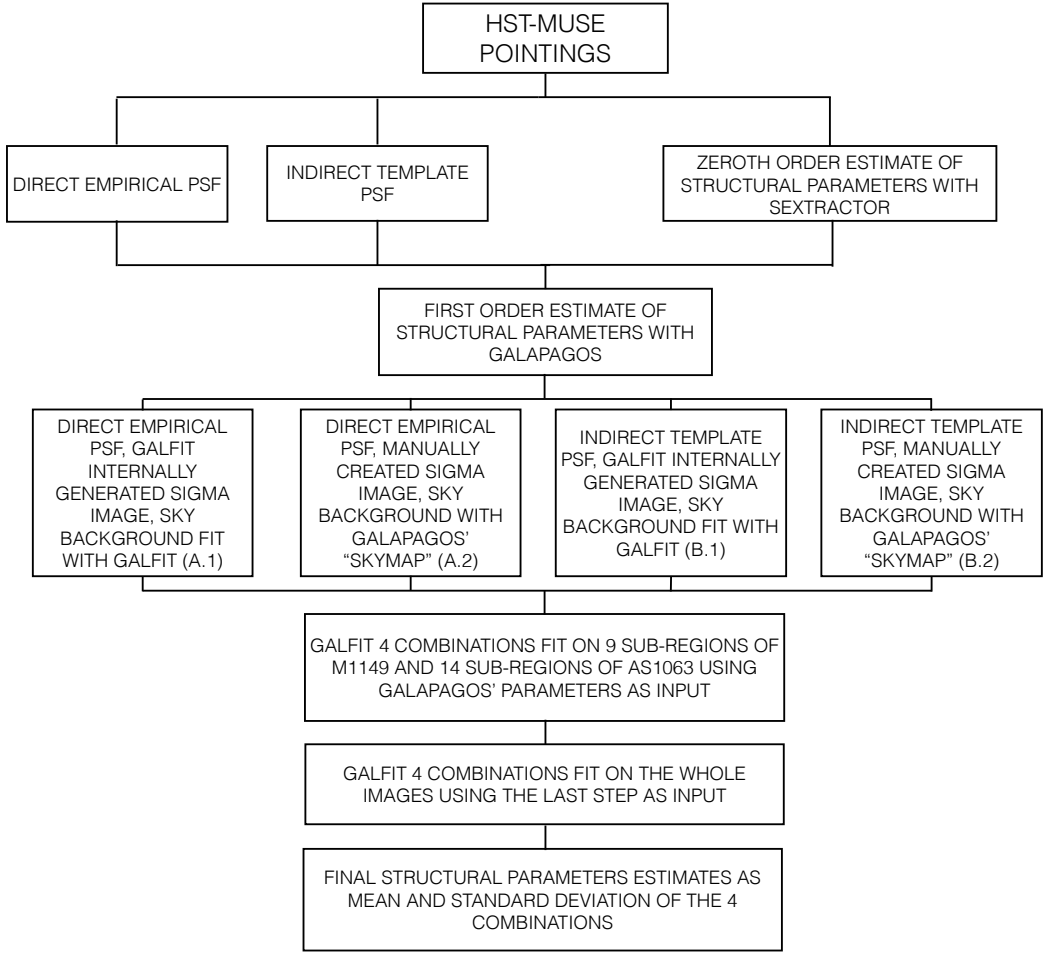}
\caption{Flowchart describing each step of the measurement of the structural parameters in each HST-MUSE pointing is shown. The data analysis is performed in three steps: derivation of accurate PSFs, estimatation of the sky background and optimization of the surface brightness profile fit.}
\label{fig1}
\end{figure*}

The two clusters AS1063 and M1149 are part of the Frontier Fields programme\footnote{\footnotesize http://www.stsci.edu/hst/campaigns/frontier-fields/HST-Survey} (hereafter FF, PI: J. Lotz, \citealt{Lotz2017}), aiming at combining the power of \textit{HST} and \textit{Spitzer} with the natural gravitational telescope effect of massive high-magnification clusters of galaxies to produce the deepest observations of clusters and background lensed galaxies ever obtained. Each cluster is imaged for a total of 140 orbits, divided over 7 optical/NIR bands (\textit{F435W, F606W, F814W; F105W, F125W, F140W, F160W}).

We analyse the public Hubble Space Telescope (\textit{HST}) photometric data available at the STScI MAST Archive\footnote{\footnotesize https://archive.stsci.edu/prepds/frontier/}. We use the v1.0 release of Epoch 1 images from the HST FF ID 14037 for the cluster AS1063 and the v1.0 release, Epoch 2 images from the HST FF ID 13504 for the cluster M1149. The mosaics have been processed with the new "self-calibration" approach \citep{Anderson2014} to reduce low-level dark current artefacts across the detector. In orded to derive structural parameters (i.e. effective radius and surface brightness), we choose the \textit{F814W} filter which corresponds to the rest-frame \textit{R}-band for AS1063 and to the rest-frame \textit{V}-band for M1149.

AS1063 and M1149 have a wealth of spectroscopic observations. AS1063 was observed with VIMOS at the ESO VLT, as a part of the ESO Large Programme 186.A-0798 "Dark Matter Mass Distributions of Hubble Treasury Clusters and the Foundations of $\Lambda$CDM Structure Formation Models" (P.I.: Piero Rosati, \citealt{Rosati2014}, hereafter CLASH-VLT), with GMOS-S at Gemini by \citet{Gomez2012}, with the Magellan telescope (D. D. Kelson private communication), with the MUSE integral field spectrograph \citep{Karman2015,Caminha2016} as part of ESO Programme IDs 60.A-9345(A) and 095.A-0653(A) and with the WFC3-IR-GRISM of \textit{HST}, as part of the Grism Lens-Amplified Survey from Space (hereafter GLASS, GO-13459, P.I.: T. Treu, \citealt{Treu-GLASS2016}). M1149 has been observed with WFC3-IR-GRISM, as a part of GLASS \citep{Treu-GLASS2016}, and with the MUSE integral field spectrograph as part of ESO Programme ID 294.A-5032(A). The total sample consists of 3850 spectra for AS1063 and 429 spectra for M1149, of which 175 \citep{Karman2015, Karman2016} and 117 \citep{Grillo2016} have been observed by MUSE, respectively. MUSE data include two pointings of the cluster core of AS1063, a North-Eastern (NE) one and a South-Western (SW) one, with respect to the position of the BCG \citep{Karman2016} and one pointing for the cluster core of M1149 (\citealt{Grillo2016}, see fig. 1 in \citealt{Karman2015} and fig. 2 in \citealt{Grillo2016}, respectively). Differently from other spectroscopic datasets, the cluster members for which MUSE IFU spectra are available allow us to be fully complete down to magnitude 22.5 in the \textit{F814W} waveband \citep{Caminha2016}. This limit corresponds roughly to a stellar mass value of M$_{*}$ $\sim 10^{9.8}$ M$_{\odot}$ and  M$_{*}$ $\sim 10^{10.0}$ M$_{\odot}$ for Abell\,S1063 and MACS\,J1149.5+2223, respectively, for the typical SED of the sources we are interested in and considering a Salpeter IMF \citep{Salpeter1955}. For this reason, we use only galaxies in MUSE pointings with magnitudes below this threshold to derive the KR for both clusters. In the following, we refer to the \textit{HST} images of those regions as HST-MUSE pointings. 

In order to define the membership for these two clusters, we analyse the 1-D velocity distribution (e.g., Fig. 1 in \citealt{Grillo2016}  for M1149) of the total sample of spectra. Spectroscopically confirmed cluster members, for which MUSE IFU spectra are available, are $95$ for AS1063 (z = 0.348), having redshifts in the range $0.335 \le$ z $\le 0.361$ \citep{Karman2015}, and 68 with redshifts $0.52 \le$ z $\le 0.57$ \citep{Grillo2016} for M1149 (z = 0.542). 60 out of 95 and 42 out of 68 galaxies have magnitudes brighter than the completness limit we adopted in \textit{F814W} waveband for AS1063 and M1149, respectively. These galaxies constitute our starting sample on which we apply the different selection criteria.

\section{Data Analysis}

\begin{figure*}
\centering
\includegraphics[scale=0.4]{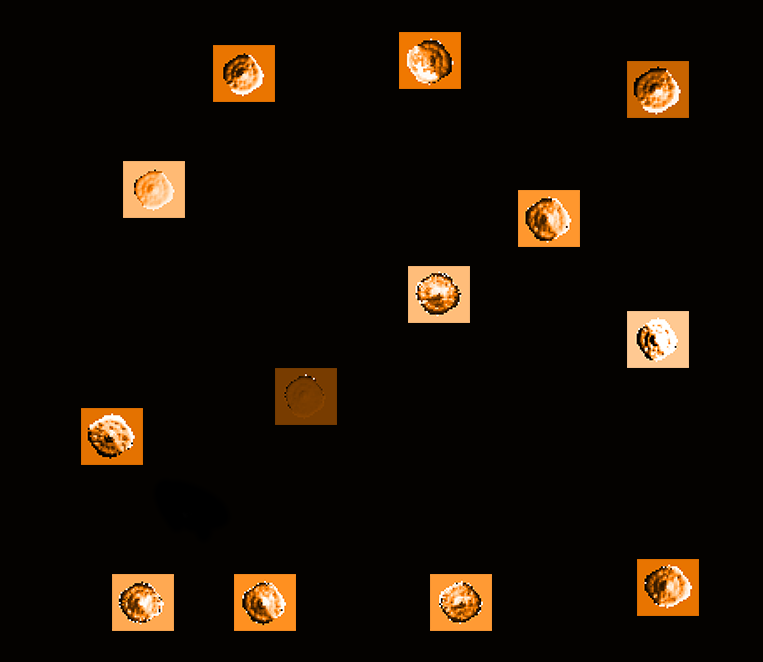}
\caption{Relative residuals between the PSF at the centre of the field and the PSFs in other positions superimposed on a blank ACS field are shown. The PSFs are simulated in different positions in order to sample the entire ACS field-of-view. In the adopted linear colorscale, black and white pixels represent negative and positive residuals, respectively, while brown represents the zero level. As we can see, moving towards the edges of the detector leads to more asymmetric residuals due to the ACS spatial distortion. Differences are of the order of few percent in terms of FWHM and degree of symmetry.}
\label{fig2}
\end{figure*}

In order to build the KR, we derive the structural parameters of the spectroscopically confirmed member galaxies, performing a parametric fit of the light distribution using a single S\'ersic profile, since it gives us the versatility of fitting both spheroids and discs. Since what we observe is the light profile convolved with the PSF, in order to recover the intrinsic light profiles, an accurate modeling of the real PSF is mandatory. Furthermore, the central region of a galaxy cluster has two other complications: it is a crowded environment and it is characterized by the presence of the Intracluster light (ICL). The latter constitutes a contribution in flux that is added on top of the sky background, while the former makes the determination of galaxy light profiles highly dependent on the simultaneous fit of the nearest neighbours. These reasons lead us to design a new methodology to estimate the structural parameters based on an iterative approach which analyse images of increasing size (to deal with nearest neighbours) and on multiple background estimation (to deal with the ICL contamination in flux).

In detail, the data analysis is performed in three steps (Fig. \ref{fig1}): in the first one, we derive accurate PSFs using both direct (from the data themselves) and indirect (from external libraries) empirical templates (Sect. 3.1). The second one is the estimate of the sky background. The last step is the optimization of the surface brightness profile fit. This step starts with an automatic fit on small stamps containing only a few galaxies (\texttt{SExtractor}, \citealt{Bertin1996}, and \texttt{GALAPAGOS}, \citealt{Barden2012}) and ends with a simultaneous fit of all the galaxies in the image performed with \texttt{GALFIT} \citep{Peng2011}. The outputs of \texttt{GALAPAGOS} are used, in turn, as guesses for the inputs of the last run. The last step is repeated using different realizations of the PSFs and sky estimates (Sect. 3.2). The final estimates of the structural parameters are the average over all the different realizations (Sect. 3.3).

\subsection{PSF measurement}

The direct PSF is measured extracting and combining observed stars from the scientific images. This method has the advantage of obtaining a PSF directly from stars observed in exactly the same conditions of the analysed galaxies. 

First, we run \texttt{SExtractor} on the whole images of the two clusters in order to select point-like sources in the field. Sources are classified as stars if the CLASS\_STAR parameter from \texttt{SExtractor} is greater than 0.9. From this sample, we reject saturated stars and select isolated ones through a visual inspection. Finally, we refine our sample considering only S/N > 100 sources in the ACS field. We average these and perform a gaussian fit of the resulting PSF to check if the effective FWHM (FWHM$_{\mathrm{eff}} = 0.105^{''}$) is close to the nominal one of HST/ACS ($0.1^{''}$).

The indirect PSF is obtained by adding stars extracted from the Anderson library of empirical PSFs \citep{Anderson2006} on raw images in different detector positions, following the same reduction procedure of real data to account for the spatial distortion of ACS, as detailed below.

The \texttt{Multiking HST/ACS simulator}\footnote{http://people.na.infn.it/~paolillo/MyWebSite/Software.html} \citep{Paolillo2011,Goudfrooij2012,Puzia2014} is used to add empirical Anderson PSFs to HST/ACS observations, simulating the effect of dithering on the final drizzled and stacked image. Anderson empirical PSFs are designed to be added to original images that have been bias subtracted and flat-fielded, but not resampled or drizzled (hereafter, \texttt{flt}). We first align the original \texttt{flt} images and then we run the \texttt{Multiking} tool in order to add the PSFs on each individual aligned \texttt{flt} image and, finally, combine them. The procedure sets the Astrodrizzle \citep{Gonzaga2012} parameters in order to reproduce the dithering and projection pattern used in real data, in particular those of the FF images. Thanks to the possibility of generating PSFs on different positions on the detector, we are able to investigate how different in terms of FWHM and degree of symmetry are the generated PSFs across the field (Fig. \ref{fig2}). We find that differences are of the order of few percents. We then tested the effect of such differences on some of the brightest galaxies in the HST-MUSE pointings, performing a fit of the galaxy surface brightness using PSFs generated by \texttt{Multiking} at different positions on the detector. We find differences of the order of $\sim$ 5 per cent in the estimate of structural parameters when using PSFs within the HST-MUSE pointings. On the contrary, using PSFs generated outside or at the edges of the HST-MUSE pointings, leads to bigger differences ($\sim$ 10--15 per cent). So the use of the \texttt{Multiking HST/ACS simulator} is necessary in order to take into account the spatial distortion of the PSF due to the effect of detector plus reduction procedures and to estimate the degree of precision with which one derives the structural parameters. Moreover, the tool is fundamental if it is not possible to find stars in the field suitable to measure the PSF.

\subsection{Background estimates and Fit Optimization}

\texttt{GALFIT} and \texttt{GALAPAGOS} request 7 guess parameters (x,y positions, position angle, axis ratio, major-axis effective radius, S\'ersic index and total magnitude) to inizialize the fit of the Sersic profile. Those initial values are obtained from \texttt{SExtractor} runs on each HST-MUSE pointing of the two clusters separately. Furthermore, \texttt{GALFIT} uses a "bad pixel map" to select pixels belonging to sources that one wants to include in the fit. Therefore, also \texttt{SExtractor} "Segmentation maps" are provided to the latter, so that spectroscopically confirmed cluster members and their neighbours are fitted. The quality of the Segmentation map depends on the choice of \texttt{SExtractor} input parameters. Their determination is crucial since, in both fields, we deal with small galaxies embedded in the halo of larger and brighter sources, and this is especially true if we consider the contribution of the ICL. Indeed, we use the "hot" and "cold" mode of \texttt{SExtractor} (see \citealt{Barden2012}) and we perform different runs to assess which is the best combination that produces reliable output structural parameters, following also the prescriptions highlighted in \citet{Annunziatella2013,Annunziatella2016}.

\texttt{SExtractor} parameters are then used by \texttt{GALAPAGOS} as starting guesses for the light profile fitting. \texttt{GALAPAGOS} cuts the science images into smaller sections centred on the individual sources in order to significantly reduce the total fitting time. \texttt{GALAPAGOS} does not allow a free fit of the sky background in each stamp, but it estimates a value before the fitting. To do this, \texttt{GALAPAGOS} creates a "sky-map", i.e. a copy of the input images where the pixel values indicate the nature of the measured flux. In the sky-map a pixel value of 0 stands for blank background sky, while positive numbers indicate the presence of a source. The extension of each object is then estimated using a flux growth method (see \citealt{Barden2012} for more details).

In our approach, \texttt{GALAPAGOS} parameters constitute a first-order approximation of reliable structural parameters for the sample of galaxies. We refer to Sect. 3.3 for a comparison between the final structural parameters we use to build the KR and \texttt{GALAPAGOS} estimates of them.

Continuing the approach of increasing images size, we perform a fit of the light-profile of the sample of galaxies with \texttt{GALFIT}, using parameters provided by \texttt{GALAPAGOS} as starting point, enlarging step by step the image sizes and thus the number of galaxies simultaneously fitted (typically 15--20 sources versus 3--4 in \texttt{GALAPAGOS}). We cut stamps of each HST-MUSE pointing and relative Segmentation map, whose dimensions are chosen to contain the number of sources mentioned above and paying attention to consider regions in which the sky background has a quite homogeneous value. We divide the HST-MUSE pointing of M1149 in 9 sub-regions and the NE and SW pointings of AS1063 in 7 sub-regions each. The typical area on the sky spanned by each sub-region is 25 $\times$ 25 arcsec$^2$.

To optimize the fitting method, both on stamps and later on the whole images, and to account for systematics, we perform 4 sub-analysis by using different combinations of PSFs, Sigma images (e.g., see \citet{Peng2011} for a definition) and sky background values. The structural parameters obtained through the different combinations are then compared and used to estimate systematic errors (Sect. 3.3). The combinations of PSFs, Sigma images and sky background estimates used are:
\begin{itemize}
\item Direct empirical PSF, Sigma image internally generated by \texttt{GALFIT}, sky background value as free parameter in galaxy fitting. Hereafter, we refer to this combination as A.1.
\item Direct empirical PSF, Sigma image generated through inverse variance map of FF images, fixed sky background value estimated from \texttt{GALAPAGOS} skymap. Hereafter, A.2.
\item Indirect template PSF, Sigma image internally generated by \texttt{GALFIT}, sky background value as free parameter in galaxy fitting. Hereafter, B.1.
\item Indirect template PSF, Sigma image generated through inverse variance map of FF images, fixed sky background value estimated from \texttt{GALAPAGOS} skymap. Hereafter, B.2.
\end{itemize}

For the sky background value estimate, we follow two approach:
\begin{itemize}
\item We mask all sources in each pointing using \texttt{SExtractor} Segmentation map and we fit the sky background value with \texttt{GALFIT}, using as starting point for the fit the mean value of the \texttt{SExtractor} BACKGROUND output parameter. The resulting value, in turn, is used as a starting point for the sky free parameter in combinations A.1 and B.1.
\item We use \texttt{GALAPAGOS} skymap. In correspondence of the regions where the skymap is blank (pixel value is zero), we cut stamps of the original image, sampling the whole pointing as much as we can. Using the task \texttt{imstat} in \texttt{IRAF}, we calculate the mean of pixel values in each stamps, obtaining an estimate of the sky background suitable for the whole HST-MUSE pointing. The estimated value of the sky background is held fixed in combinations A.2 and B.2.
\end{itemize}

The Sigma image used in \texttt{GALFIT} can be provided by the user or internally generated by the program. We perform galaxy fitting considering both cases:
\begin{itemize}
\item We let \texttt{GALFIT} internally generate the Sigma Image. In order to do that, we transform the units of the images from electrons/s to ADU, using the exposure time and the GAIN of the detector (in units of \#electrons ADU$^{-1}$), read from the headers of the FF images, according to the following
\begin{equation}
[\mathrm{ADU}] = \frac{[\mathrm{electrons\ s^{-1}}]}{[\mathrm{GAIN}]} [\mathrm{EXPTIME}]
\end{equation}
This transformation is required since \texttt{GALFIT} can perfectly deal with the Sigma image only if [GAIN] $\times$ [ADU] = total electrons collected at each pixel. This Sigma image is used in combinations A.1 and B.1.
\item We generate our own Sigma image following the prescription in the \texttt{GALFIT}'s User Manual\footnote{users.obs.carnegiescience.edu/peng/work/galfit/README.pdf}. The flux variance of the data $f_{\mathrm{data}}$(x, y) at each pixel is obtained from the flux of the drizzled science image (drz), the mean value of the sky $\left \langle \mathrm{sky}_{\mathrm{estim}} \right \rangle$ is that estimated with the \texttt{GALAPAGOS} skymap, the GAIN is actually an effective GAIN equal to the exposure time EXPTIME (if we are using the science images in electrons s$^{-1}$) read from the image header, and the sky RMS is given by the pixel per pixel value of the inverse variance weight image (wht), which contains all of the noise sources in each exposure, including read-noise, dark current, sky background and Poisson noise:
\begin{equation}
\sigma = \sqrt{(\mathrm{drz} - \left \langle \mathrm{sky}_{\mathrm{estim}} \right \rangle) / \mathrm{EXPTIME} + 1/\mathrm{wht}}
\end{equation}
This Sigma image is used in combination A.2 and B.2.
\end{itemize}

\begin{figure}
\centering
\includegraphics[scale=0.1505]{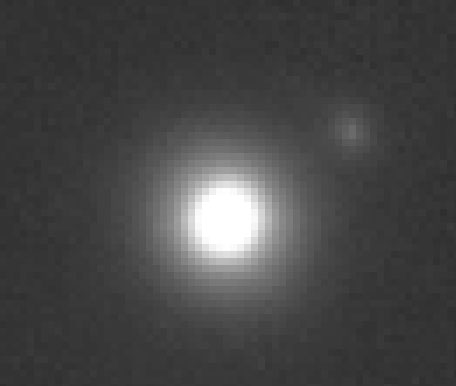} 
\includegraphics[scale=0.1295]{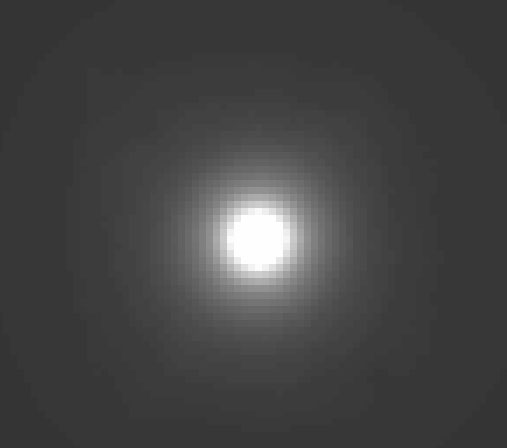}
\includegraphics[scale=0.13]{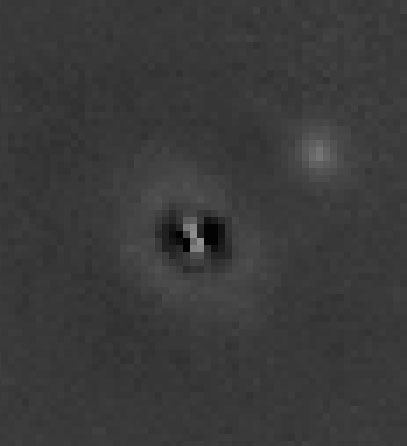} \\
\includegraphics[scale=0.47]{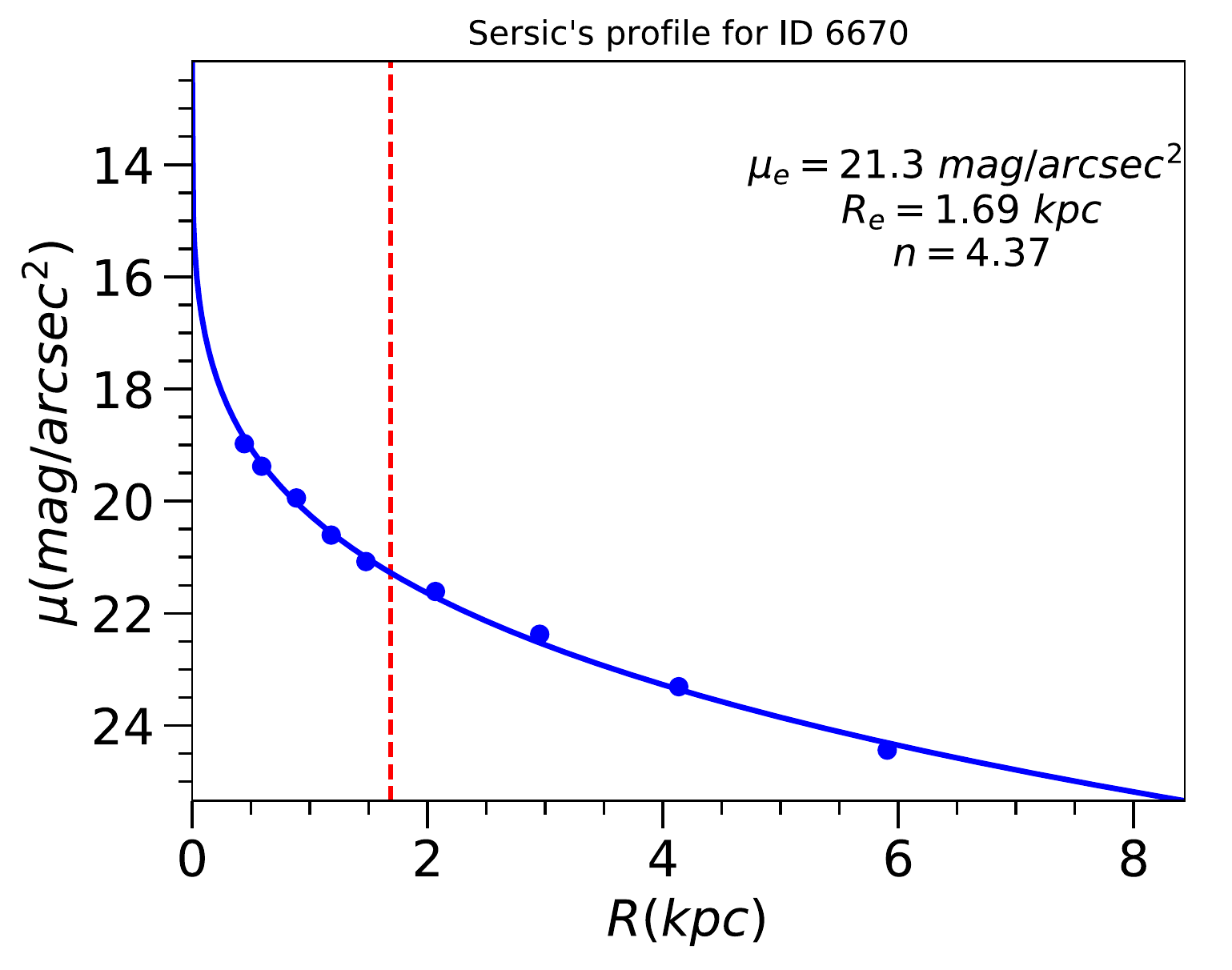}
\caption{Observed image (top left panel), model generated by \texttt{GALFIT} (top central panel), residuals of the fit (top right panel) and S\'ersic profile (bottom panel) of a spectrophotometrically selected ETG (see Sect. 4 for sample definitions) are shown. The fitted S\'ersic profile and the surface brightnesses obtained with aperture magnitudes (blue points in the central panel) are not corrected for the cosmological dimming effect. The structures visible in the residual image are negligible in terms of flux with respect to the real and the model image, although they seem enhanced due to the scale choice. Indeed, the ratio between the residuals and the same regions in the real and model image is $\sim$ 1\%.}
\label{fig3}
\end{figure}

The last step we perform in order to optimize the structural parameters of galaxies is the simultaneous fit of sources on the whole HST-MUSE pointings. In particular, the fit is performed only on spectroscopically selected cluster members and their nearby sources (defined as sources for which the elliptical aperture, within which we measure the Kron magnitude, overlaps with that of spectroscopically selected members), masking the remaining ones in order not to increase the number of free parameters in the fit. In order not to affect the fit by a different choice of the mask, we use the same Segmentation maps that we have previously cut in stamps. For this last step, we use the same combination of PSFs, sky background values and Sigma Images of the stamps fitting. This is done in order to discuss (Sect. 3.3) the systematic errors of the structural parameters on the basis of the 4 different methods used.

\begin{figure*}
\centering
\includegraphics[scale=0.5]{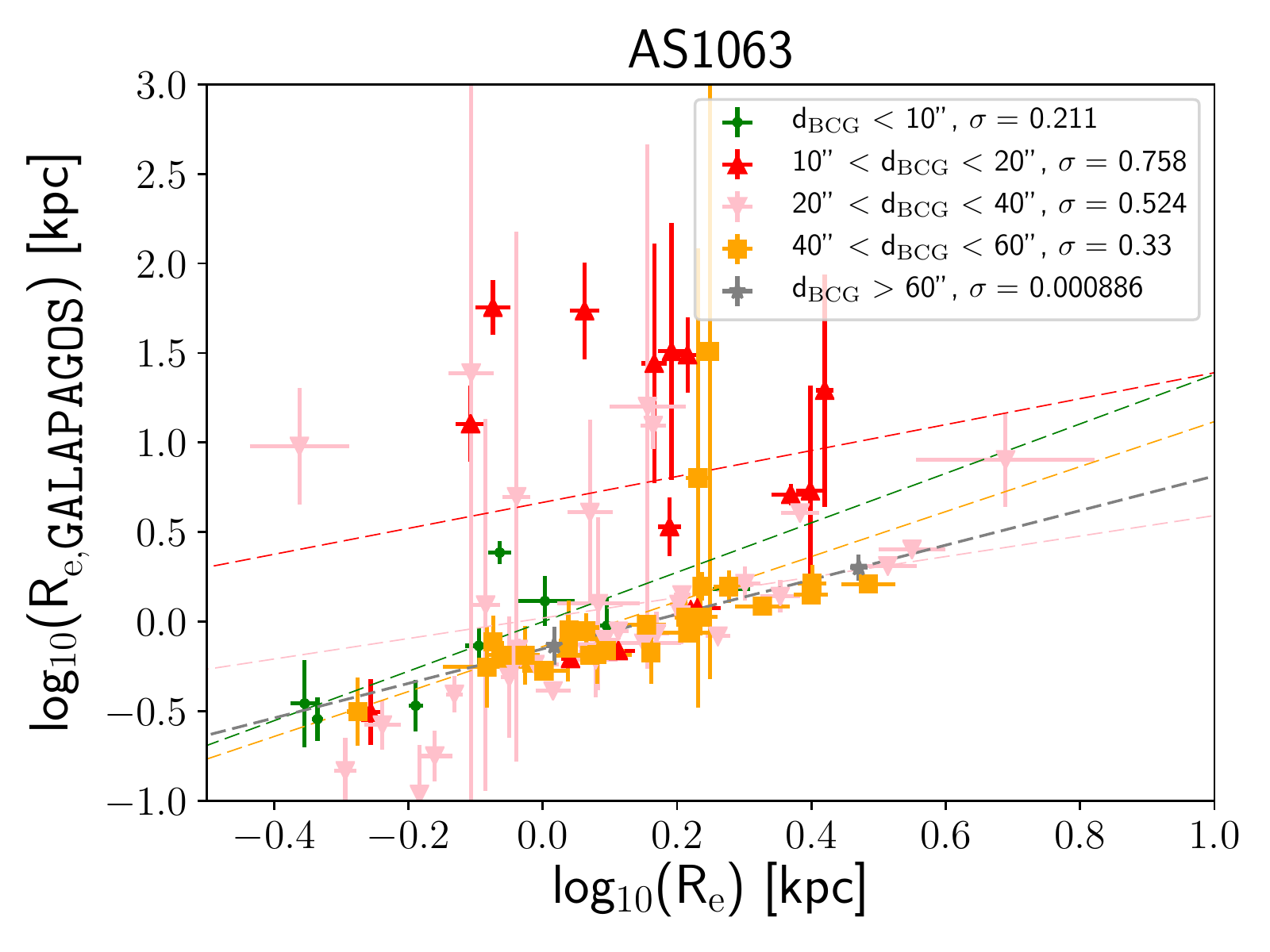}
\includegraphics[scale=0.5]{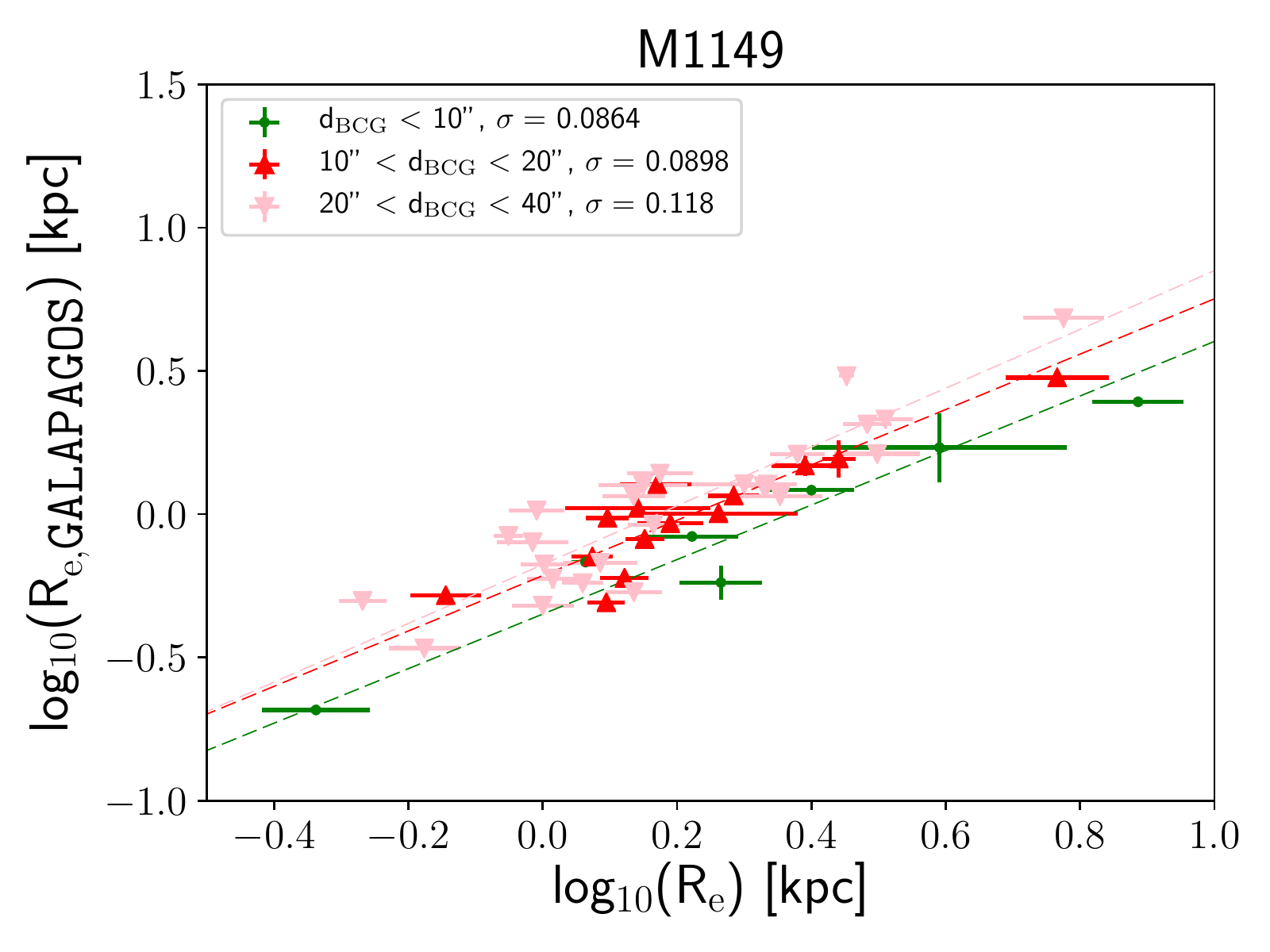}
\includegraphics[scale=0.5]{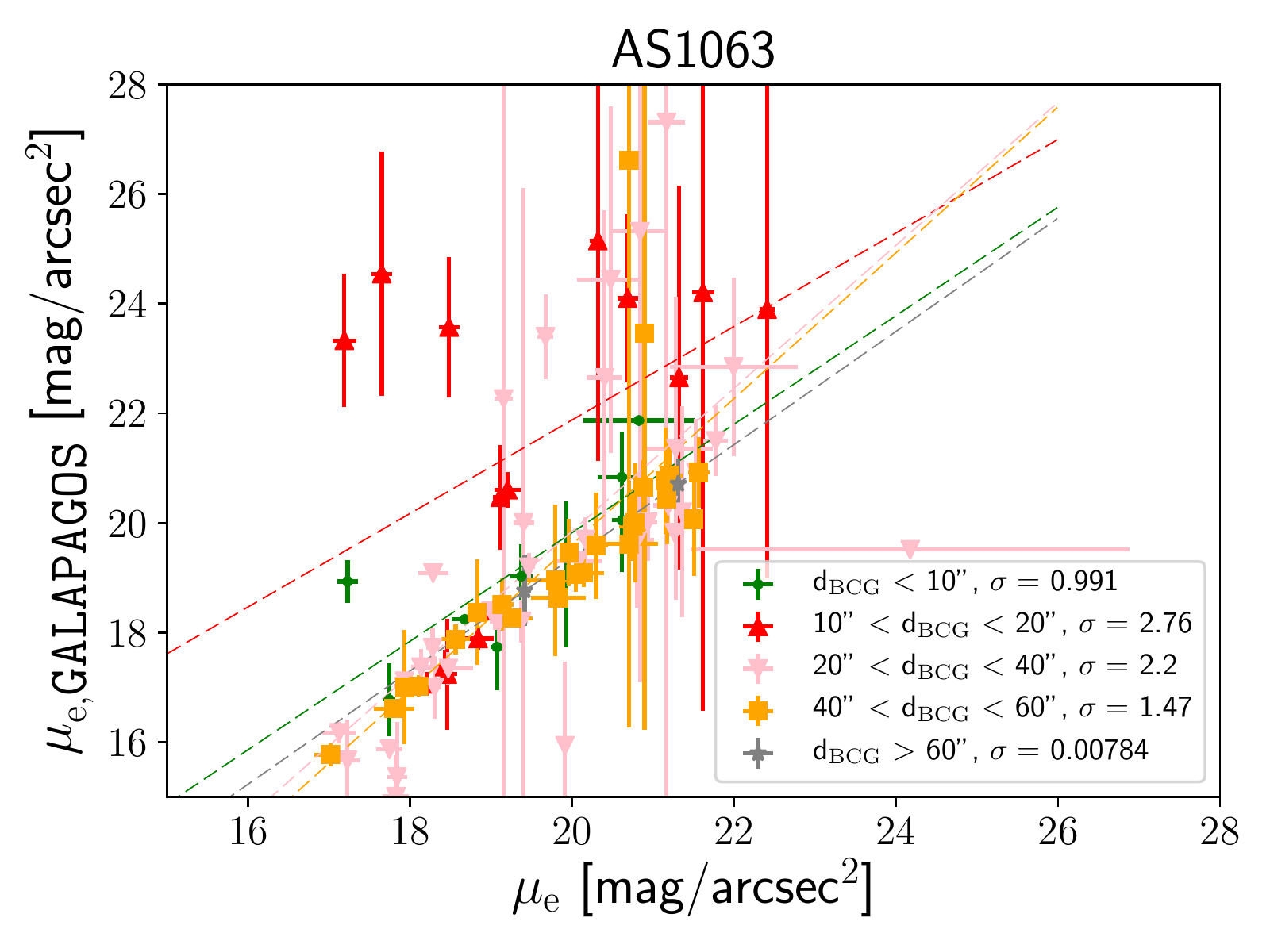}
\includegraphics[scale=0.5]{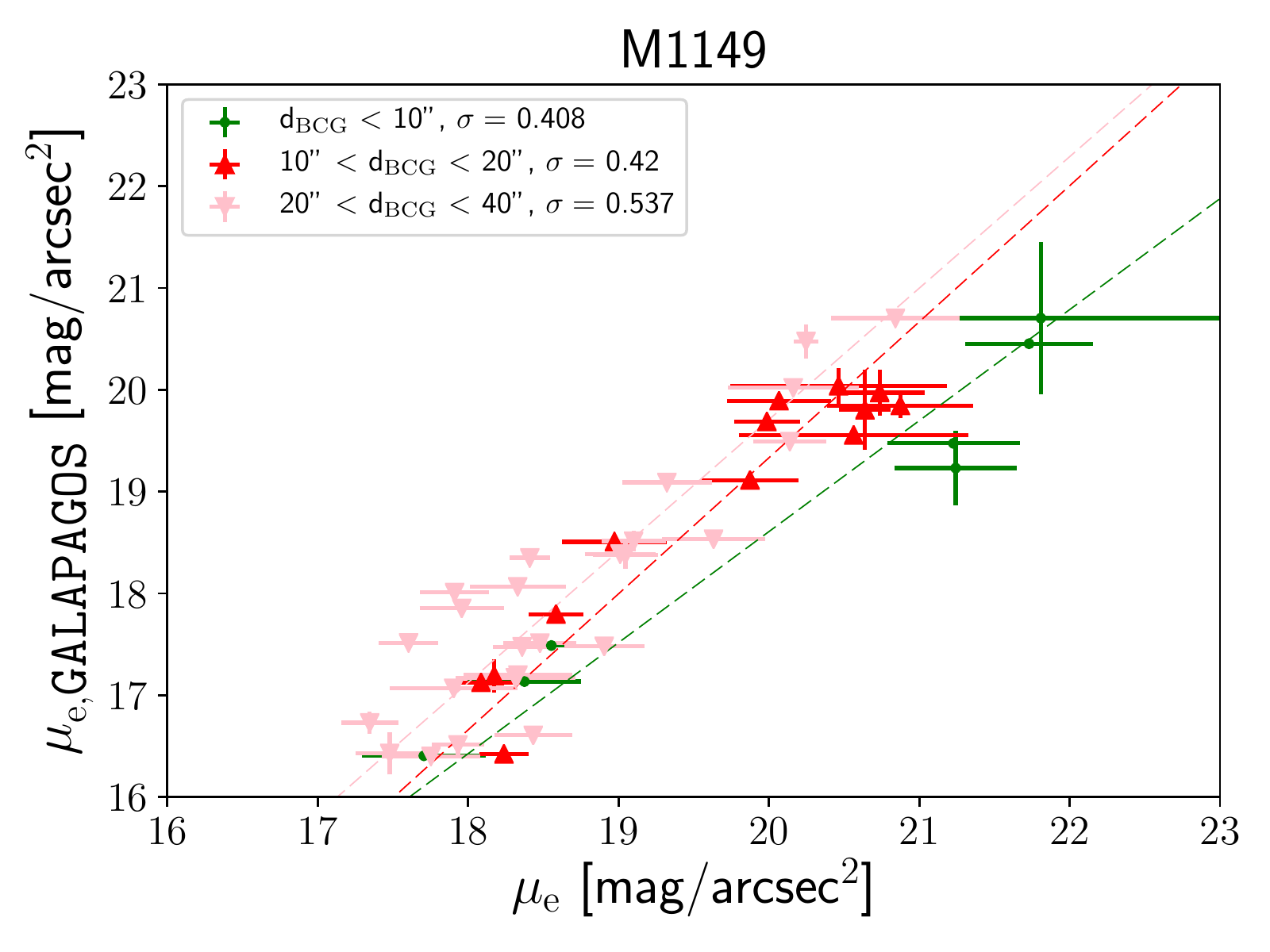}
\caption{Comparisons of R$_{\mathrm{e}}$ and $\left \langle \mu \right \rangle_{\mathrm{e}}$ between a \texttt{GALAPAGOS} blind run and one with our new methodology are shown. Left panels refer to AS1063, while right panels to M1149. Green circles, red up triangles, pink down triangles, orange squares and grey stars refer to galaxies having different projected distances (see legend), from less than 10 arcsec to more than 60 arcsec from the BCGs. Dashed lines represent the best-fitting linear relations between the two parameters, colour-coded by the distance from the BCG. Plot legends show also the scatter of the points from each corresponding best-fitting linear relation.}
\label{fig4}
\end{figure*}

\subsection{Final Estimates of the Structural Parameters}

\begin{figure*}
\centering
\includegraphics[scale=0.172]{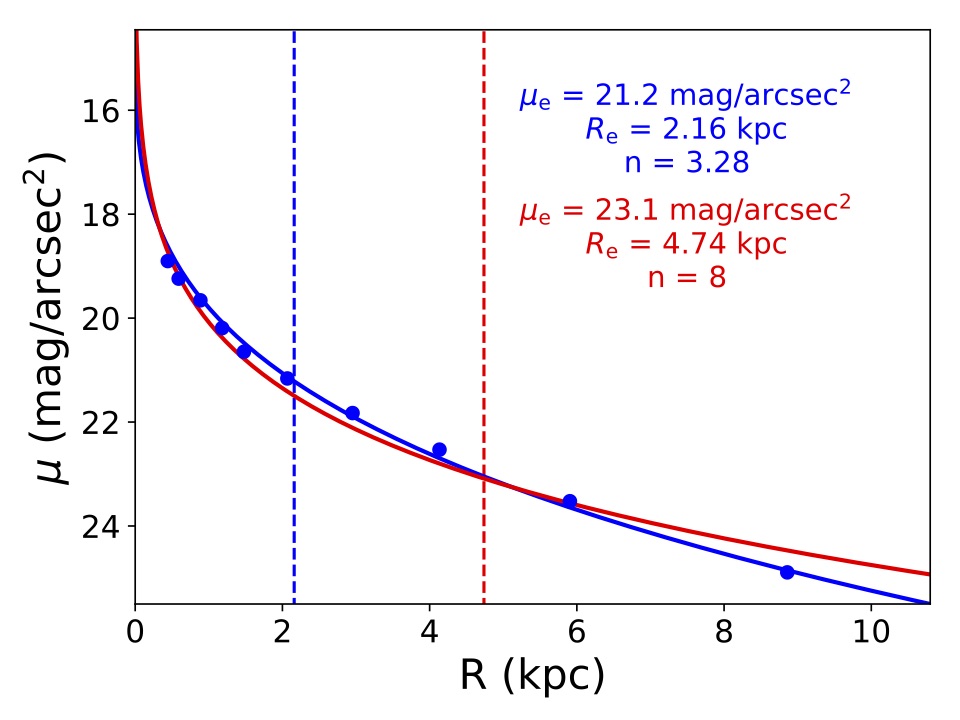}
\includegraphics[scale=0.172]{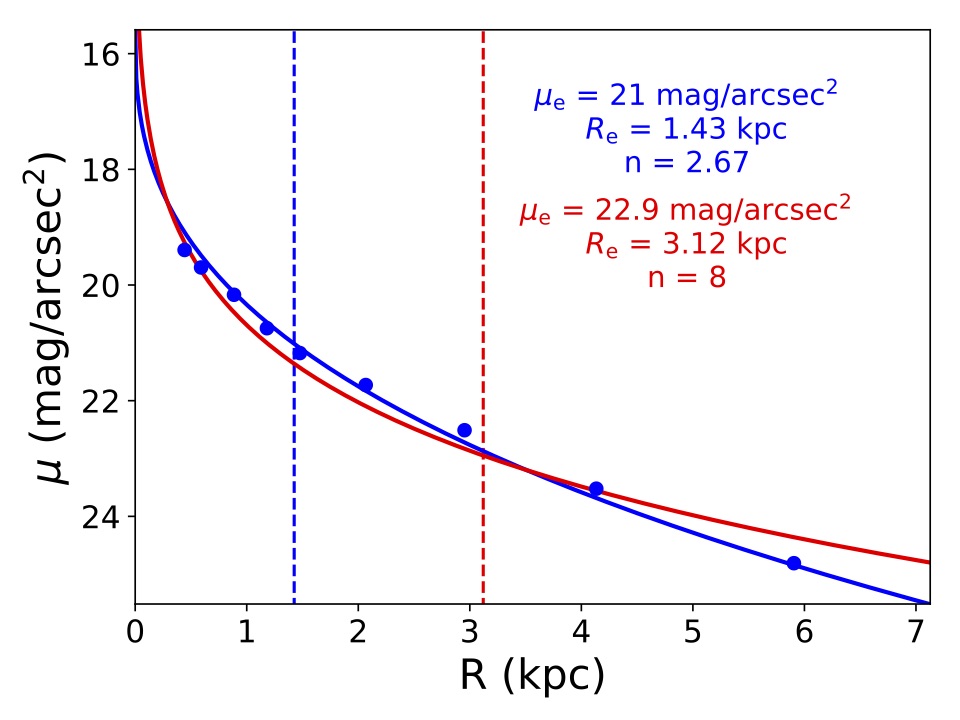}
\includegraphics[scale=0.172]{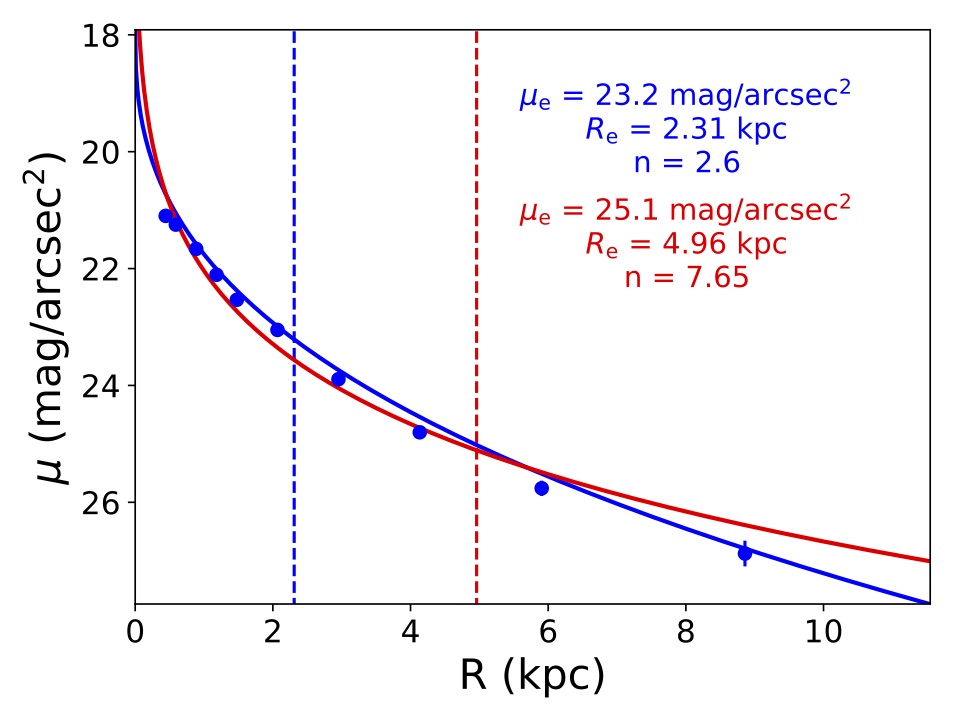}
\includegraphics[scale=0.172]{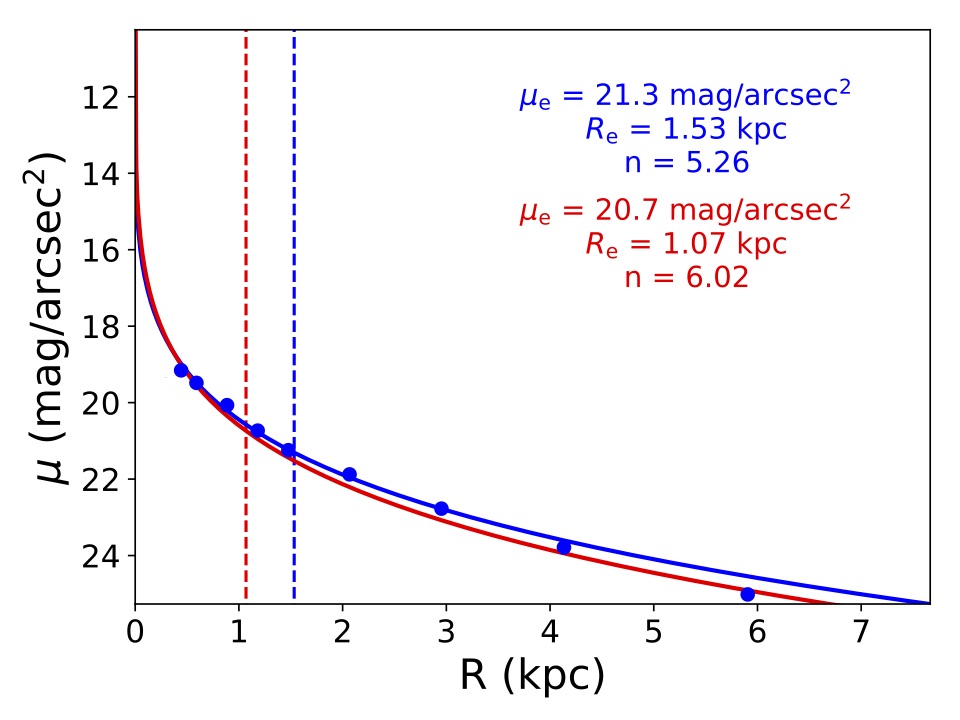}
\includegraphics[scale=0.172]{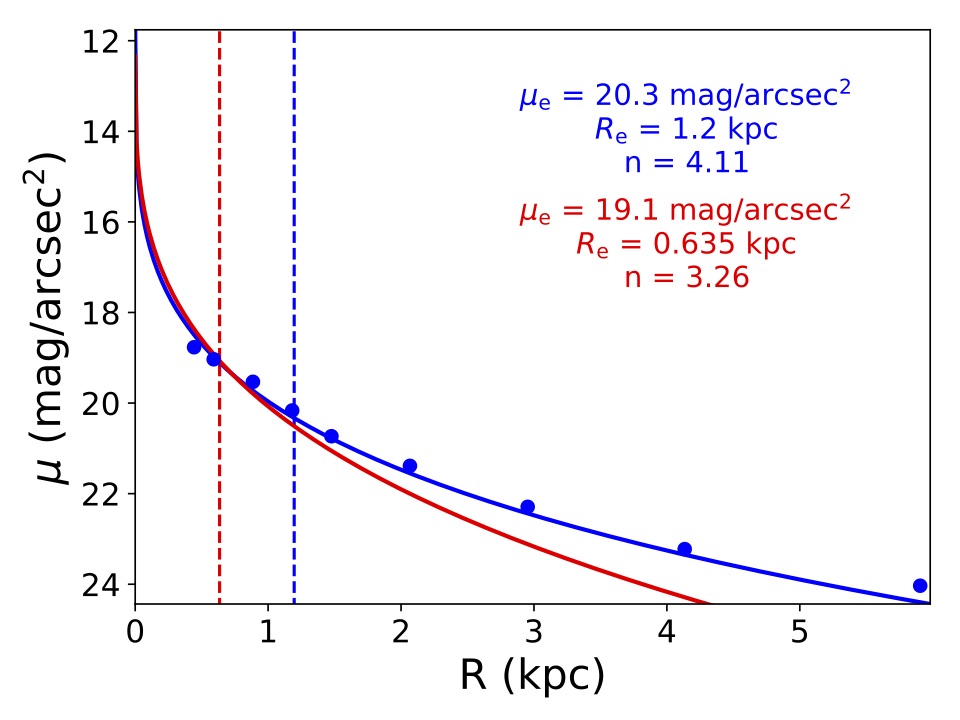}
\includegraphics[scale=0.172]{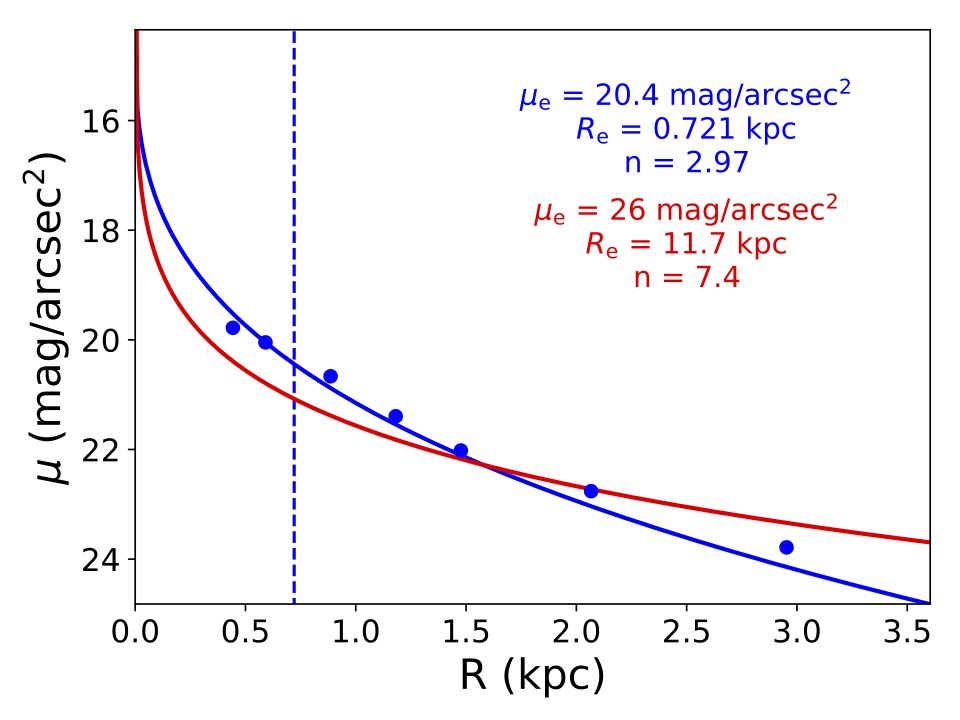}
\caption{S\'ersic profiles of the 6 galaxies showing the highest scatter with respect to the best-fitting linear relations in Fig. \ref{fig4} are shown. Blue solid lines are the S\'ersic profiles built with the best-fitting parameters we estimate in Sect. 3.3. Red solid lines are the S\'ersic profiles built with the \texttt{GALAPAGOS} run best-fitting parameters. Blue points are aperture magnitudes in the \textit{F814W} waveband. Colour-coded dashed lines mark the estimated effective radii.  On the top right of each plot, the best-fitting parameters colour-coded by the method are shown.}
\label{fig5}
\end{figure*}

For every individual galaxy, we have 4 different measurements of the structural parameters. Due to the small number statistics, we use the classical standard deviation as the most reliable scale estimator \citep{Beers1990}. Following this approach, we use the mean of the total magnitude m$_{\mathrm{tot}}$ in the \textit{F814W} band, of the mean surface brightness $\left \langle \mu \right \rangle_{\mathrm{e}}$, of the effective radius in kpc (R$_\mathrm{e}$) and of the S\'ersic index n, for each spectroscopically confirmed cluster galaxy, as best estimates of the structural parameters. The errors on the latter, instead, are estimated using the standard deviation, since the formal errors computed by \texttt{GALFIT} account only for statistical uncertainties in the flux \citep{Peng2011}.  We report in Table \ref{appendixa1} and \ref{appendixa2} in Appendix A the final structural parameters for all the spectroscopically confirmed members of both clusters (both below and above 22.5 ABmag in the \textit{F814W} waveband). 

We also visually inspect the residuals of the fit after subtracting the model galaxy. We find that two galaxies (ID 64 in AS1063 and ID 25 in M1149) show very complex morphologies (they are spatially overlapping with a combination of gravitational arcs) and that the S\'ersic fit does not converge for one galaxy (ID 16 in AS1063). We also find that these three galaxies are also those which have relative errors greater than 30 per cent in effective radius. So, we exclude them from the final sample analysis. For the remaining sample, the average relative uncertainties for R$_{\mathrm{e}}$ are below 15 per cent for 95 (84) per cent of AS1063 (M1149) cluster members, while, for $\left \langle \mu \right \rangle_{\mathrm{e}}$, all galaxies have uncertainties below 15 per cent in both clusters. The typical error on the effective radius R$_\mathrm{e}$ (for galaxies with average relative uncertainties below 15 per cent) is $0.09$ kpc for AS1063 and 0.27 kpc for M1149, while, on the surface brightness within R$_\mathrm{e}$, $\left \langle \mu \right \rangle_{\mathrm{e}}$, they are $0.19$ and $0.26$ mag arcsec$^{-2}$ for AS1063 and M1149, respectively. These values and the relative uncertainties for galaxies up to the completeness magnitude limit are used to derive the KR (Sect. 5).

We show in Fig. \ref{fig3} a typical example of a spectrophotometrically selected ETG (\textit{HST} image in the top left panel). The model of the galaxy (top central panel) is accurate as demonstrated by the residual image (top right panel).
The structural parameters are reported in the central panel: aperture magnitudes centred on the galaxy are very well fitted by the mean S\'ersic profile computed with the best-fitting structural parameters.

In Fig. \ref{fig4}, we compare the estimates of R$_{\mathrm{e}}$ and $\left \langle \mu \right \rangle_{\mathrm{e}}$ for both clusters, obtained from our procedure, with those obtained with a blind run of \texttt{GALAPAGOS}, as a function of the distance from the BCGs. \texttt{GALAPAGOS} errors refer to those computed with its internal \texttt{GALFIT} run, while errors on the other axis are the ones described above. We show the best-fitting linear relations between \texttt{GALAPAGOS} and the best estimates of the structural parameters as dashed lines, colour-coded according to the distance from the BCG.

In Fig. \ref{fig4}, we observe a different behaviour between the two clusters. The left panels show that in AS1063 the largest scatter can be found for sources having projected distance from the BCG between 10\arcsec\ and 20\arcsec\ (red up triangles). This could be due to the fact that \texttt{GALAPAGOS} fit in small stamps is strongly dependent on the different radial contributions of the BCG diffuse component. Indeed, for sources close to the BCG (< 10\arcsec, green points), the background is dominated by its diffuse component, which uniformely covers the stamp without any significant gradient. Therefore, \texttt{GALAPAGOS} is able to correctly determine the size of the object. From 10\arcsec\ on, the contribution of the BCG diffuse component starts to radially decrease. Therefore, the background value is not uniform throughout the stamp. This has the effect of artificially enlarge the effective radius of the fitted galaxy. The latter effect tends to decrease for sources farther than 20\arcsec\ from the BCG (pink down triangles, orange squares and grey points). The right panels show that in M1149 the scatter is lower in absolute value with respect to AS1063, but similar in each projected distance range. Indeed, in this cluster the contamination from the diffuse component of the BCG is less dramatic with respect to AS1063. This suggests that contamination may vary from cluster to cluster and so, in turns, the performance of \texttt{GALAPAGOS} in crowded environments.

We show in Fig. \ref{fig5} an example of the comparison between the best-fitting S\'ersic profiles obtained with the final measurements (blue solid lines) and the \texttt{GALAPAGOS} run (red solid lines). The 6 panels refer to the 6 galaxies (red triangles in the top left panel of Fig. \ref{fig4}) which show the highest scatter with respect to the best-fitting linear relations (dashed lines in Fig. \ref{fig4}). The observed surface brightness profiles (blue points) are very well fitted by the best-fitting S\'ersic profiles obtained with the final measurements. On the contrary, \texttt{GALAPAGOS} light profiles do not agree with the blue points, showing a large discrepancy of the best-fitting parameters with those obtained with our methodology. 

Other than that, the \texttt{GALAPAGOS} run provides a converging fit only for 91 out of 95 and 50 out of 68 galaxies (see Sect. 2) in AS1063 and M1149, respectively. We also try to enlarge the dimension of the stamps defined by \texttt{GALAPAGOS} to try to improve the fit. Nonetheless, we were not able to obtain reliable results on structural parameters, in particular close to the BCG, thus suggesting that the fit with \texttt{GALAPAGOS} could be tricky in very crowded fields.

\section{Galaxy Samples}

\begin{table*}
\centering
\begin{tabular}{l|c|c|c|c|c}
\hline
\hline
Cluster&Early-type&Ellipticals&Red galaxies (at $1 \sigma$) & Red galaxies (at $3 \sigma$)&Passive \\
\hline
Abell S1063& 37 (41) & 35 (39)& 41 (50) & 46 (57) & 37 (42)\\
MACSJ1149.5+2223& 36 (51)& 32 (47)& 32 (45) & 39 (57)& 29 (29)\\
\hline
\hline
\end{tabular}
\caption{We report the number of cluster members in the sample of early-type (col. 2), ellipticals (col. 3), red at $1 \sigma$ (col. 4) and $3 \sigma$ (col. 5) from the best-fitting CM relation and passive (col. 6), with m $\le$ 22.5 ABmag in the \textit{F814W} waveband. The total number of galaxies of each sample is reported in parenthesis.}
\label{table1}
\end{table*}

As previously mentioned, the KR defines an observational correlation between effective radius and surface brightness of ETGs. The latter are classified in many different ways: through morphologies, colours, spectra, S\'ersic indices. We analyze 4 of them, the ones that are more versatile and robust for the widest range in redshift. In this section, we define and compare samples selected according to to different galaxy properties, in order to investigate the impact of the sample selection on the KR (Sect. 5).

\begin{figure*}
\centering
\includegraphics[scale=0.37]{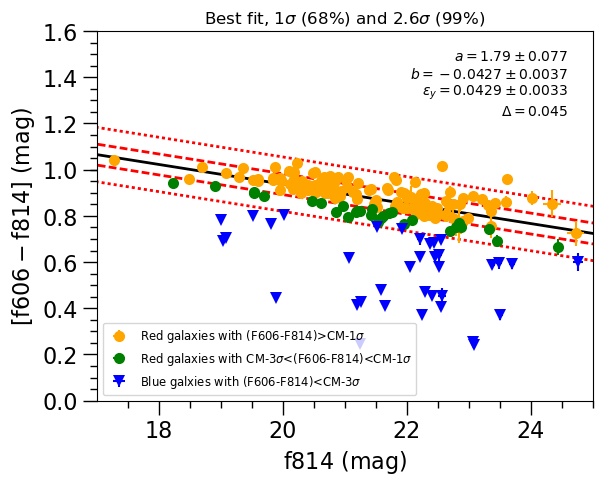}
\includegraphics[scale=0.37]{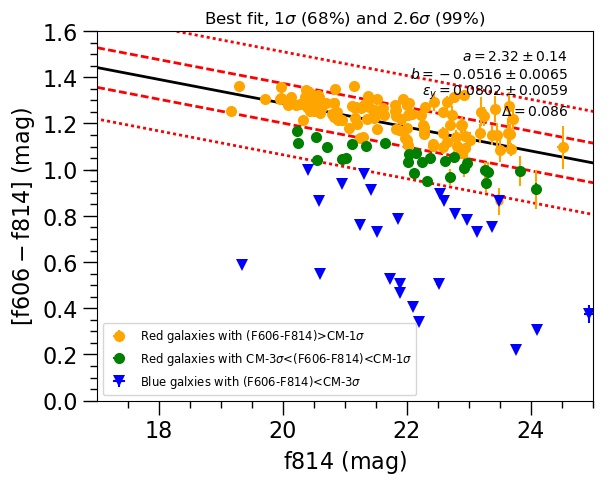}
\caption{The CM diagram and the best-fitting relation for the clusters AS1063 (left panel) and M1149 (right panel) are shown. The coefficients of the best-fitting CM relation (b for the slope, a for the zero-point) and the corresponding observed scatter $\Delta$ in y are reported at the top left of the plots. The two red dashed and dotted lines mark the $1 \sigma$ bands (enclosing 68 per cent of the values for a Gaussian distribution) and $2.6 \sigma$ (99 per cent), respectively. Orange points represent red galaxies with colour \textit{F606W} - \textit{F814W} greater than the best-fitting CM minus $1 \sigma$ from the relation, green points represent red galaxies with colour between the best-fitting CM minus $1 \sigma$ and $3 \sigma$ from the relation and blue points represent blue galaxies, i.e. with colour lower than the best-fitting CM minus $3 \sigma$ from the relation. Galaxies shown in these figures are all the spectroscopically confirmed members in \textit{HST} images for which we have photometric informations, both with and without MUSE spectra.}
\label{fig6}
\centering
\end{figure*}

\begin{figure}
\centering
\includegraphics[scale=0.25]{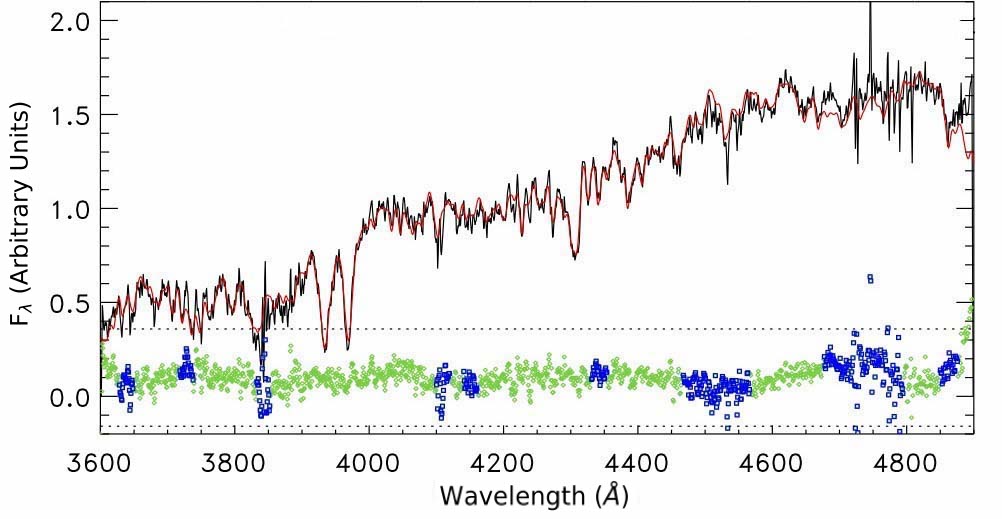}
\caption{Rest-frame spectrum of the spectrophotometrically selected ETG of Fig. \ref{fig3} is shown. Green points are the residuals between the observed spectrum (black solid line) and the model (red solid line) obtained using MILES stellar population library \citep{MILES2011}. Blue points are regions excluded from the fit due to the contribution of sky lines in the observed spectrum.}
\label{fig7}
\end{figure}

\subsection{Samples Classification}

\begin{figure*}
\centering
\includegraphics[scale=0.54]{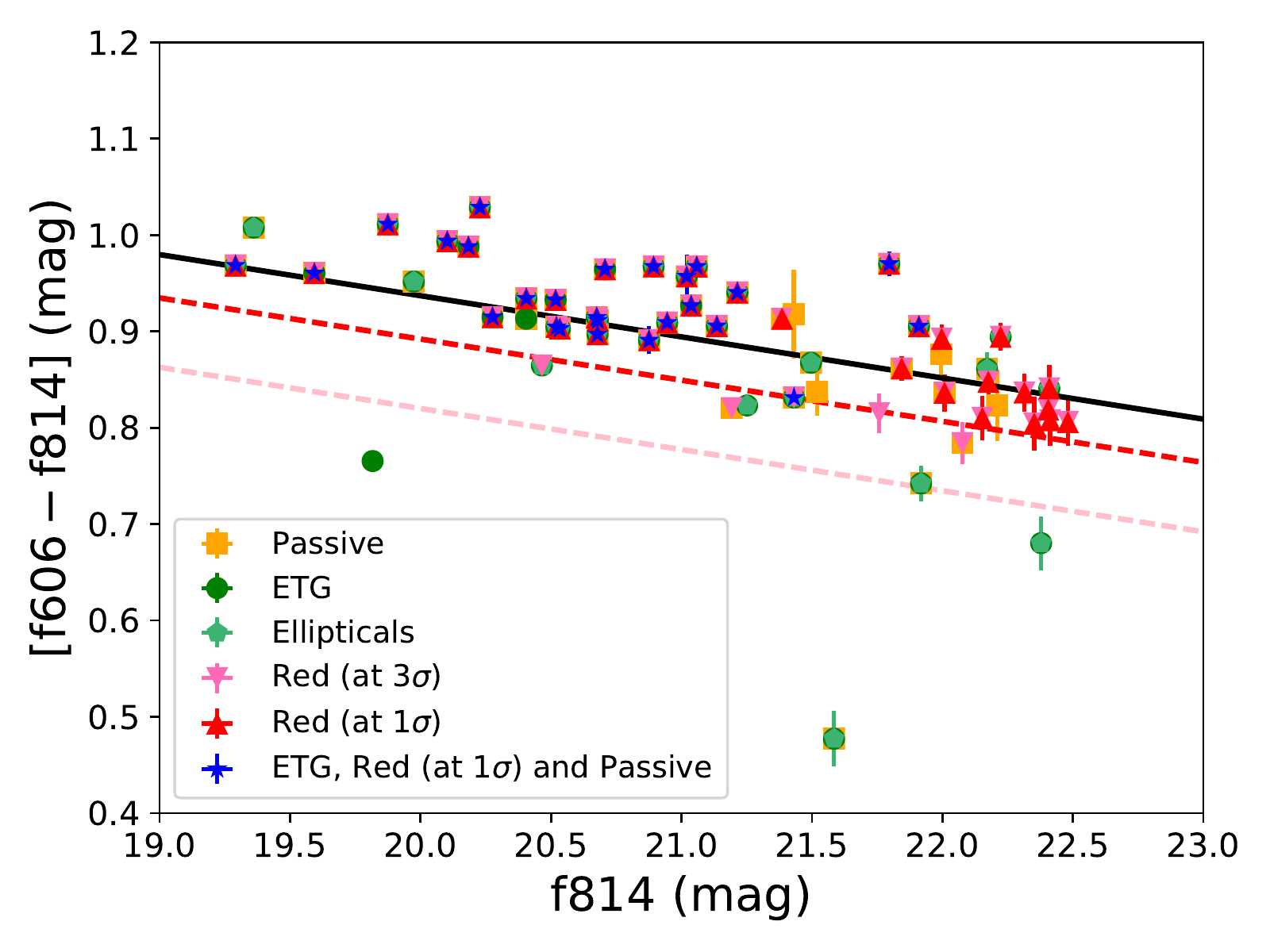}
\includegraphics[scale=0.54]{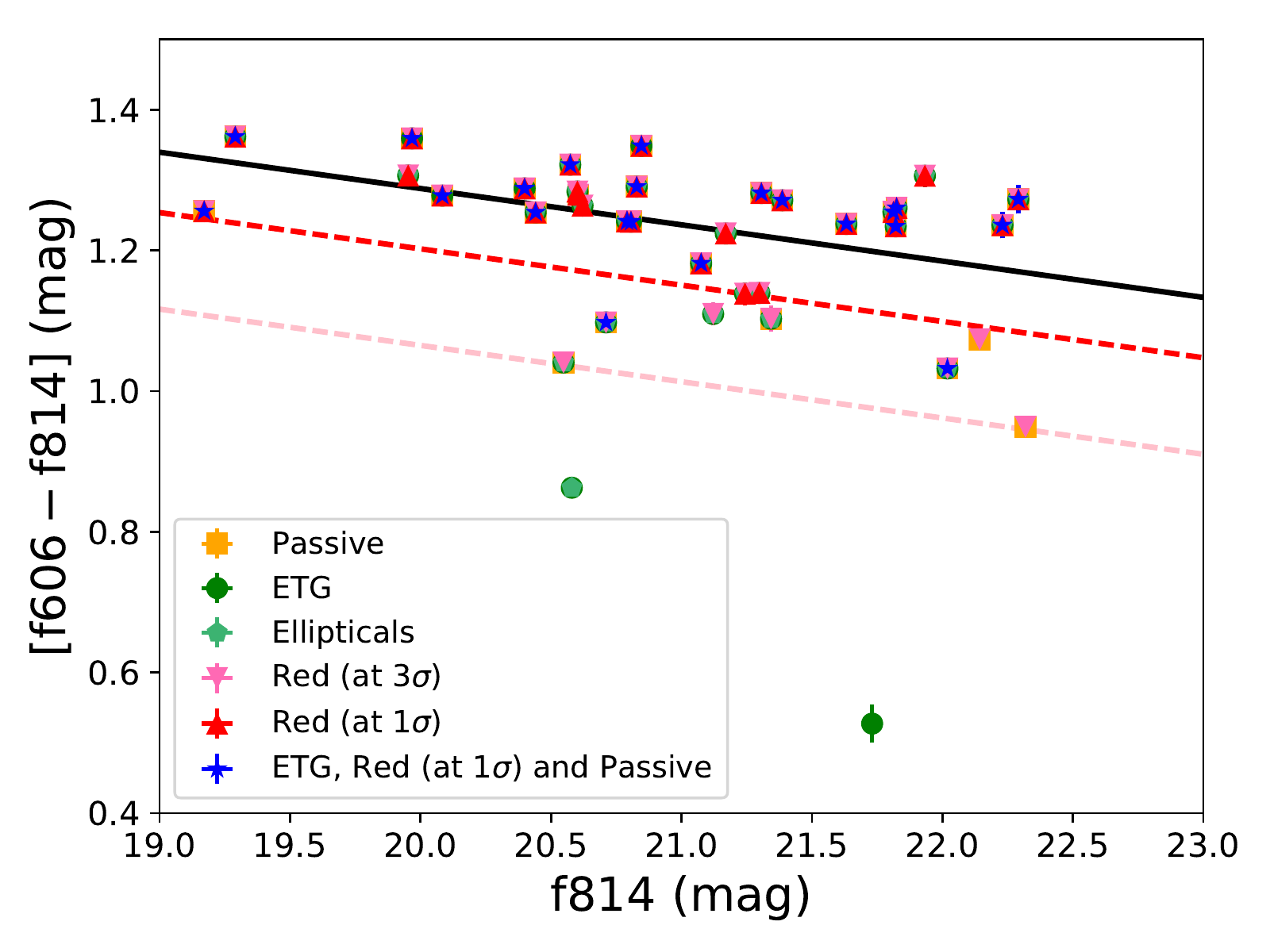}
\caption{The CM diagram and the best-fitting relation for the clusters AS1063 (left panel) and M1149 (right panel) for all samples are shown. The black solid lines are the best-fitting CM relations shown in Fig. \ref{fig6}. The red and pink dashed lines mark the $1 \sigma$ and $3 \sigma$ thresholds, respectively. Green circles, sea green pentagons, pink down triangles, red up triangles, orange squares and blue stars are ETG, ellipticals, red (at 3$\sigma$), red (at 1$\sigma$), passive and spectrophotometric selected samples, respectively.}
\label{fig8}
\end{figure*}

We define the KR for four different samples of spectroscopically confirmed cluster members, which have magnitudes brighter than the adopted completeness limit. We select galaxies according to their surface brightness profiles, visual inspection, their positions on the colour-magnitude diagram and their spectra in:
\begin{itemize}
\item \textit{ETG}: we fit the light profile of each galaxy with a S\'ersic law. We define as ETGs, those galaxies which have S\'ersic indices n $\ge$ 2.5 (e.g., \citealt{Blanton2003, Bell2004, McIntosh2005}). $37$ galaxies are selected as ETGs for AS1063, while $36$ for M1149 (column 2 of Table \ref{table1}).
\item \textit{Ellipticals}: we perform  a morphological classification based on the visual inspection of the galaxy images and the residuals after the best-fitting model subtraction, infact, as demonstrated in literature (e.g., \citealt{Mei2012,Tamburri2014}), genuine morphologically selected ellipticals can show S\'ersic indeces n < 2.5. 35 galaxies are morphologically selected ellipticals for AS1063, while 32 for M1149 (column 3 of Table \ref{table1}).
\item \textit{Red}: we use the Colour-Magnitude (CM) relation to separate red and blue cluster members. The Kron magnitudes in the \textit{F814W} band are used as estimates of the total galaxy magnitudes, while aperture magnitudes in the \textit{F606W} and \textit{F814W} bands are used to derive colours. The CM relation is built with all the spectroscopically confirmed member galaxies for which we have photometric informations, both with and without MUSE spectra. We show in Fig. \ref{fig6} the CM diagram for AS1063 (left panel) and M1149 (right panel). The fit to the relation is obtained using the Least Trimmed Squares (LTS) technique implemented by \citet{Cappellari2013}. Fig. \ref{fig6} shows the red-sequence with measured scatter of $0.043\pm0.003$ for AS1063 and $0.080\pm0.006$ for M1149. We consider two samples of red galaxies: one containing those which are redder than the best-fitting CM relation minus $1 \sigma$ (hereafter, red at 1$\sigma$; orange points in Fig. \ref{fig6}) and another one containing those which are redder than the best-fitting CM relation minus $3 \sigma$ (hereafter, red at 3$\sigma$; green+orange points in Fig. \ref{fig6}), where the $1 \sigma$ band from the best-fitting relation encloses 68 per cent of the values for a Gaussian distribution \citep{Cappellari2013}. 41 and 32 galaxies are selected as red at 1 $\sigma$ for AS1063 and M1149, respectively, while 46 and 39 as red at 3 $\sigma$ (columns 4 and 5 of Table \ref{table1}).
\item \textit{Passive}: an useful tool to classify passively evolving galaxies is the analysis of the presence of emission lines in their spectra plus the investigation of the equivalent width of the H$\delta$ absorption line (EW(H$\delta$), e.g., \citealt{Worthey1994}). To select passive galaxies, we perform a two step analysis: first, we run a full spectral fitting of MUSE spectra with the \texttt{pPXF} code \citep{Cappellari2004,Cappellari2017} using MILES stellar population library \citep{MILES2011}. We check the residuals and we select, for the next step, only those galaxies in which the residuals of the emission lines are lower than the variance of the spectrum (dashed black lines in Fig. \ref{fig7}). In the second step, we measure the EW of the H$\delta$ absorption line among the galaxies selected in the previous step. Thus, we define as passive those galaxies with no detectable emission lines and EW(H$\delta$) < 3 \AA\ \citep{Mercurio2004}. These criteria are chosen to select galaxies without signs of ongoing or recent star-formation. The adopted criteria also allows us to exclude the presence of bright AGNs that would modify the nuclear profile. Very Low Luminosity or obscured AGNs could still be present in our sample, but those will not contribute significantly to the galaxy light. 37 and 29 galaxies are classified as passive for AS1063 and M1149, respectively (column 6 of Table \ref{table1}).
\end{itemize}

Sometimes in literature the passivity criterion, which distinguish between star-forming and passive galaxies, is based also on the specific star-formation rate, i.e. the star-formation rate per unit stellar mass, sSFR = SFR/M$_{*}$. In order to compare our selection method with that based on sSFR only, we run the SED fitting code \texttt{MAGPHYS} \citep{dacunha2008} using the 16 CLASH HST wavebands available for the cluster members. We measure the total magnitudes using the \texttt{SExtractor} code. For two galaxies in AS1063 we are not able to measure their \textit{F225W}, \textit{F275W}, \textit{F336W} and \textit{F390W} magnitudes, so we do not fit these galaxies. The \texttt{MAGPHYS} procedure provides a measurement of the sSFR. Following \citealt{annunziatella2014} (and references therein) we can use the value sSFR = 10$^{-10}$ yr$^{-1}$ to identify the population of passive galaxies. We find that all the galaxies selected as passive using spectra are also passive according to the sSFR criterion. Only one passive galaxy in M1149 has $\log$(sSFR) = - 9.675 $\pm$ 0.05 yr$^{-1}$, which could be related to the presence of a close blue arc (see Sect. 4.2) that can contaminate the magnitude measurement. We also find one galaxy in each cluster that, despite having sSFR < 10$^{-10}$ yr$^{-1}$, shows signs of emission lines in its spectrum.

\begin{figure*}
\centering
\includegraphics[scale=0.435]{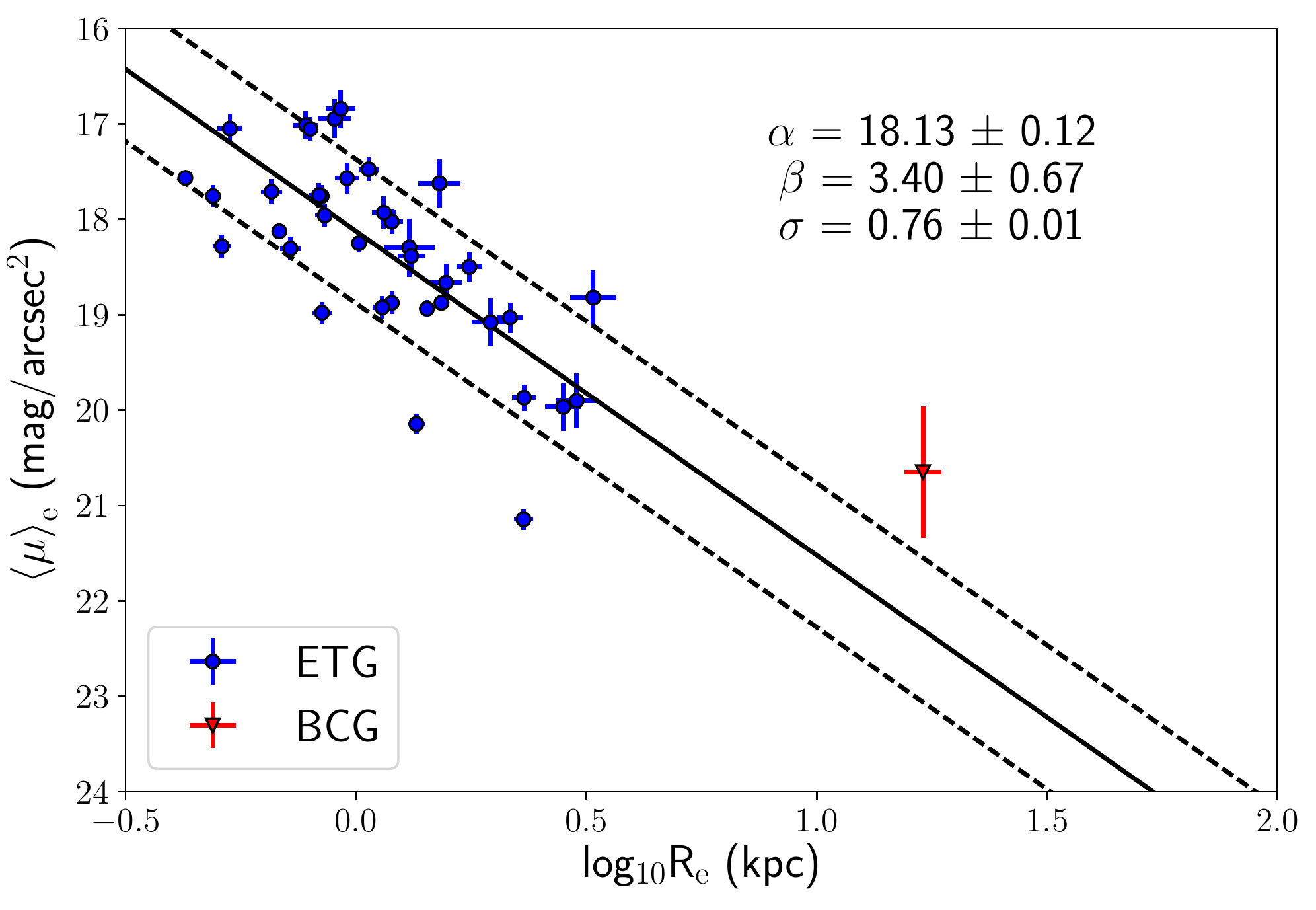}
\includegraphics[scale=0.435]{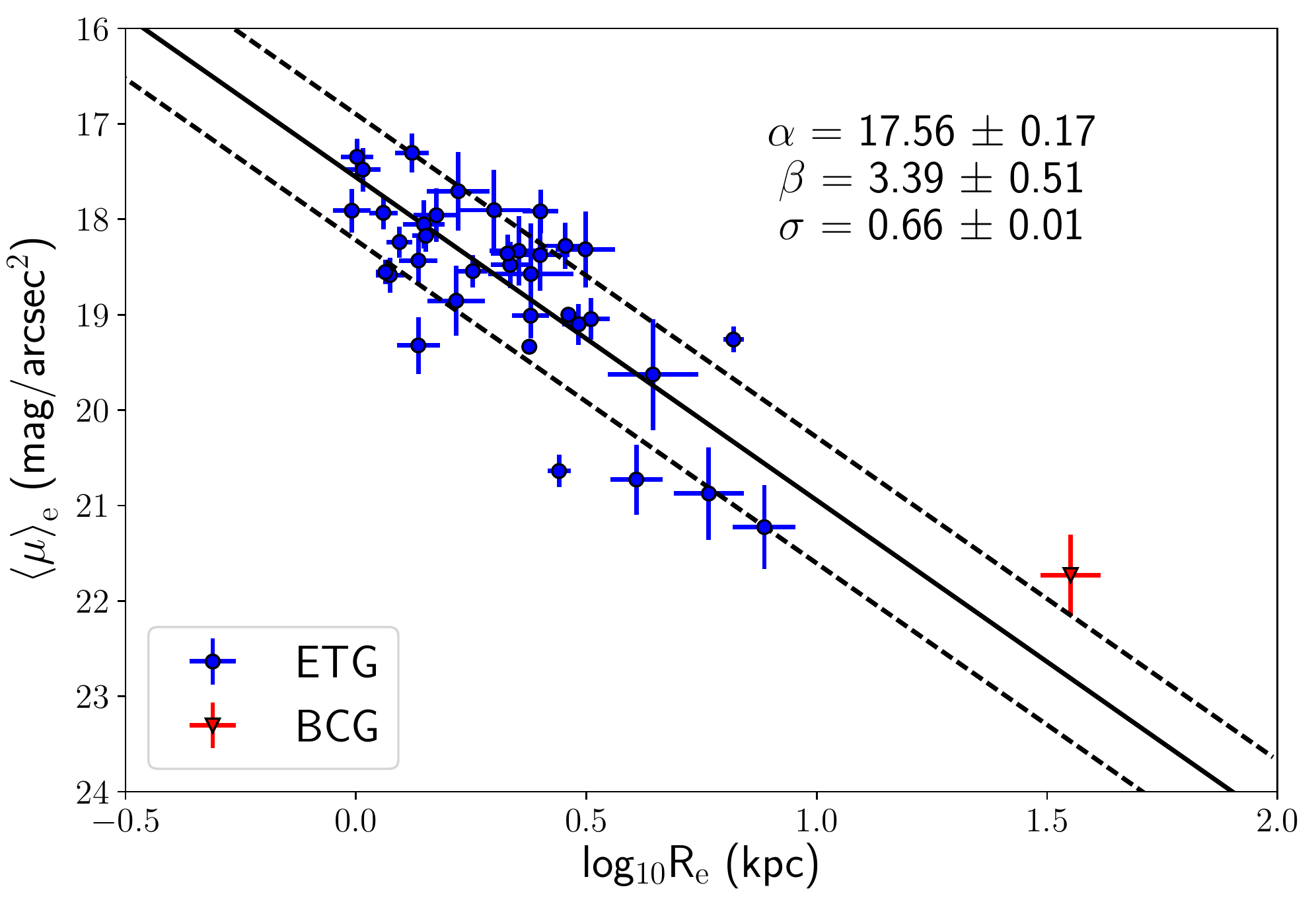}
\caption{The KRs obtained with the ETG samples for AS1063 (left panel) and M1149 (right panel) are shown. The coefficients of the best-fitting relation ($\beta$ for the slope, $\alpha$ for the zero-point) and the corresponding observed scatter $\sigma$ are shown at the top right of the plot. The solid black line represents the best-fitting KR. The two black dashed lines are the $1 \sigma$ limits of the relation. Blue points represent galaxies brighter than the adopted completeness limit (m$_{\mathrm{F814W}}$ $\le$ 22.5 ABmag), while the red triangle is the BCG of each cluster. This fit and the following ones with the other samples are performed without taking into account the BCGs, since it is difficult to determine their global light profile, being characterized by a diffuse and faint envelope.}
\label{fig9}
\end{figure*}

\begin{figure*}
\centering
\includegraphics[scale=0.435]{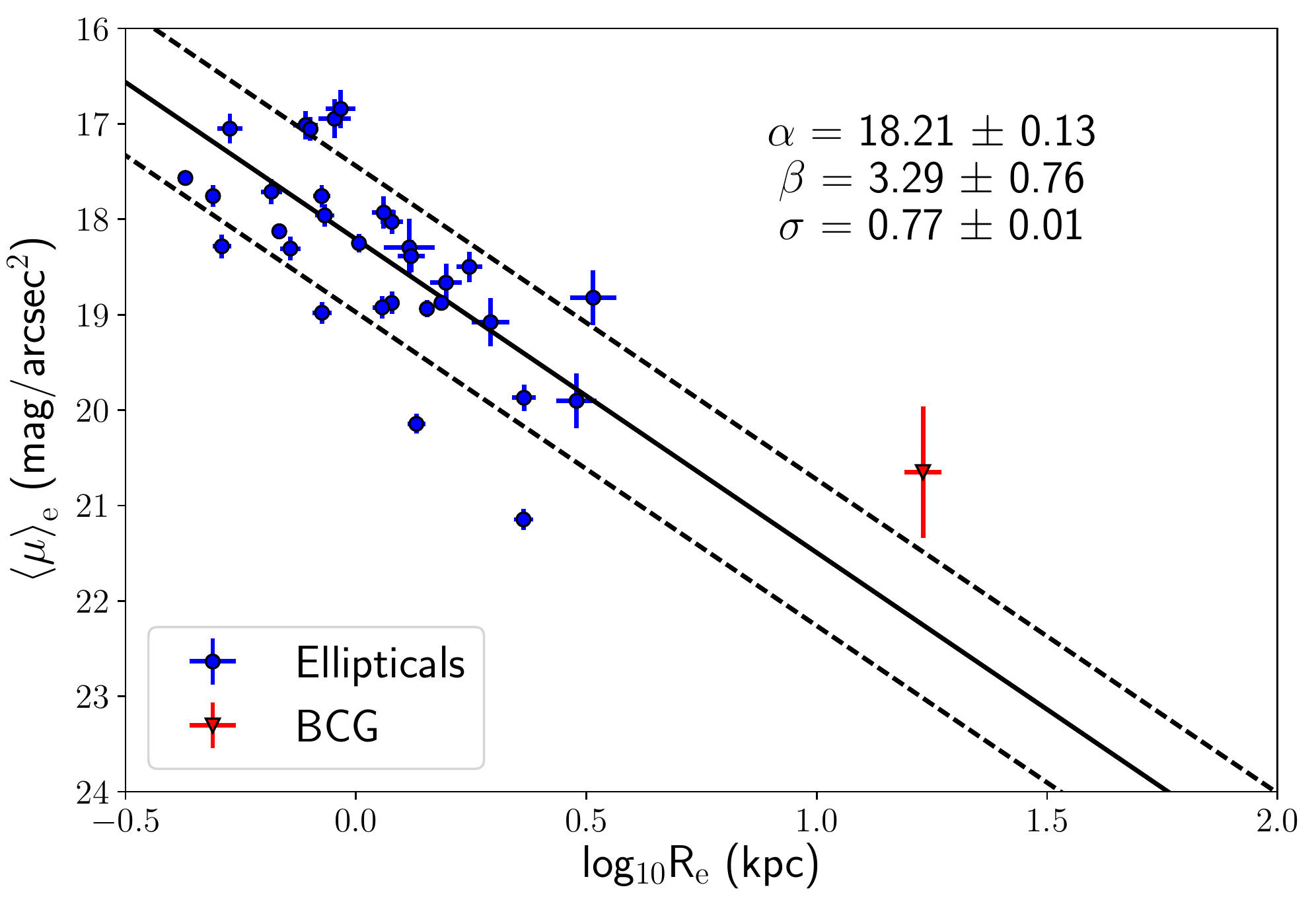}
\includegraphics[scale=0.435]{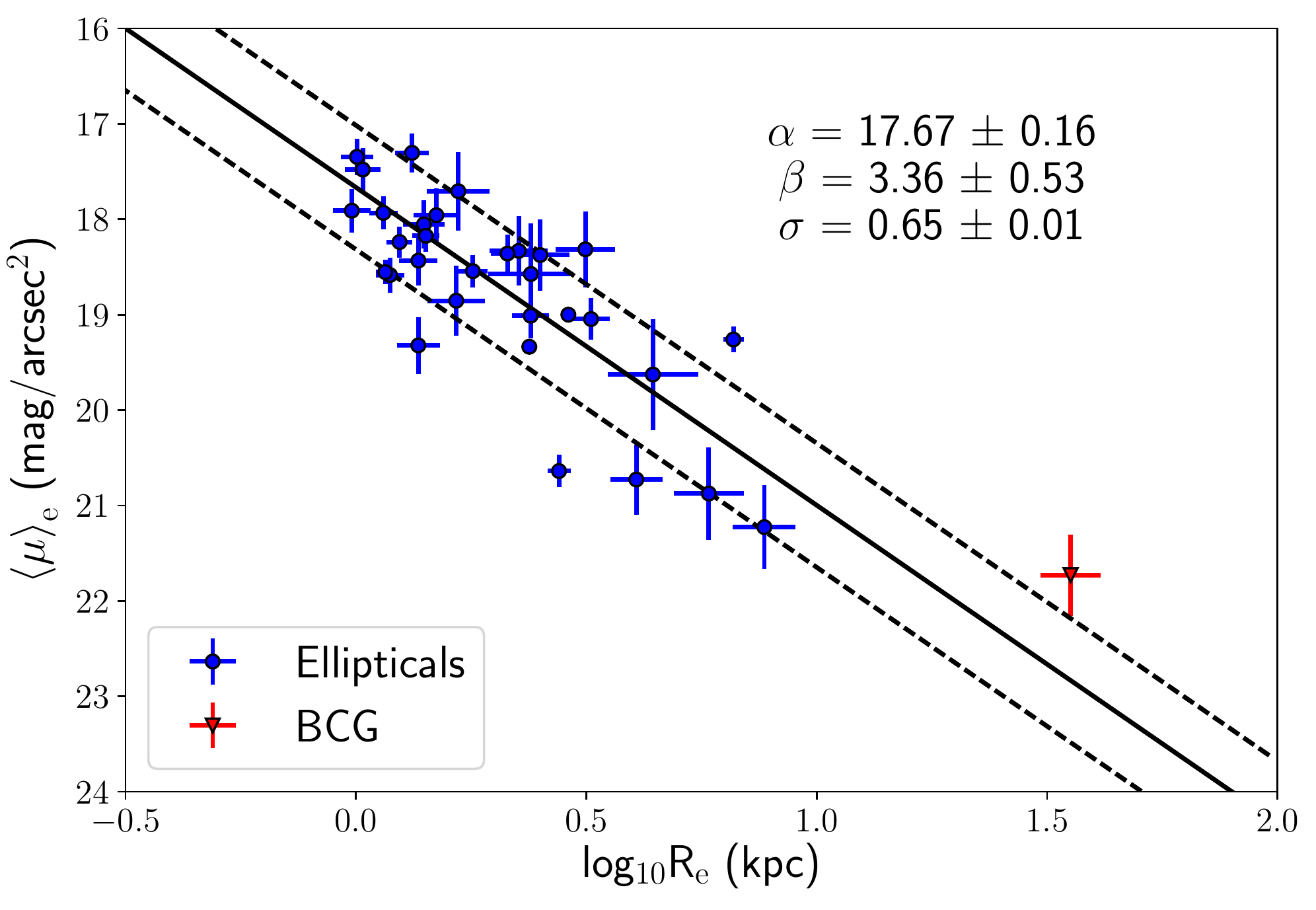}
\caption{The KRs obtained with the elliptical samples for AS1063 (left panel) and M1149 (right panel) are shown. The coefficients of the best-fitting relation ($\beta$ for the slope, $\alpha$ for the zero-point) and the corresponding observed scatter $\sigma$ are shown at the top right of the plot. The solid black line represents the best-fitting KR. The two black dashed lines are the $1 \sigma$ limits of the relation. Blue points represent galaxies with m$_{\mathrm{F814W}}$ $\le$ 22.5 ABmag, while the red triangle is the BCG of each cluster.}
\label{fig10}
\end{figure*}

We report in Fig. \ref{fig7}, as an example, the spectrum of the same galaxy shown in Fig. \ref{fig3}. The spectrum shows no emission lines, except at $\sim 4750$ \AA\, where residuals from the sky emission lines subtraction in the observed frame are visible (this region is excluded from the fit as highlighted by blue points in the lower panel of Fig. \ref{fig7}), and it shows the $4000$ \AA\ break, with a weak H$\delta$ absorption line (EW(H$\delta$) < 3 \AA). The S/N of the spectrum is $18.8$ and the measured stellar velocity dispersion is $\sigma_{*} = (192 \pm 9)$ km s$^{-1}$.

Considering the whole sample of 37 and 36 ETGs for AS1063 and M1149, 27 and 24 are also selected as ellipticals, red (at 1$\sigma$) and passive, respectively. We discuss in the next section the effects of these differences in the sample selection on the derivation of the KR.

\subsection{Sample Comparison}

\begin{figure*}
\centering
\includegraphics[scale=0.435]{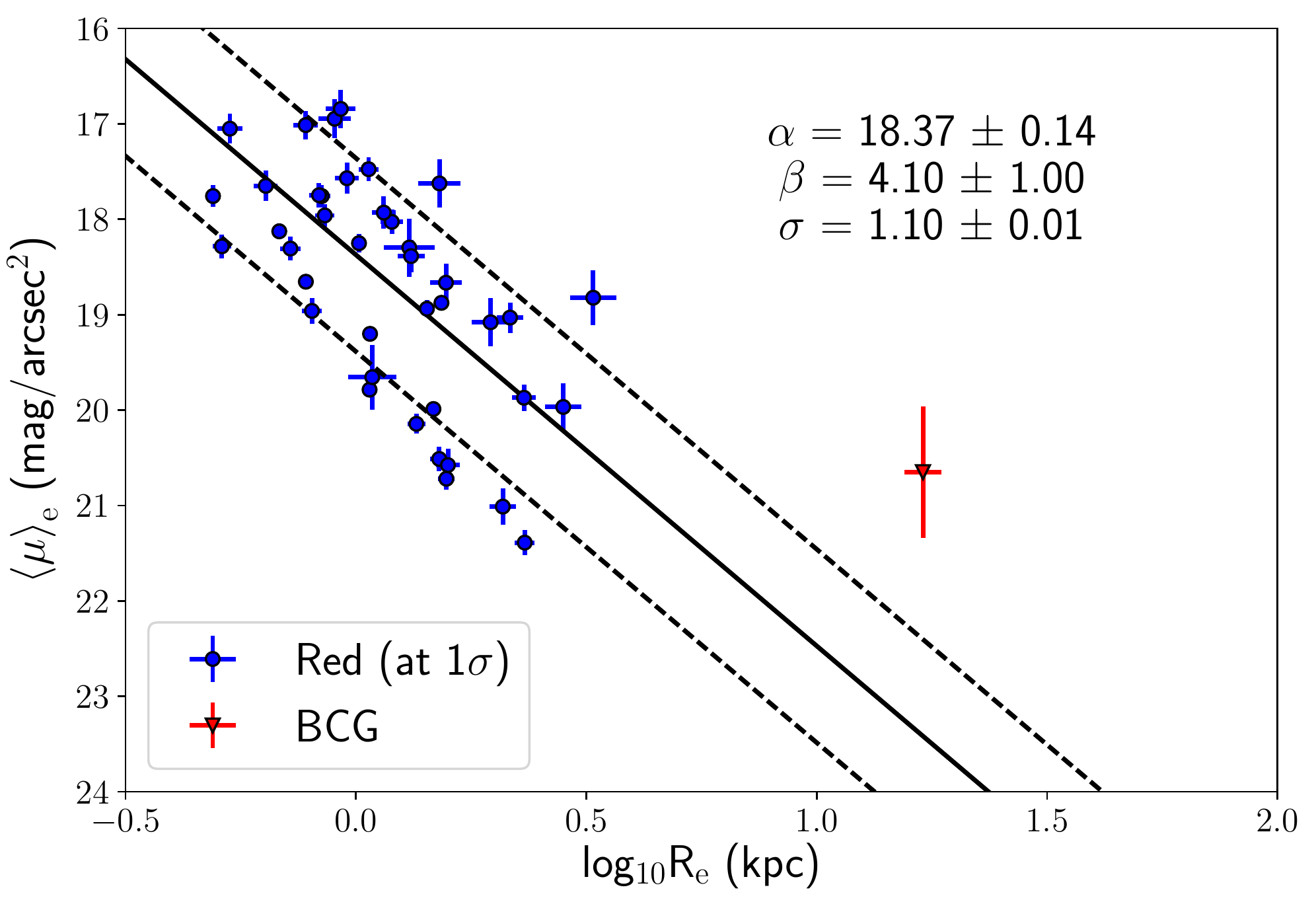}
\includegraphics[scale=0.435]{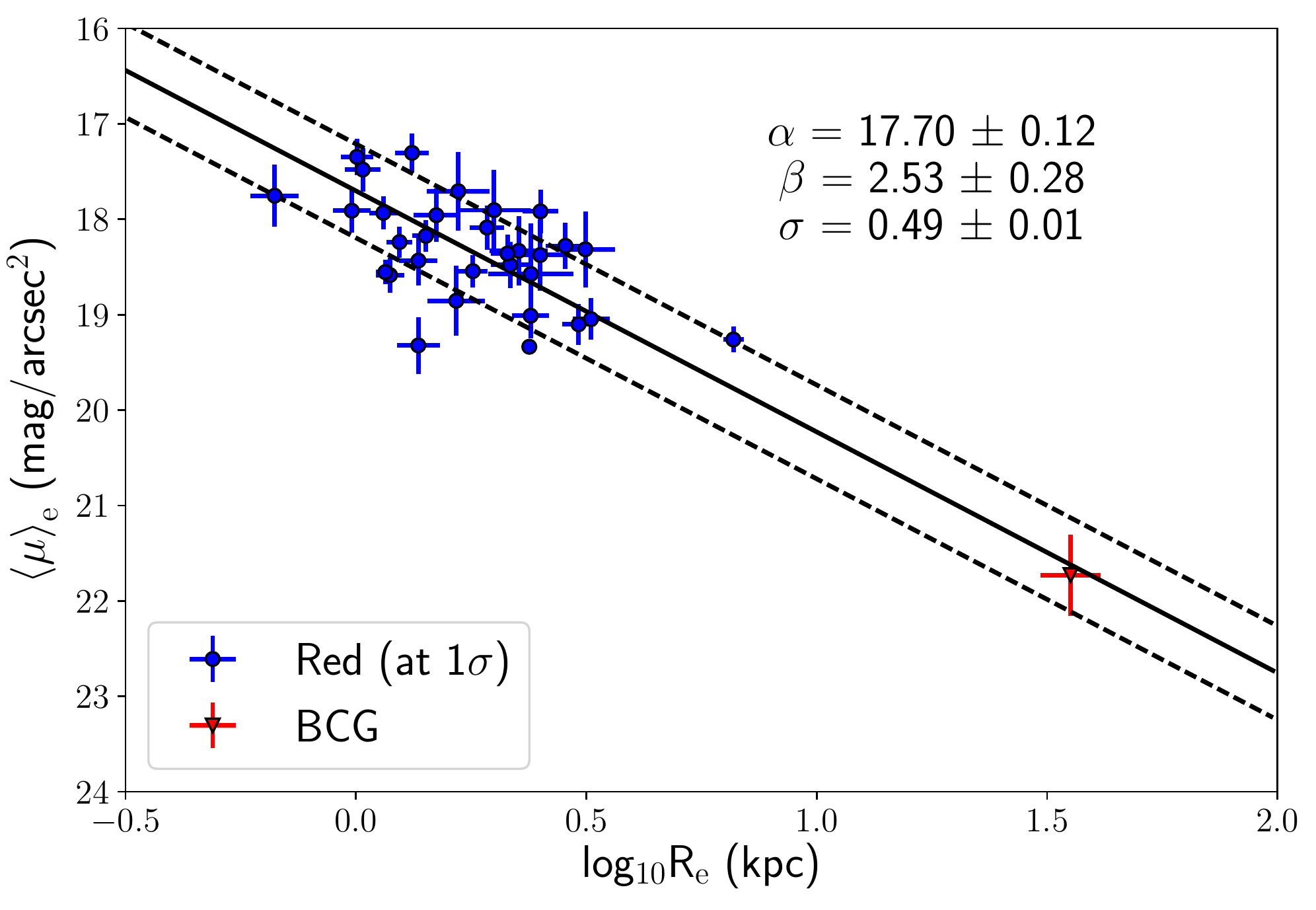}
\includegraphics[scale=0.435]{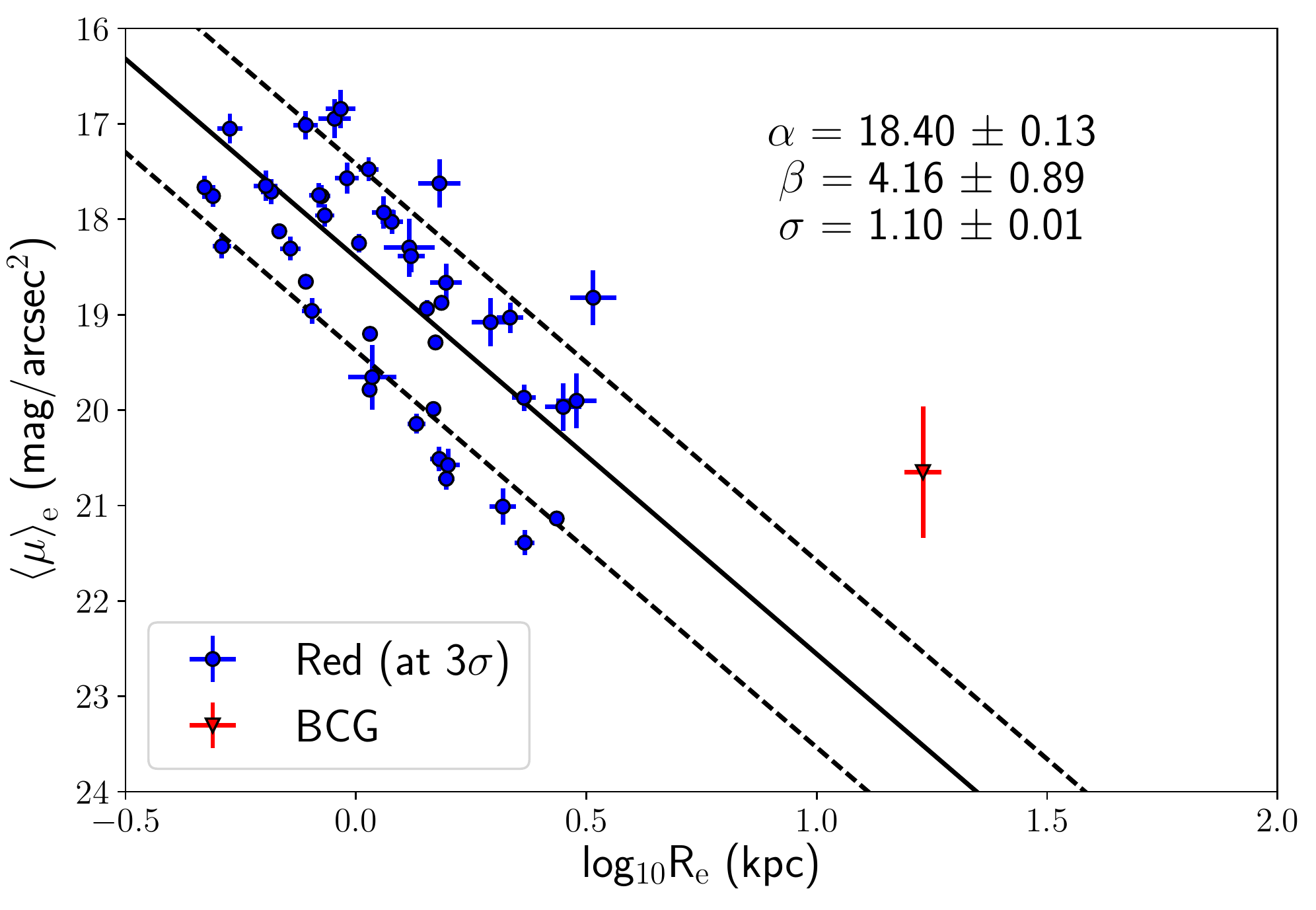}
\includegraphics[scale=0.435]{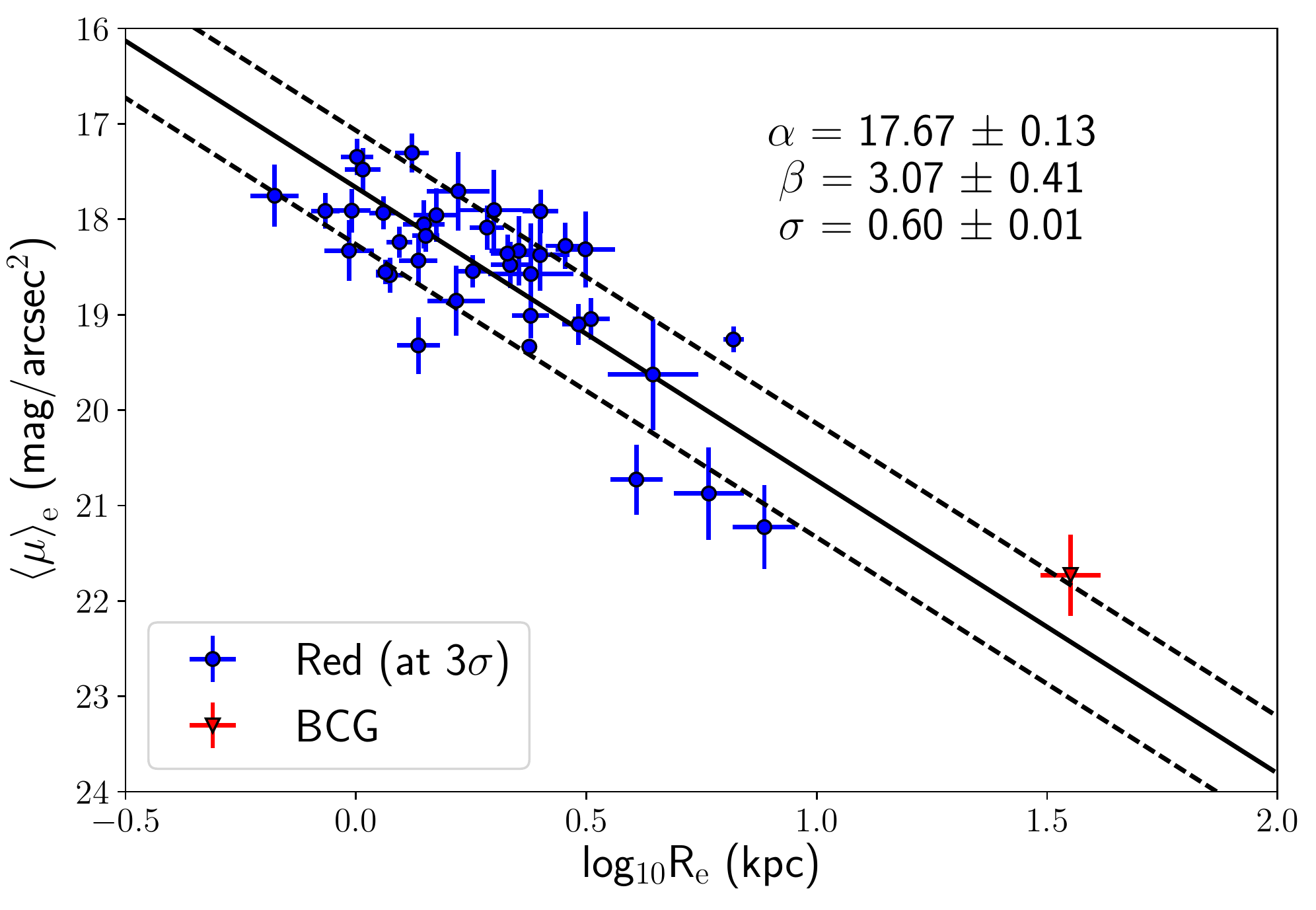}
\caption{The best-fitting KRs of the red samples for AS1063 (left panels) and M1149 (right panels) are shown. Upper panels are those relative to red galaxies at $1 \sigma$ from the best-fitting CM relation, while lower panels are those relative to red galaxies at $3 \sigma$. The coefficients of the best-fitting KR ($\beta$ for the slope, $\alpha$ for the zero-point) and the corresponding observed scatter $\sigma$ are shown at the top right of the plot. The two black dashed lines are the $1 \sigma$ limits of the relation. Blue points represent galaxies with m$_{\mathrm{F814W}}$ $\le$ 22.5 ABmag, while the red triangle is the BCG of each cluster.}
\label{fig11}
\end{figure*}

Here we compare the samples of galaxies classified according to their light profiles, morphologies, colour and spectral properties, as described in Sect. 4.1.

First, we compare the sample of galaxies classified as ETGs and ellipticals. 31 and 32 galaxies are classified both as ETGs and ellipticals for AS1063 and M1149, respectively. Furthermore, 4 galaxies having S\'ersic index n < 2.5 show an elliptical morphology in the inspection of residuals in AS1063, while all the galaxies classified as ellipticals in M1149 have also S\'ersic index n $\ge$ 2.5. The remaining 6 and 4 ETGs, which are not classified as ellipticals in AS1063 and M1149, respectively, show late-type morphology in the inspection of residuals after the best-fitting model subtraction. This suggests that there is a 10\% difference between the visual and the S\'ersic index classification, although this number varies from cluster to cluster.

Then, we compare the sample of galaxies classified both as ETG and red. Only 29 and 31 ETGs in AS1063 are classified also as red galaxies at $1 \sigma$ and at $3 \sigma$ from the best-fitting CM relation, respectively, while, in M1149, they are 30 and 35. Thus, 93 and 89 per cent of the sample of red at $1 \sigma$ and at $3 \sigma$ are also ETG for M1149, while, for AS1063, they are the 70 and 67 per cent (Fig. \ref{fig8}). A visual inspection of the residuals of the fit reveals that, in both clusters, the ETGs having blue colours could be contaminated either by gravitational arcs or by nearby blue galaxies, which are visible only after subtracting the model galaxy, while red galaxies, which are not ETGs, show a late-type morphology, suggesting that they could have had their star-formation truncated by environmental effects or they could appear red due to dust reddening.

The comparison of the passively evolving sample with the ETG sample shows the presence of 6 and 7 ETGs, which are not classified as passive, and $5$ and 3 passively evolving galaxies, which are not classified as ETGs, in AS1063 and M1149, respectively. As for the first sub-sample, 2 galaxies for each cluster show emission lines in their spectra, which could be due to a recent burst in their star-formation activity. For 4 galaxies in AS1063 and 2 galaxies in M1149, we could have missed weak emission lines, since the S/N of the spectrum is $\sim$ 5. For the remaining 3 in M1149, these are excluded from the passive sample, since the fit of the spectrum with \texttt{pPXF} does not converge due to its low S/N. The second sub-sample is constituted by passive galaxies with S\'ersic index n < 2.5. From the visual inspection of the residuals of the surface brightness fit for these sources, they seem to be, indeed, late-type galaxies, so they could be "dusty" spirals or late-type galaxies in which the star-formation has been quenched recently.

Both for the passively evolving and the ETG sample, from Fig. \ref{fig8}, it is clear how the colour selection misses few ETGs and passive galaxies. More interestingly, it is evident the presence of a number of red galaxies (both at 1$\sigma$ and at 3$\sigma$) that are classified neither as ETGs nor as passives. This suggest how a selection based only on the CM relation may contaminate the sample from which one derives the KR.

Considering the four classifications, $28$ galaxies are classified simultaneously as early, ellipticals, red at 1$\sigma$ and passively evolving in AS1063, while $30$ in M1149.

\begin{figure*}
\centering
\includegraphics[scale=0.435]{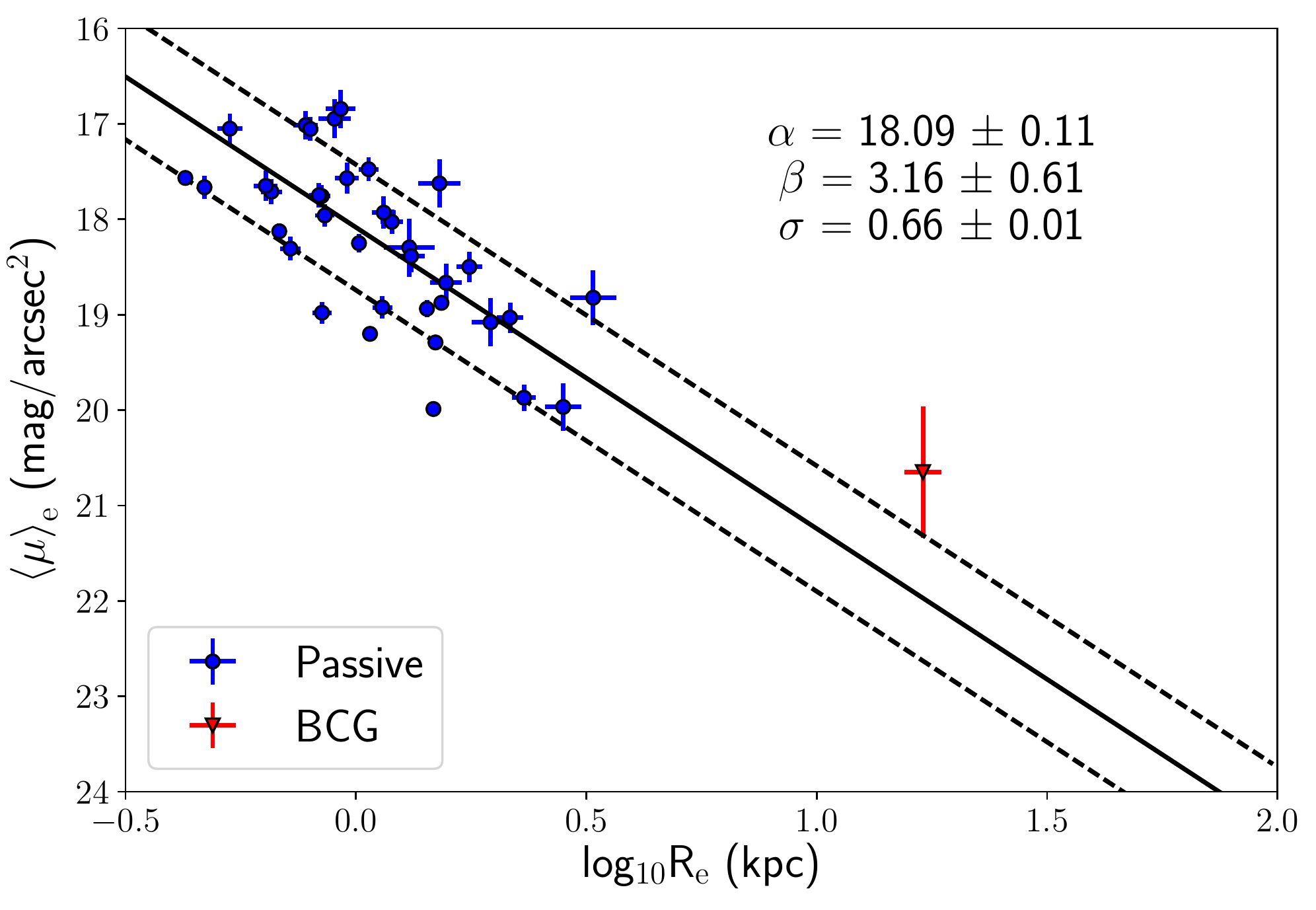}
\includegraphics[scale=0.435]{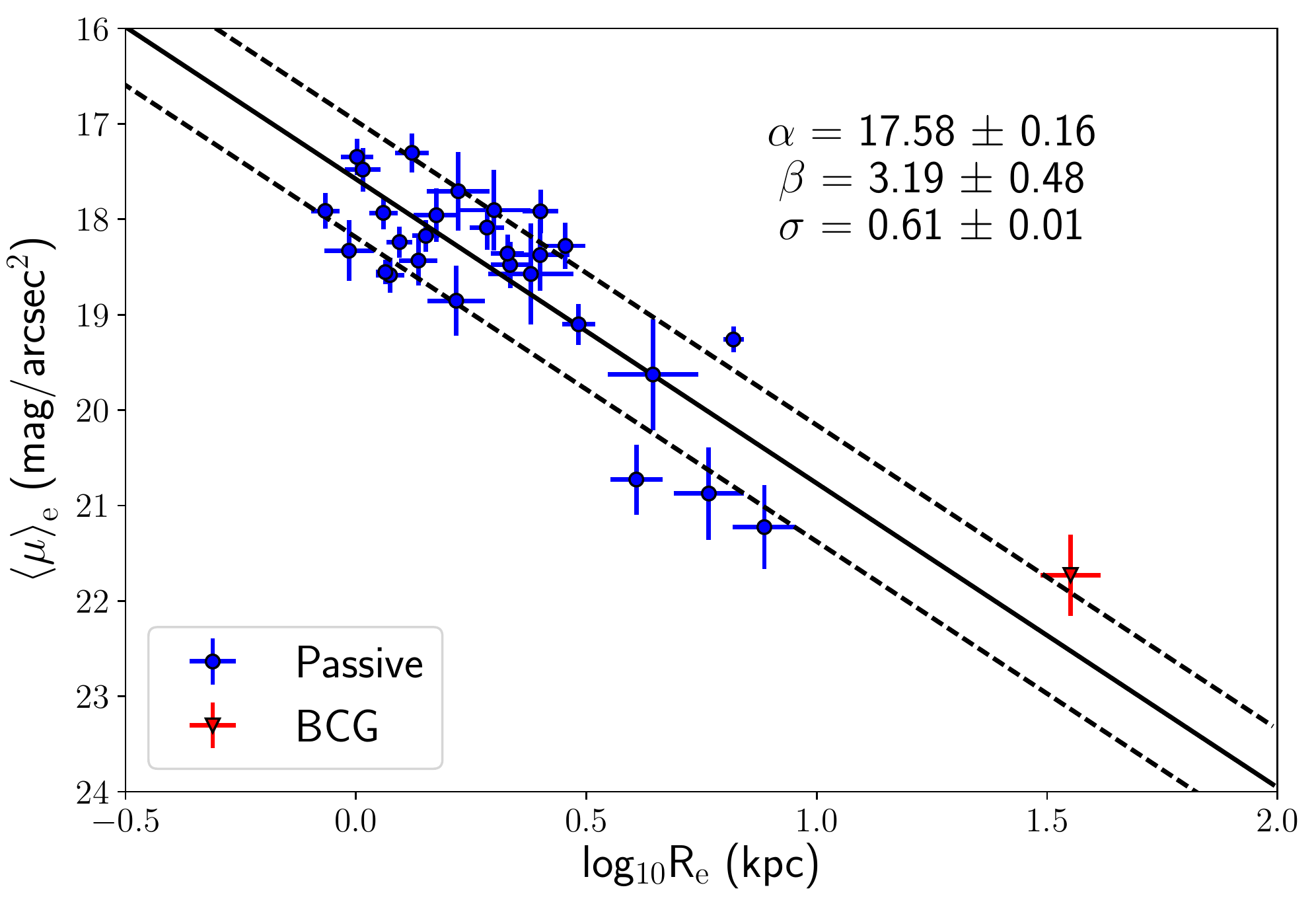}
\caption{The best-fitting KRs of the samples of passively evolving galaxies for AS1063 (left panel) and M1149 (right panel) are shown. The coefficients of the best-fitting KR ($\beta$ for the slope, $\alpha$ for the zero-point) and the corresponding observed scatter $\sigma$ are shown at the top right of the plot. The two black dashed lines mark the $1 \sigma$ band. Blue points represent galaxies with  m$_{\mathrm{F814W}}$ $\le$ 22.5 ABmag, while the red triangle is the BCG of each cluster. All galaxies classified as passively evolving have magnitudes brighter than the adopted completeness limit in the \textit{F814W} waveband in M1149.}
\label{fig12}
\end{figure*}

\begin{figure*}
\centering
\includegraphics[scale=0.435]{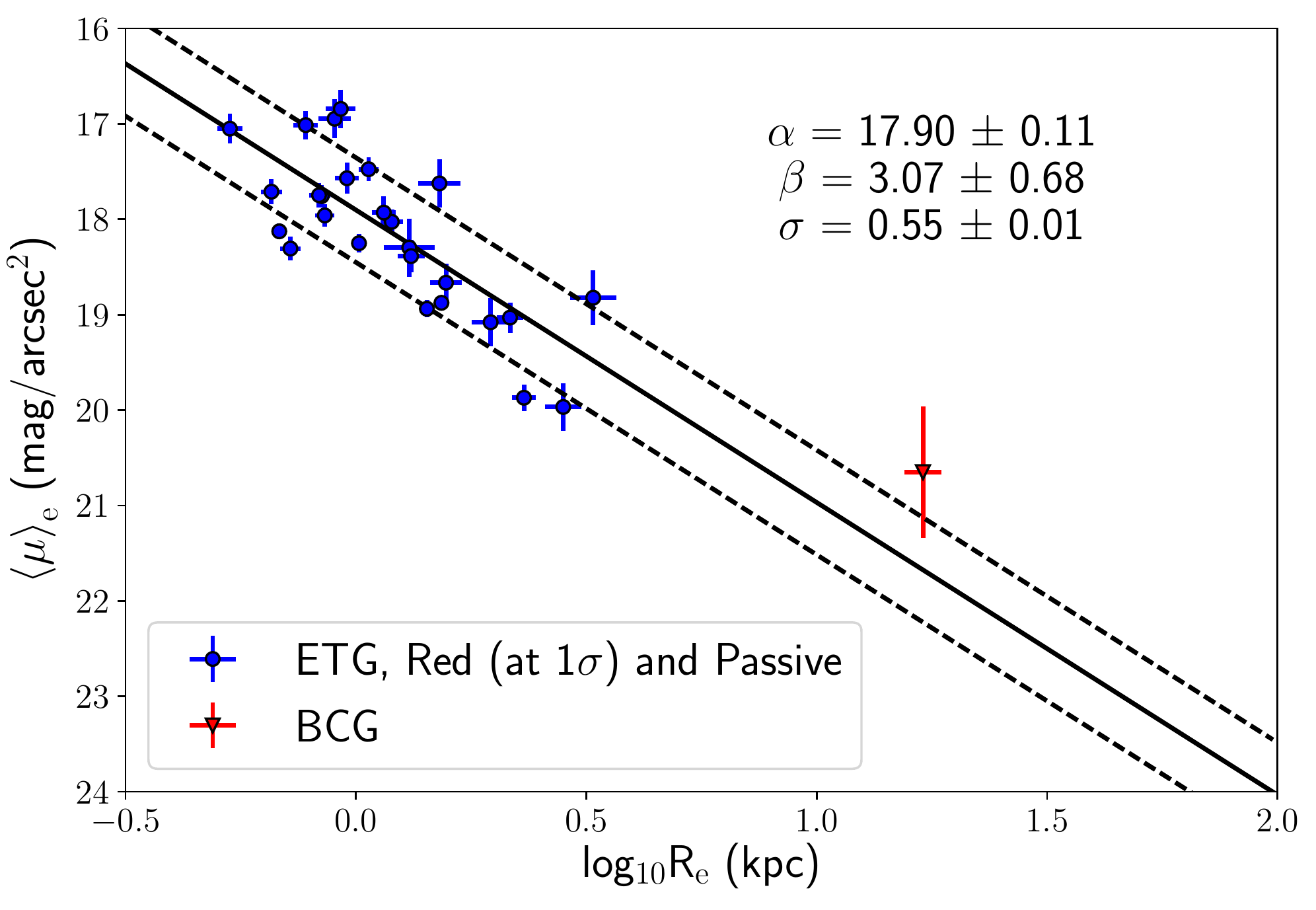}
\includegraphics[scale=0.435]{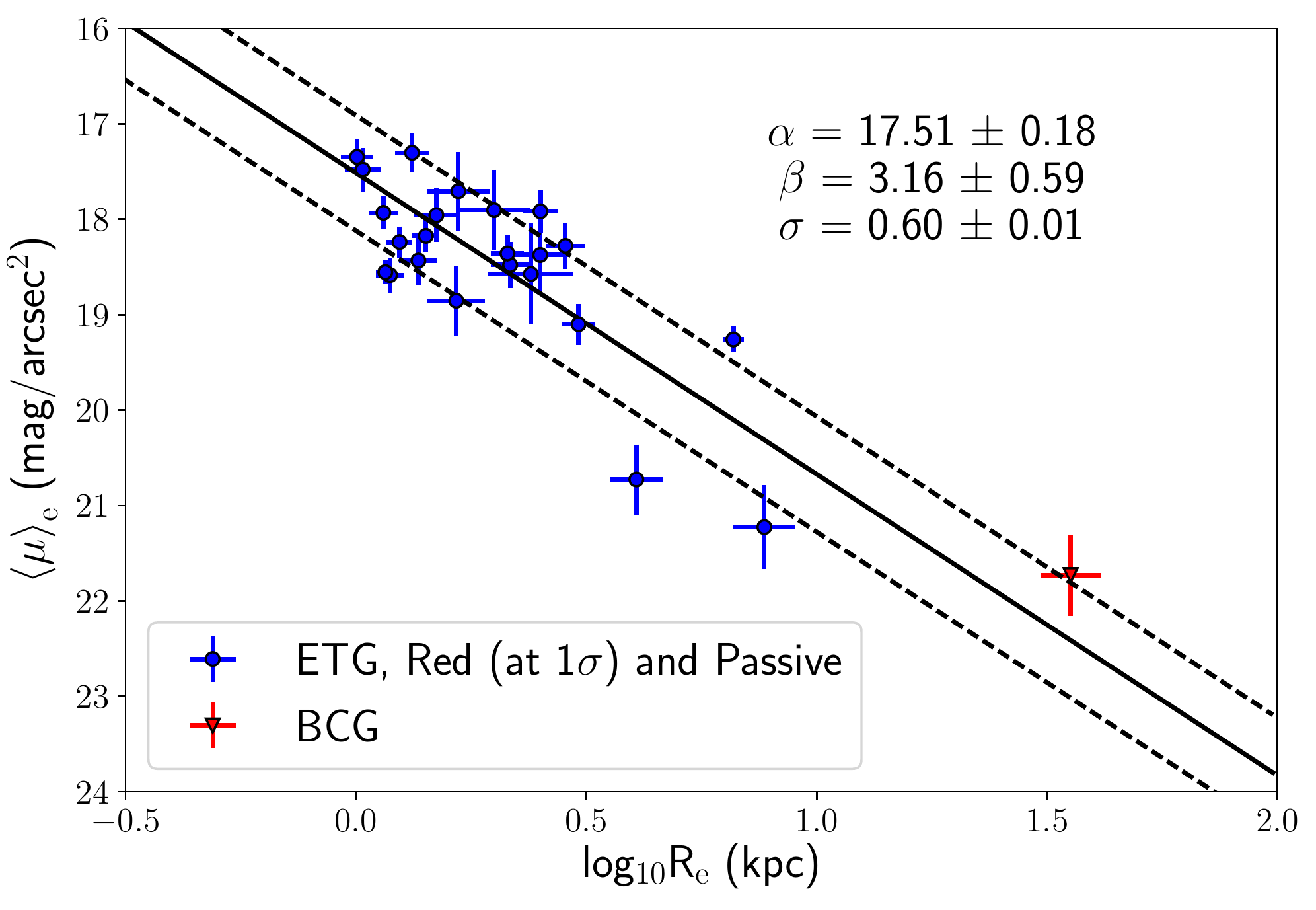}
\caption{The best-fitting KRs of the spectrophotometrically selected samples for AS1063 (left panel) and M1149 (right panel) are shown. The coefficients of the best-fitting KR ($\beta$ for the slope, $\alpha$ for the zero-point) and the corresponding observed scatter $\sigma$ are shown at the top right of the plot. The two black dashed lines mark the $1 \sigma$ band. Blue points represent galaxies with  m$_{\mathrm{F814W}}$ $\le$ 22.5 ABmag, while the red triangle represents the BCG of each cluster. All galaxies spectrophotometrically selected have magnitudes brighter than the adopted completeness limit in the \textit{F814W} band for both clusters.}
\label{fig13}
\end{figure*}

\section{The Kormendy Relation as a Function of Samples}

In this section, we discuss the analysis of the KR, $\left \langle \mu \right \rangle_{\mathrm{e}} = \alpha + \beta \log{\mathrm{R}_\mathrm{e}}$, for the two clusters AS1063 and M1149, considering the four samples defined in the previous section. In particular, we compare the KRs as a function of the different samples, while, in the next section, we compare the KRs of the present work with the literature. As discussed in Sect. 2, we limit our analysis to galaxies brighter than the completeness limit (m$_{\mathrm{F814W}}$ $\le$ 22.5 ABmag). The linear regression analysis is carried out using the method of the Bivariate Correlated Errors and intrinsic Scatter estimator (BCES) described in \citet{Akritas1996}. The BCES is a direct extension of the ordinary least square fitting. It has the advantage that it allows to consider the case in which both variables are affected by errors and in which the errors on the two variables are not independent. From now on, the scatter we refer to is the combination of the intrinsic scatter of the relation plus the photometric errors. All the results presented in this section are in physical units, i.e. $\left \langle \mu \right \rangle_{\mathrm{e}}$ in mag arcsec$^{-2}$ and R$_{\mathrm{e}}$ in kpc. We correct the surface brightness for cosmological dimming effect.

In Fig. \ref{fig9}, the KRs for the ETG samples of AS1063 (left panel) and M1149 (right panel) are shown. The slopes of the two KRs (first row in Table \ref{table2}) are consistent within errors.

In Fig. \ref{fig10} we show the KRs for the elliptical samples of AS1063 (left panel) and M1149 (right panel). The slopes of the two KRs (second row in Table \ref{table2}) are consistent within errors between themselves and with those obtained with the ETG sample.

\begin{figure*}
\centering
\includegraphics[scale=0.439]{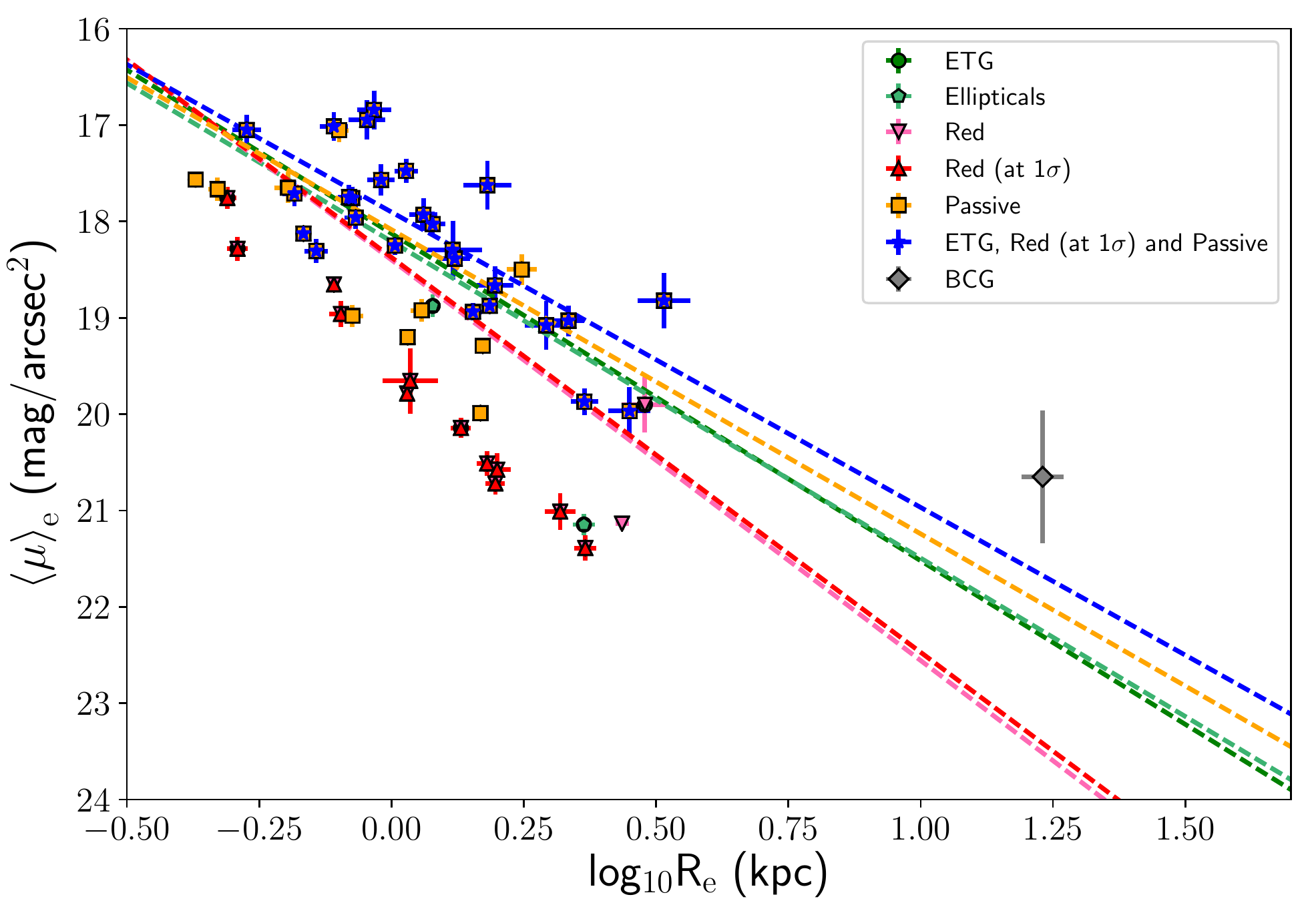}
\includegraphics[scale=0.439]{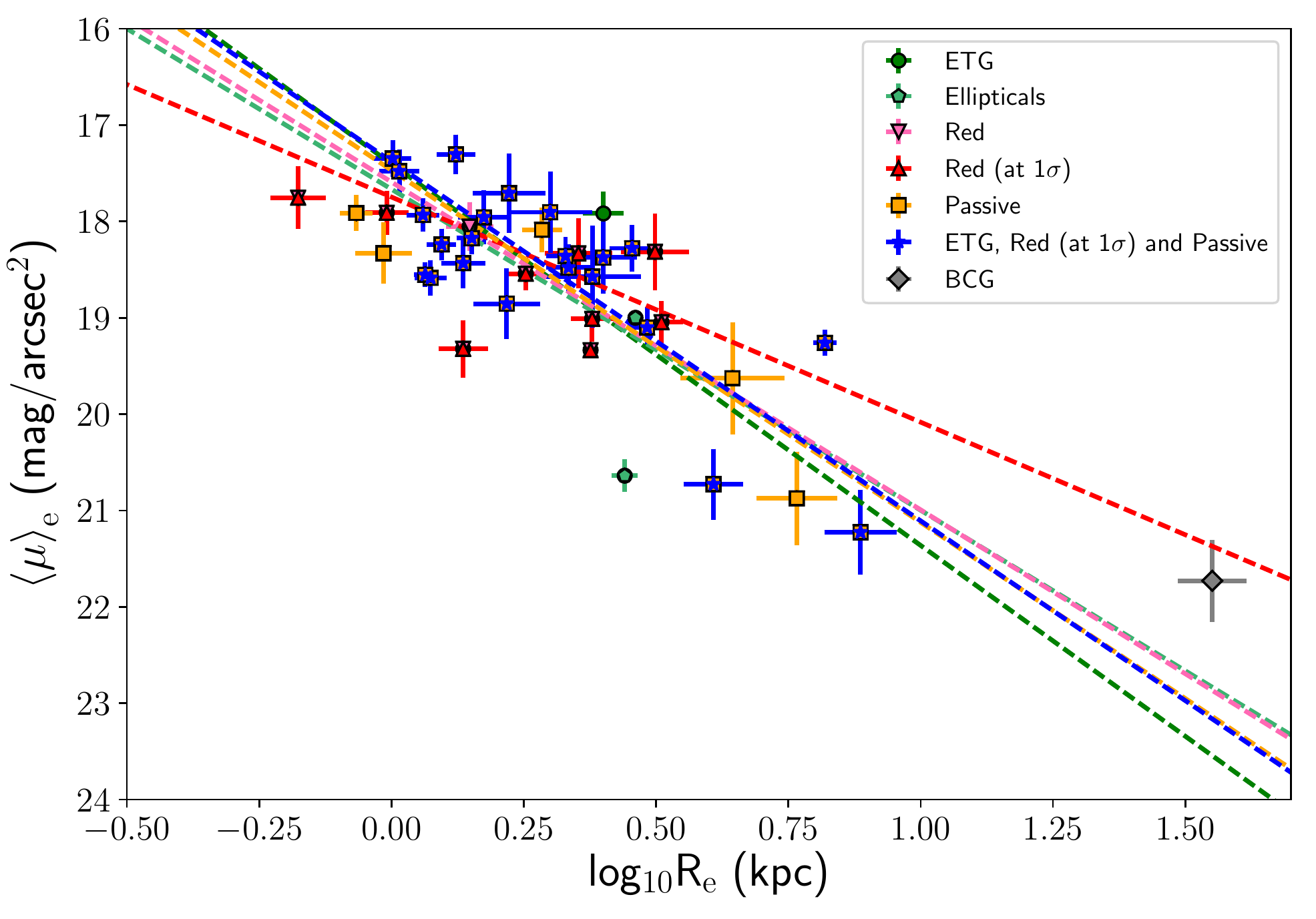}
\caption{The best-fitting KRs of all analysed samples with m $\le$ 22.5 in the \textit{F814W} waveband are shown for the two clusters AS1063 and M1149 (left and right panels, respectively). Green circles, sea green pentagons, pink down triangles, red up triangles, orange squares and blue stars are ETG, ellipticals, red (at 3$\sigma$), red (at 1$\sigma$), passive and spectrophotometric selected samples, respectively. Grey diamond is the BCG. Dashed lines are the best-fitting KRs colour-coded according to the sample. It is worth noticing how, for both clusters, the best-fitting KRs for the ETG, ellipticals, passive and spectrophotometric selected samples show the same behaviour, while for the red samples not.}
\label{fig14}
\end{figure*}

\begin{table*}
\centering
\begin{tabular}{l|c|c|c|c|c|c|c}
\hline
\hline
&&AS1063&&&M1149&\\
\hline
& $\alpha$ & $\beta$ & $\sigma$ & $\alpha$ & $\beta$ & $\sigma$\\
ETG&18.13 $\pm$ 0.12&3.40 $\pm$ 0.67&0.76 $\pm$ 0.01&17.56 $\pm$ 0.17&3.39 $\pm$ 0.51&0.66 $\pm$ 0.01\\
Ellipticals& 18.21 $\pm$ 0.13 & 3.29 $\pm$ 0.76 & 0.77 $\pm$ 0.01 & 17.67 $\pm$ 0.16 & 3.36 $\pm$ 0.53 & 0.65 $\pm$ 0.01\\
Red (at 1 $\sigma$)&18.37 $\pm$ 0.14&4.10 $\pm$ 1.00&1.10 $\pm$ 0.01&17.70 $\pm$ 0.12&2.53 $\pm$ 0.28&0.49 $\pm$ 0.01\\
Red (at 3 $\sigma$)&18.40 $\pm$ 0.13&4.16 $\pm$ 0.89&1.10 $\pm$ 0.01&17.67 $\pm$ 0.13&3.07 $\pm$ 0.41&0.60 $\pm$ 0.01\\
Passive&18.09 $\pm$ 0.11&3.16 $\pm$ 0.61&0.66 $\pm$ 0.01&17.58 $\pm$ 0.16&3.19 $\pm$ 0.47&0.61 $\pm$ 0.01\\
ETG, Red \& Passive &17.90 $\pm$ 0.11&3.07 $\pm$ 0.68&0.55 $\pm$ 0.01&17.51 $\pm$ 0.18&3.16 $\pm$ 0.59&0.60 $\pm$ 0.01\\
\hline
\hline
\end{tabular}
\caption{We report the coefficients of the best-fitting KR for the samples defined in Sect. 4. Column 1 shows the sample used to derive the KR, while columns 2, 3, 4 and 5, 6, 7, are the zero-points, slopes and scatters of the relations for AS1063 and M1149, respectively.}
\label{table2}
\end{table*}

In Fig. \ref{fig11}, the KR for the sample of red galaxies at $1 \sigma$ (upper panels) and at $3 \sigma$ (lower panels) from the best-fitting CM relation of AS1063 (left panels) and M1149 (right panels) are shown. The two fits for AS1063 show larger scatters ($\sigma = 1.10$) than those obtained for ETGs ($\sigma = 0.76$) and for ellipticals ($\sigma = 0.77$) and the best-fitting slopes have higher values with the respect to the ETG and the ellipticals sample (first 3 best-fitting parameters columns of rows 1, 2, 3 and 4 in Table \ref{table2}). Although the slopes are consistent with those of the ETG and the ellipticals sample, the large scatter combined with the large error on the slope ($\sim$ 25 \% relative error) suggests that this sample could be contaminated by galaxies which are not ETGs. As for M1149, instead, the scatter of the relation is lower than that found for the ETGs and the ellipticals and it grows from the $1 \sigma$ to the $3 \sigma$ sample, as we expect from the less stringent colour selection going from 1 to 3 $\sigma$. The slopes of the KRs for the 1$\sigma$ and $3 \sigma$ sample of M1149 are consistent with each other, but only the red at $3 \sigma$ sample is also consistent with the one obtained for AS1063 (third and fourth row in Table \ref{table2}). Furthermore, the slope of the red at $1 \sigma$ sample of M1149 is not consistent, within errors, neither with the one of the same sample in AS1063, nor with the slope obtained with the ETG and the ellipticals sample in M1149. The discrepancies in the KRs parameters obtained with the two samples could be due to the presence of galaxies which appear red in colours, but lack an early-type morphology, as highlighted in Sect. 4.

The KRs for passive galaxies are shown in Fig. \ref{fig12}. The slopes of the KRs for the two clusters are consistent within the errors (fifth row in Table \ref{table2}). Furthermore, they are consistent with those obtained with the ETG and ellipticals samples. Also in this case, there is marginal consistency within errors of the passive sample slope with the red sample ones in AS1063, and with the red at $1 \sigma$ sample slope in M1149. This indicates that there is a better overlap between the samples of ETGs, ellipticals and passively evolving galaxies, rather than between ETGs, ellipticals and red galaxies (Table \ref{table2}).

Considering galaxies which are ETGs, ellipticals, red and passive (labeled ETG, Red (at 1 $\sigma$) and Passive in plots and tables), the KRs for these spectrophotometrically selected samples are shown in Fig. \ref{fig13}. This selection produces KRs which have slopes which are consistent within errors with those of the ETG, ellipticals and passive samples of both clusters (last row in Table \ref{table2}). Furthermore, as for the passive and ETG samples, there is only marginal consistency= with the red at $1 \sigma$ sample of M1149. The scatter of the KRs is also the lowest for AS1063 and as low as the others for M1149.

\begin{figure*}
\centering
\includegraphics[scale=0.475]{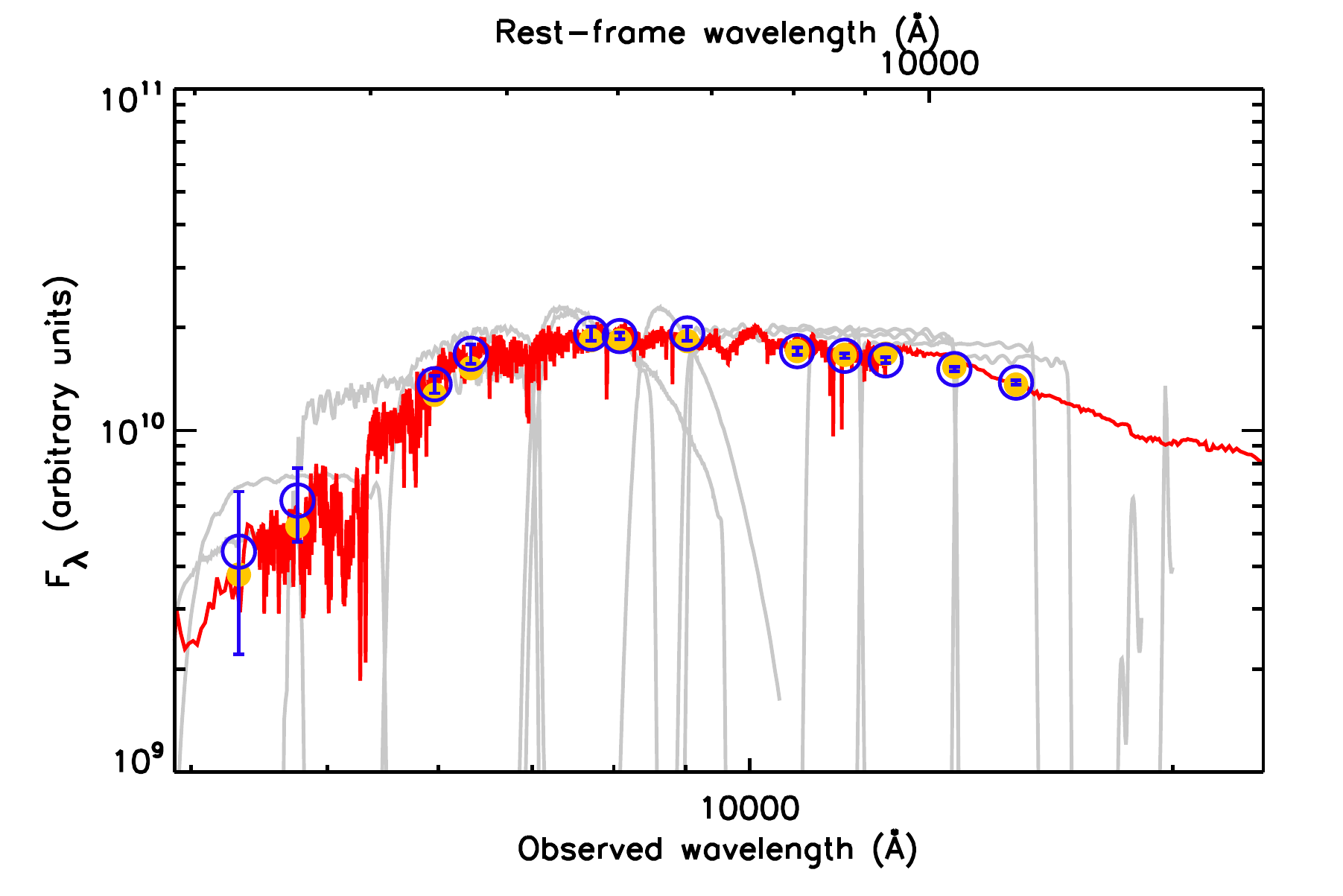} 
\includegraphics[scale=0.475]{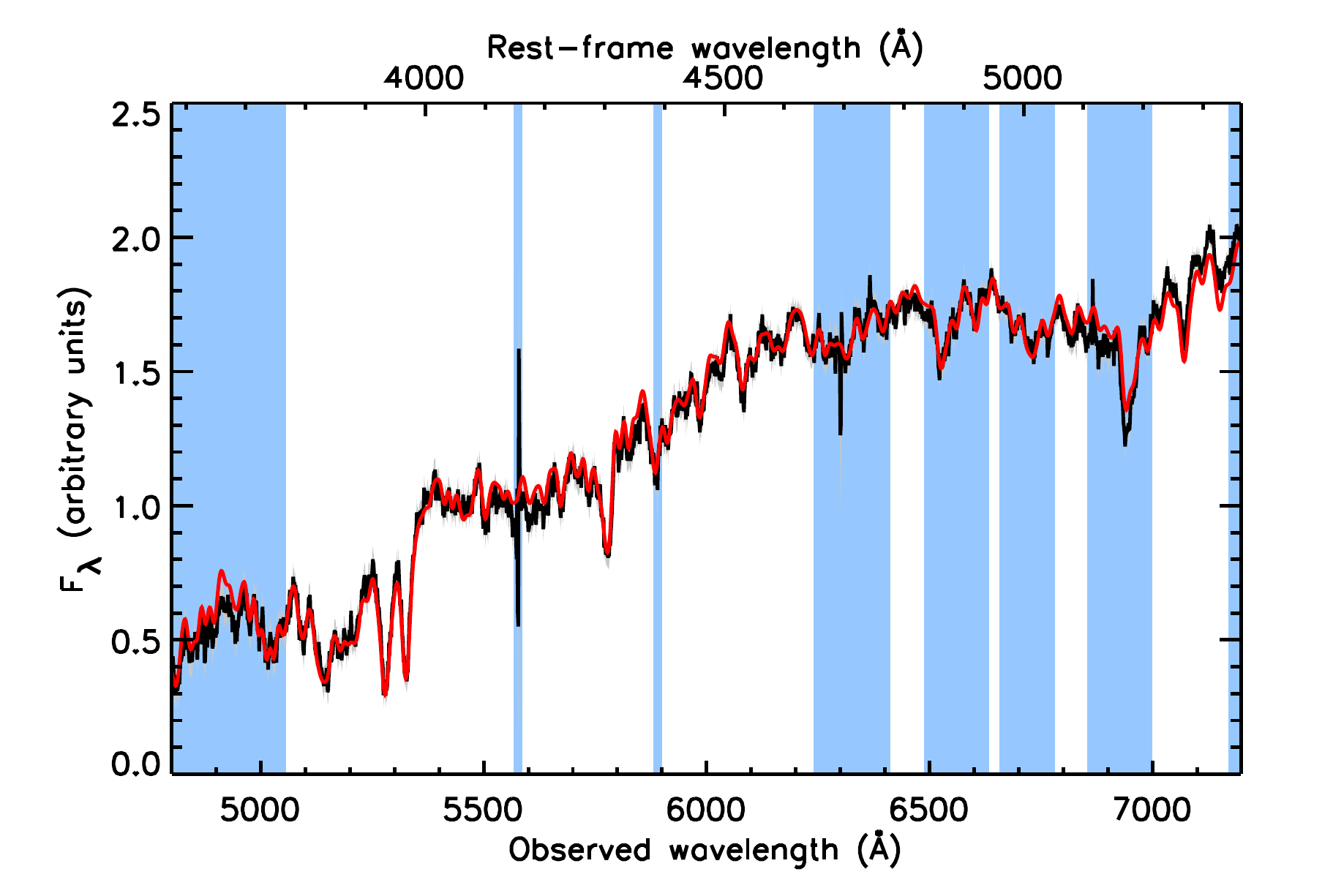} \\
\caption{Simultaneous fit of SED and spectrum is shown. Left panel: best-fitting composite stellar population model \citet{bc03} of the 12 \textit{HST} photometric band observed by CLASH \citep{CLASH}, with the filter transmission curves in light gray. Observed fluxes with 1$\sigma$ errors are represented with blue empty circles and bars, model-predicted fluxes are shown as orange filled circles. Right panel: best-fitting composite stellar population model (in red) of the observed spectrum (in black). Light blue regions highlight the masked regions, where we expect residual of sky subtraction emission lines or sky absorptions. In both panels the best-fitting model is in red.}
\label{fig15}
\end{figure*}

The comparison of the zero-points between samples leads to conclusions similar to those highlighted above. At fixed cluster, the KRs zero-points of ETG, passive and spectrophotometrically selected samples are consistent within errors, while they are only marginally consistent with both the red (at 1$\sigma$) and red (at 3$\sigma$) samples (first and fourth rows of Table \ref{table2}).

In Fig. \ref{fig14}, we show the best-fitting KRs for the different samples on the same plot. For AS1063, all samples span the same range in R$_{\mathrm{e}}$, but there is a clear difference in $\left \langle \mu \right \rangle_{\mathrm{e}}$ between them. At fixed R$_{\mathrm{e}}$, red galaxies are systematically fainter (higher surface brightness) than ETGs, ellipticals, passive and spectrophotometrically selected samples, and the trend is even more marked going towards larger galaxies (increasing R$_e$). This is consistent with the steeper KR obtained from the red sample. Furthermore, the ETG, ellipticals and passive samples have remarkably similar slope and intercept of their KRs, suggesting that they constitute an homogeneous population. The classification according to the colour leads also to a more scattered relation. In M1149, the samples seem to be more homogeneous, indeed the best-fitting KRs are remarkably similar, but for that of the red (at 1$\sigma$) sample which is characterized by more compact galaxies, lacking those with $\log_{10}$R$_{\mathrm{e}}$ > 0.50 kpc.

\section{Literature Comparison}

\begin{figure*}
\centering
\includegraphics[scale=0.435]{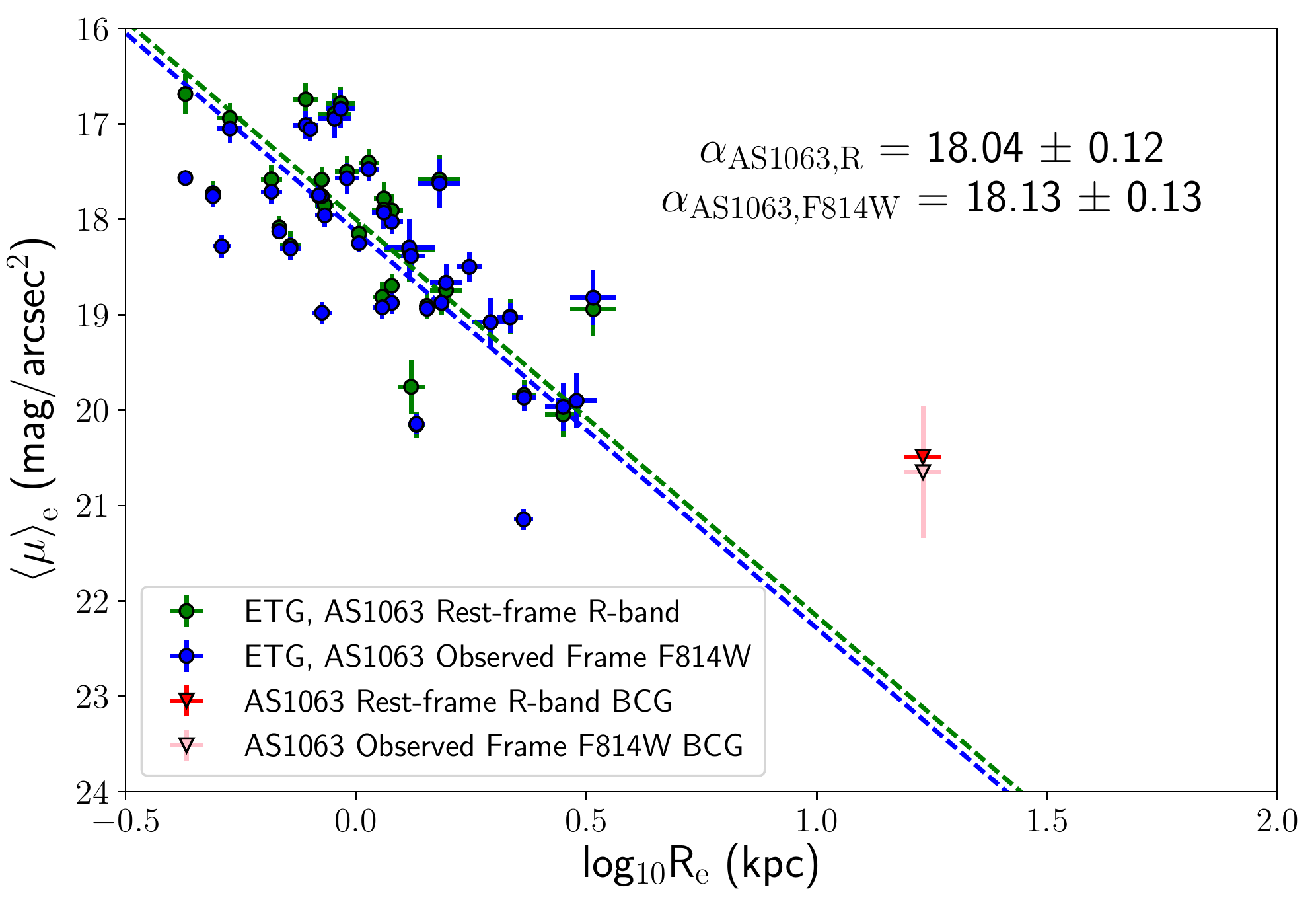}
\includegraphics[scale=0.435]{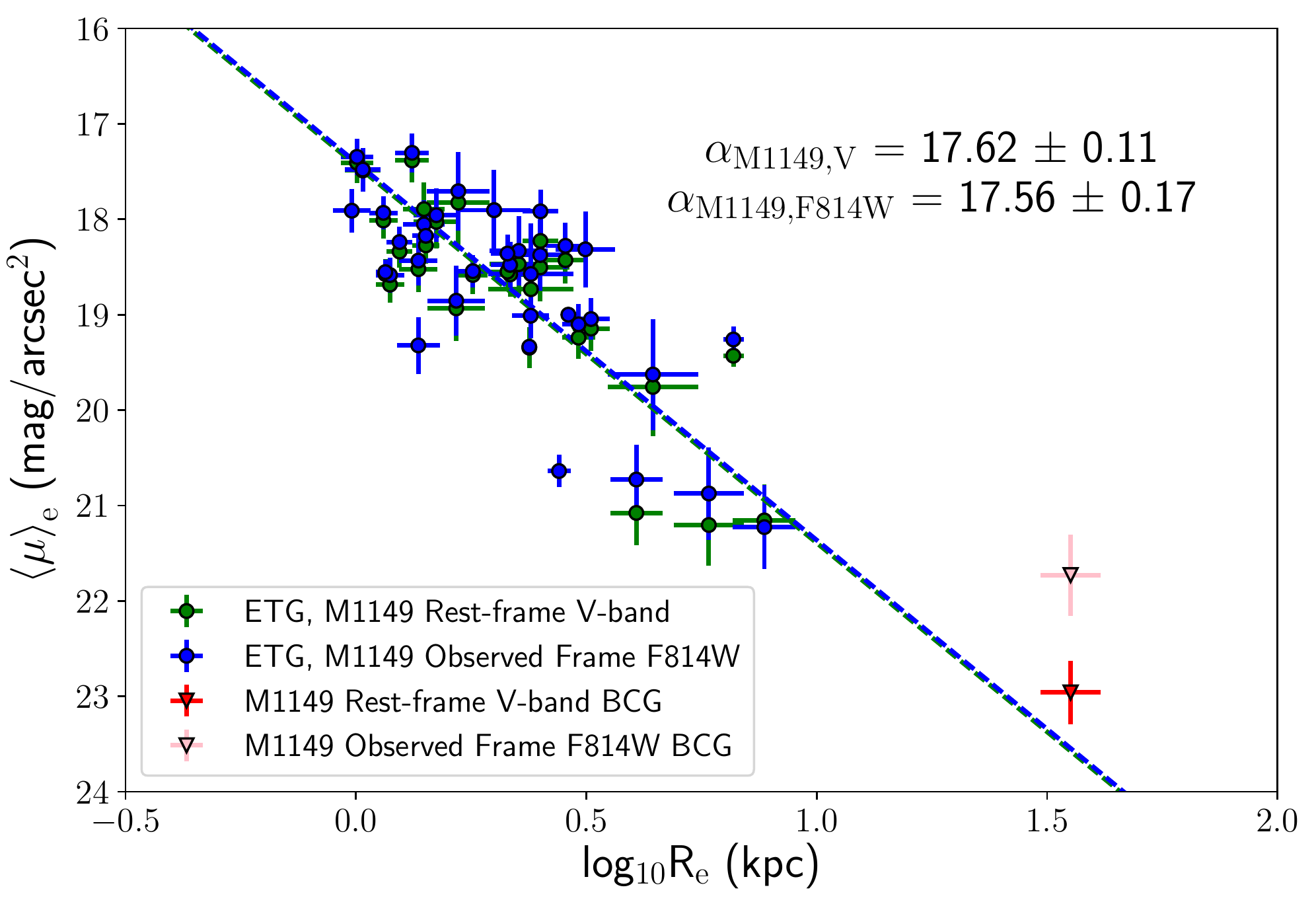}
\caption{The KRs obtained in the rest-frame and in the observed frame bands with the ETG samples for AS1063 (left panel) and M1149 (right panel) are shown. Left panel shows the comparison of the zero-points of the KRs for observed frame \textit{F814W} band and the corresponding rest-frame \textit{R}-band of AS1063, while the right panel shows the same, but for the rest-frame \textit{V}-band of M1149. The coefficients of the best-fitting zero-points $\alpha$ are shown at the top right of the plots. The dashed lines represent the best-fitting KRs colour-coded by the bands. Red and pink down triangles are the BCGs of each cluster, but in different bands.}
\label{fig16}
\end{figure*}

The KR has been investigated by different authors in various photometric bands, with different sample selections and for clusters spanning a large range in redshift, from z $\sim$ 0 to z $\gtrsim$ 1. This section is devoted to the comparison of the slope and zero-point of the KR with those obtained from studies conducted both at similar redshifts and at higher redshifts, by using the same sample selection.

\citet{LaBarbera2003} investigated the KR for a sample of four clusters at z $\sim$ 0, 0.21, 0.31, 0.64, for a total of N = 228 spheroidal galaxies. They selected their sample according to the S\'ersic index, thus we compare this work with our results obtained with the ETG sample. They found a value of the slope that does not change from z = 0.64 to the present epoch and ranges between 2.74 and 3.04, with a typical uncertainty of 0.2 (1 $\sigma$ standard interval). In particular, the z = 0.31 cluster has a slope of $\beta$ = 2.74 $\pm$ 0.16, while the z = 0.64 cluster has $\beta$ = 3.04 $\pm$ 0.13. Comparing the latter with M1149 and the former with AS1063 ETG samples, we find consistency within errors for the values of the slopes. Same conclusions apply to the ellipticals, passive and spectrophotometrically selected samples. In the case of the red samples, only the KR of galaxies red at 3 $\sigma$ for M1149 has slope consistent within errors with that of \citet{LaBarbera2003} for the z = 0.64 cluster. M1149 red at 1$\sigma$ sample and AS1063 red at 1 and 3$\sigma$ sample KRs have instead slopes which are not consistent within errors with that of the z = 0.64 and z = 0.31 clusters, respectively. This confirms that the comparison between KR parameters obtained using different selection criteria could be biased and this seems to be true in particular for the selection based on colours. \citet{LaBarbera2003} found an observed scatter of 0.60 at z $\sim$ 0.31 and 0.42 at z $\sim$ 0.64, while our ETG samples show higher dispersions at comparable redshifts.

\begin{figure*}
\centering
\includegraphics[scale=0.55]{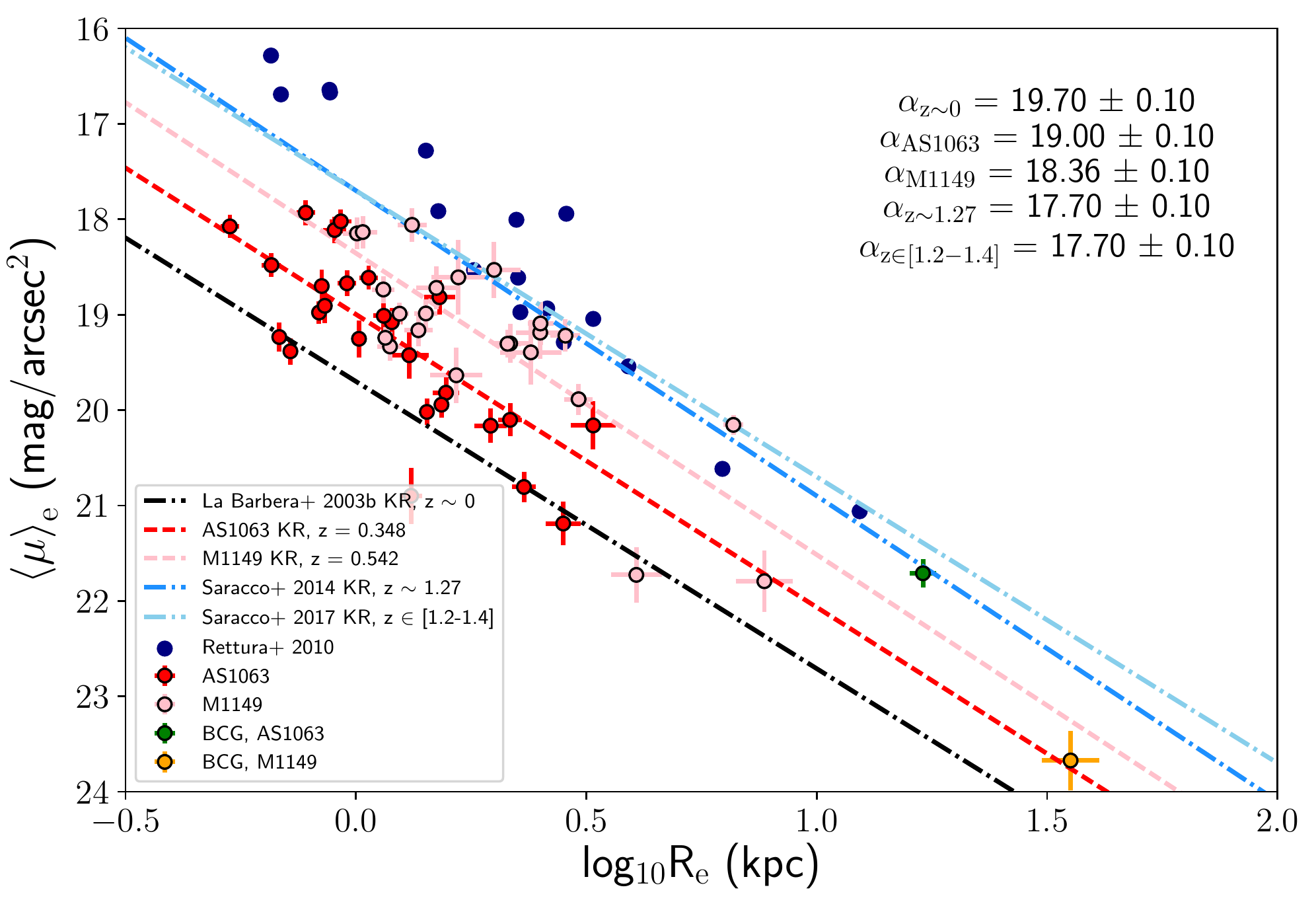}
\caption{The comparison between the KRs of \citet{Rettura2010}, \citet{Saracco2014,Saracco2017} and ours in the rest-frame \textit{B}-band are shown. Red, pink and navy blue points represent ETGs in AS1063, M1149 and RDCSJ0848+4453, respectively, while the green and the orange point represents the BCG of the first two clusters. Red and pink dashed lines refer to AS1063 and M1149 best-fittingKRs, respectively. The latter are computed fixing the slope to the Coma value obtained in \citet{LaBarbera2003}. Black, navy blue, skyblue and dodger blue dash-dotted lines refer to the z $\sim$ 0 \citep{LaBarbera2003}, z $\sim$ 1.24 \citep{Rettura2010}, z $\sim$ 1.27 \citep{Saracco2014} and z $\in$ [1.2-1.4] \citep{Saracco2017} KRs, respectively.  In the upper right part of the plot, the best-fittingzero-points are reported.}
\label{fig17}
\end{figure*}

\citet{Rettura2010} and \citet{Saracco2014,Saracco2017} carried out studies of the KR at z $\gtrsim$ 1. The first work studied a sample of 27 massive ETGs in the RDCS1252.9-2927 cluster at z = 1.237. The sample was selected with a similar approach to that of this work, i.e. through morphologies, colours and spectra. In \citet{Rettura2010}, they do not calculate the KR, but they just compare the location of their data in the plane of the KR found by \citet{LaBarbera2003} at z = 0. Following the same approach, we compared the location of the \citet{Rettura2010} data (blue points in Fig. \ref{fig17}) with our rest-frame \textit{B}-band ETGs data points (red and pink points in Fig. \ref{fig17}). We find agreement in the location of our points with respect to \citet{Rettura2010} data, although, as expected, there is a difference in the zero-points of the relations (see below).

\citet{Saracco2014} investigated the KR for a sample of 16 morphologically selected elliptical galaxies belonging to the cluster RDCSJ0848+4453 at z $\sim$ 1.27. They found a slope of $\beta_{\mathrm{B}}$ = 3.2 $\pm$ 0.5 in the rest-frame \textit{B}-band and $\beta_{\mathrm{R}}$ = 2.6 $\pm$ 0.7 in the rest-frame \textit{R}-band. Both are consistent within errors with the KR slope of the ellipticals sample of AS1063 and M1149. \citep{Saracco2017} analysed 56 cluster ellipticals in three clusters in the redshift range 1.2 < z < 1.4, selected according to their morphologies. They found $\beta$ = 3.0 $\pm$ 0.2, which is consistent within errors with the ellipticals samples of both clusters we analysed.

Despite the fact that the study of the evolution of the KR zero-points with redshift is beyond the scope of this paper, for sake of completeness we compare our zero-points with those at z $\sim$ 0 and at high redshifts in the rest-frame \textit{B}-band and \textit{R}-band.

In order to consistently compare the zero-points of both clusters with themselves and with the literature, we have to derive the KRs in the same rest-frame wavebands. To do that, we model the multicolour photometry, composed by 12 \textit{HST} bands (optical-NIR) observed by CLASH \citep{CLASH}, plus the spectra, observed by MUSE, of the member galaxies, to obtain mainly the rest-frame photometry and the stellar mass of these objects. We consider composite stellar populations (CSP) based on \citet{bc03} models with Z $\in$ [0.5,1.5] Z$_\odot$, with a Salpeter stellar IMF \citep{Salpeter1955}, delayed exponential star formation histories (SFHs) and no reddening. In Fig. \ref{fig15}, we show a typical example of the SED (right panel) plus spectral fitting (left panel) of an ETG galaxy. In the left panel, the blue empty circles and the bars are the observed magnitudes with 1$\sigma$ errors. Orange filled circles are the magnitudes measured on the best-fitting model (in red) using the filter transmission curves (in light gray). In the right panel, the best-fitting CSP model (in red) is shown, superimposed on the observed spectrum (in black). From those best-fitting models, we compute the rest-frame \textit{B}-band and \textit{R}-band magnitudes for all the galaxies in our samples. We reserve the study of the physical properties of those sources to a future work.

To further check the robustness of our SED plus spectral fitting, we compare the observed frame KRs with those obtained with the corresponding rest-frame bands for both clusters. In Fig. \ref{fig16}, the rest-frame \textit{R}-band and the rest-frame \textit{V}-band are compared to the observed frame \textit{F814W} band (blue points) for AS1063 (left panel) and M1149 (right panel), respectively. By fixing the slope of the rest-frame KRs to that of the observed frame KRs, we find agreement of the zero-points within errors for both clusters. The small difference of the values is due to the different shape of the observed and rest-frame filters, confirming what we already highlighted in Sect. 2.

In Fig. \ref{fig17}, we compare the rest-frame B-band zero-points of our KRs with those of \citet{Rettura2010} and \citet{Saracco2014,Saracco2017}. The zero-points of both AS1063 and M1149, as expected, are fainter than those of the z = 1.237 \citep{Rettura2010}, z $\sim$ 1.27 \citep{Saracco2014} and z $\in$ [1.2-1.4] \citep{Saracco2017} clusters, but brighter than the z $\sim$ 0 Coma cluster in \citet{LaBarbera2003} K-corrected to the B-band.

\begin{figure*}
\centering
\includegraphics[scale=0.55]{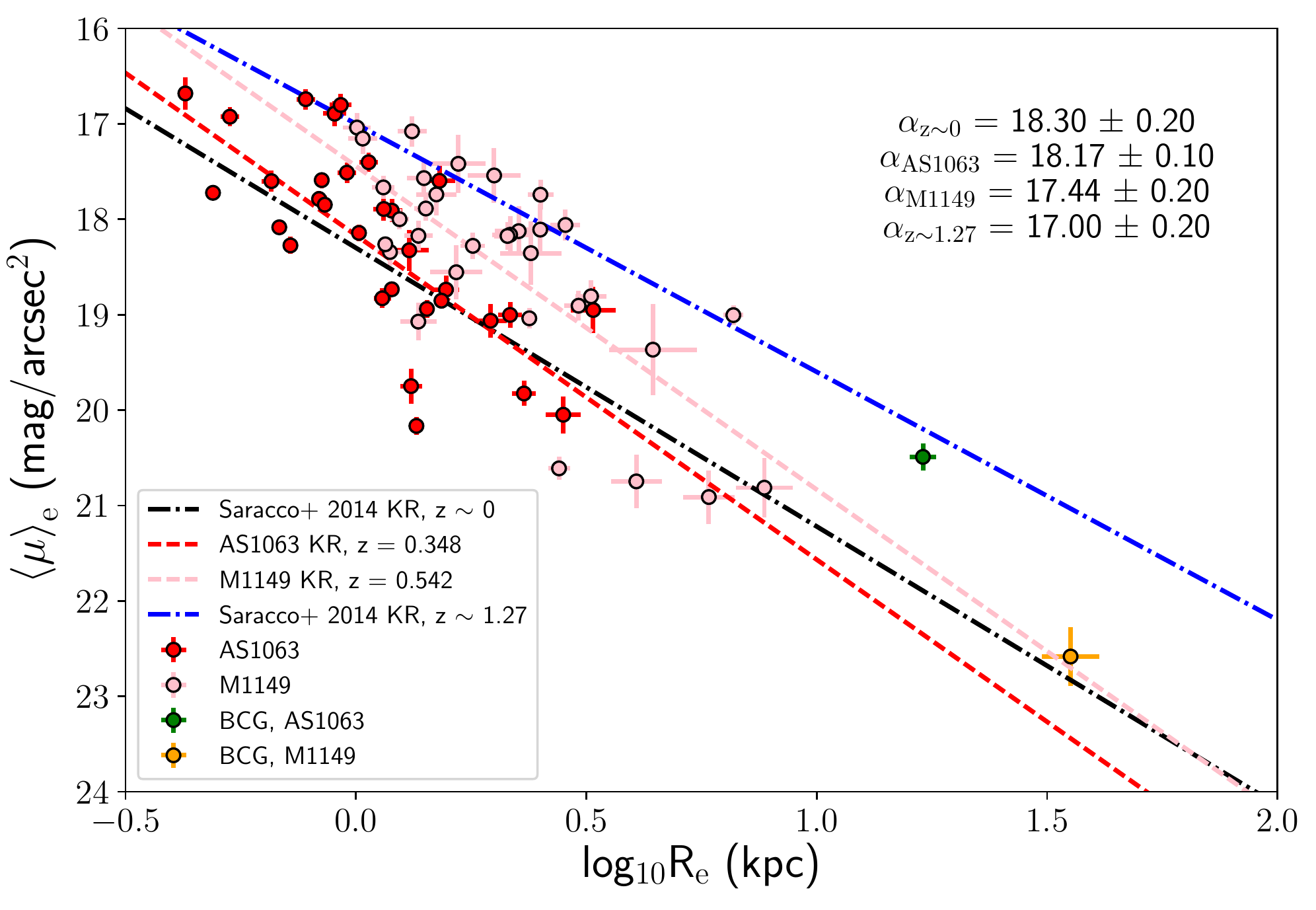}
\caption{The comparison between the KRs of \citet{Saracco2014} and ours in the rest-frame \textit{R}-band is shown. Red and pink points represent ETGs in AS1063 and M1149, respectively, while the green and the orange point represents the BCG of each cluster. Red and pink dashed lines refer to AS1063 and M1149 best-fittingKRs, respectively. The latter are computed fixing the slope to the R-band z $\sim$ 0 value obtained in \citet{Saracco2014}. Black and blue dash-dotted lines refer to the z $\sim$ 0 and z $\sim$ 1.27 \citep{Saracco2014} KRs, respectively. In the upper right part of the plots, the best-fittingzero-points in the rest-frame \textit{R}-band are reported.}
\label{fig18}
\end{figure*}

Fig. \ref{fig18} shows the comparison between \citet{Saracco2014} KR and ours in the rest-frame R-band. Also in this case, our zero-points lie in between the z $\sim$ 0 and the high redshift ones.

The fact that our results lie between the high redshift and the local universe ones confirm the evolution with redshift of the KR zero-point already highlighted by the mentioned works. However, we notice that in the rest-frame \textit{B}-band and \textit{R}-band the differences in zero-point between AS1063 and M1149 are $\Delta \alpha_{\mathrm{B}}$ = 0.60 $\pm$ 0.02 ABmag and $\Delta \alpha_{\mathrm{R}}$ = 0.59 $\pm$ 0.03 ABmag, respectively. This suggests that the difference in zero-points between the two redshifts cannot be fully explained by passive evolution only, since the latter accounts only for nearly half of that. However, we reserve the detailed analysis of these zero-points differences to a future work.

\section{Summary and Conclusions}

In this paper we investigate the variation of the Kormendy relation (KR) as a function of the sample selection and the influence of this variation on the studies of the KR parameters evolution. We analyse spectroscopically confirmed cluster galaxies of two \textit{HST} Frontier Fields clusters, Abell S1063 (z = 0.348) and MACSJ1149.5+2223 (z = 0.542) in the \textit{F814W} photometric band.

We propose a new methodology for the estimate of structural parameters of galaxies in crowded environments, such as that of a cluster. This methodology consists of an iterative approach which analyse images of increasing size, to deal with closely separated galaxies, and of different background estimations to deal with the ICL contamination in flux.

At high redshift, an accurate classification based only on structural parameters and/or spectral features by themselves is challenging. Therefore, the sample of galaxies for which we measure the KR is selected using both photometric and spectroscopic methods. We perform a cut in magnitude, selecting only those galaxies with m $\le$ 22.5 ABmag in the \textit{F814W} waveband to be fully complete in magnitude in both clusters. According to the best-fitting model templates, this limit roughly corresponds to a stellar mass of M$_{*} \sim 10^{9.8}\ \mathrm{M}_{\odot}$ and M$_{*} \sim 10^{10.0}\ \mathrm{M}_{\odot}$ for AS1063 and M1149, respectively, for the typical SED we are interested in and considering a Salpeter IMF.

We classify as ETGs those galaxies having S\'ersic index n $\ge$ 2.5 (37 in AS1063 and 36 in M1149), as ellipticals those galaxies showing elliptical morphologies according to the visual inspection of galaxy images and residuals after the best-fitting model subtraction (35 in AS1063 and 32 in M1149), as red galaxies those which are redder than the best-fitting CM relation of the cluster minus $1$ and $3$ standard deviation (41 and 46 in AS1063 and 32 and 39 in M1149) and as passively evolving galaxies those whose MUSE spectra present no emission lines and weak H$\delta$ absorption (37 in AS1063 and 29 in M1149). Galaxies which are classified as ETG, ellipticals, red at $1\sigma$ and passive, simultaneously, constitute the spectrophotometric sample (27 in AS1063 and 24 in M1149).

We build the KR for each of the samples described above, both for AS1063 and M1149. From the analysis of the KR as a function of samples, we find that:
\begin{itemize}
\item ETG, elliptical and passive galaxies have KRs with remarkably similar slope and intercept (at fixed wavebands) in both clusters, suggesting that they constitute an homogeneous population.
\item Using the CM diagram (Fig. \ref{fig8}), the marginal consistency between the red samples and the others is due to the presence of a large number of red galaxies which are not classified also as ETGs, ellipticals or passives. Therefore, a sample based on colour selection is contaminated by late-type galaxies.
\item On the other hand, ETG, elliptical and passive galaxies have similar distributions on the CM diagram and very few ETGs are not classified as passive objects and viceversa. Furthermore, the distribution of ETGs, ellipticals and passives is also very similar in Fig. \ref{fig14}, where we compare the KRs of all samples.
\item For AS1063, the KRs of the red samples show larger scatters ($\sigma$ = 1.10) than those obtained for ETGs ($\sigma$ = 0.76) and ellipticals ($\sigma$ = 0.77). Moreover, although the slopes are consistent, the large scatter combined with the large error on the slope ($\sim$ 25 \% relative error) suggests that the red sample could be contaminated by galaxies which are not ETGs. For M1149, the slope of the KR for the red at 1 $\sigma$ sample is not consistent within errors with the ETG or ellipticals sample. This highlights the fact that, on cluster-to-cluster basis, different selection criteria can have significant impact on the KR parameters. As a consequence, the sample selection has a crucial relevance on all of the studies aiming at constraining the galaxy evolution through the analysis of the KR at different redshifts. In fact, a not homogeneous selection over cosmic time can bias the conclusions on the luminosity and size evolution. For example, considering a fixed surface brightness $\left \langle \mu \right \rangle_{\mathrm{e}}$ = 18.5 mag arcsec$^{-2}$, according to the KR in M1149 we expect an effective radius R$_{\mathrm{e}} \sim$ 1.89 kpc for ETGs, while, considering the red at 1 $\sigma$ sample, we obtain R$_{\mathrm{e}} \sim$ 2.07 kpc, resulting in a size 10\% larger than that of ETGs.
\item Comparing our KRs with those of other authors, by using the same sample definition and the same rest-frame waveband, we find that AS1063 and M1149 slopes are consistent with literature. Furthermore, our zero-points lie between the high redshift and the local universe ones.
\end{itemize}

\section*{Acknowledgements}

The authors thank the anonymous referee for its useful and constructive comments that helped to improve the quality of this paper. We also thank Dr. Peng and Dr. Cappellari for their precious advices on how to properly manage \texttt{GALFIT} and \texttt{pPXF}, respectively, and Elisabete da Cunha for providing us with a non public library of models for the \texttt{MAGPHYS} SED fitting procedure. We acknowledge financial support from the PRIN-INAF 2014 1.05.01.94.02 (PI: M. Nonino). AM acknowledges funding from the INAF PRIN-SKA 2017 program 1.05.01.88.04. Maurizio Paolillo acknowledges support from the PRIN-INAF 2014 "Fornax Cluster Imaging and Spectroscopic Deep Survey". The CLASH Multi-Cycle Treasury Program is based on observations made with the NASA/ESA Hubble Space Telescope. The Space Telescope Science Institute is operated by the Association of Universities for Research in Astronomy, Inc., under NASA contract NAS 5-26555. Based on observations made with the European Southern Observatory Very Large Telescope (ESO/VLT) at Cerro Paranal, under programme IDs 60.A-9345(A), 095.A-0653(A), 294.A-5032(A) and 186.A-0798(A).




\bibliographystyle{mnras}
\bibliography{tortorelli_bibliography} 




\appendix

\section{Tables of Structural Parameters}

We report the tables containing the structural parameters and their errors for AS1063 (Table \ref{appendixa1}) and M1149 (Table \ref{appendixa2}). Total magnitude refers to that in the \textit{F814W} waveband, while R$_{\mathrm{e}}$ and $\left \langle \mu \right \rangle_{\mathrm{e}}$ are in units of kpc and mag arcsec$^{-2}$, respectively. Last column of the tables contains the classification of spectroscopically confirmed cluster members.

\begin{table*}
\centering
\begin{tabular}{l|c|c|c|c|c|c|c|c}
\hline
\hline
ID & RA & DEC & m$_{\mathrm{tot}}$& R$_{\mathrm{e}}$ (kpc) & $\left \langle \mu \right \rangle_{\mathrm{e}}$ (mag arcsec$^{-2}$) & n & Classification \\
\hline
1 & 342.16045 & -44.53893 & 22.09 $\pm$ 0.02  & 0.96 $\pm$ 0.03 & 19.24 $\pm$ 0.07 & 1.1 $\pm$ 0.1 & \\
2 & 342.16273 & -44.53816 & 22.32 $\pm$ 0.03 & 2.32 $\pm$ 0.11 & 21.39 $\pm$ 0.13 & 1.9 $\pm$ 0.1 & Red (3$\sigma$), Red (1$\sigma$)\\
3 & 342.16661 & -44.53482 & 21.39 $\pm$ 0.02 & 0.66 $\pm$ 0.04 & 17.71 $\pm$ 0.13 & 3.9 $\pm$ 0.2 & ETG, Elliptical, Red (3$\sigma$), Red (1$\sigma$), Passive\\
4 & 342.16825 & -44.53655 & 22.07 $\pm$ 0.03 & 0.49 $\pm$ 0.02 & 17.76 $\pm$ 0.12 & 3.5 $\pm$ 0.2 & ETG, Elliptical, Red (3$\sigma$), Red (1$\sigma$)\\
5 & 342.16903 & -44.54034 & 22.87 $\pm$ 0.01 & 1.51 $\pm$ 0.06 & 21.00 $\pm$ 0.09 & 2.1 $\pm$ 0.1 & Red (3$\sigma$), Red (1$\sigma$)\\
6 & 342.16985 & -44.53554 & 20.48 $\pm$ 0.02 & 1.31 $\pm$ 0.17 & 18.30 $\pm$ 0.31 & 4.4 $\pm$ 0.4 & ETG, Elliptical, Red (3$\sigma$), Red (1$\sigma$), Passive\\
7 & 342.17079 & -44.53583 & 23.03 $\pm$ 0.04 & 1.85 $\pm$ 0.10 & 21.60 $\pm$ 0.15 & 1.3 $\pm$ 0.1 & \\
8 & 342.17160 & -44.53951 & 22.40 $\pm$ 0.02 & 1.07 $\pm$ 0.01 & 19.79 $\pm$ 0.05 & 1.4 $\pm$ 0.1 & Red (3$\sigma$), Red (1$\sigma$)\\
9 & 342.17176 & -44.54054 & 20.48 $\pm$ 0.05 & 2.82 $\pm$ 0.26 & 19.97 $\pm$ 0.25 & 4.0 $\pm$ 0.4 & ETG, Red (3$\sigma$), Red (1$\sigma$), Passive\\
10 & 342.17217 & -44.54313 & 23.19 $\pm$ 0.04 & 0.87 $\pm$ 0.04 & 20.12 $\pm$ 0.15 & 1.8 $\pm$ 0.1 & Red (3$\sigma$), Red (1$\sigma$)\\ 
11 & 342.17368 & -44.53277 & 21.61 $\pm$ 0.01 & 1.35 $\pm$ 0.06 & 19.50 $\pm$ 0.10 & 2.1 $\pm$ 0.1& \\ 
12 & 342.17443 & -44.54622 & 22.29 $\pm$ 0.02 & 0.74 $\pm$ 0.02 & 18.88 $\pm$ 0.07 & 1.1 $\pm$ 0.1& \\ 
13 & 342.17450 & -44.52900 & 22.07 $\pm$ 0.04 & 0.47 $\pm$ 0.02 & 17.66 $\pm$ 0.12 & 2.0 $\pm$ 0.1 & Red (3$\sigma$), Passive\\ 
14 & 342.17465 & -44.53693 & 23.63 $\pm$ 0.14 & 1.12 $\pm$ 0.16 & 21.11 $\pm$ 0.45 & 1.3 $\pm$ 0.3 & \\ 
15 & 342.17492 & -44.53413 & 21.39 $\pm$ 0.03 & 0.64 $\pm$ 0.04 & 17.65 $\pm$ 0.16 & 2.4 $\pm$ 0.1 & Red (3$\sigma$), Red (1$\sigma$), Passive\\ 
16 & 342.17546 & -44.53538 & 18.71 $\pm$ 0.72 & 40.80 $\pm$ 37.00 & 24.00 $\pm$ 2.70 & 6.5 $\pm$ 3.0 & Failed Fit\\ 
17 & 342.17646 & -44.53366 & 20.89 $\pm$ 0.02 & 0.84 $\pm$ 0.04 & 17.76 $\pm$ 0.12 & 2.6 $\pm$ 0.1 & ETG, Elliptical, Red (3$\sigma$), Red (1$\sigma$), Passive\\ 
18 & 342.17686 & -44.53753 & 21.25 $\pm$ 0.04 & 1.20 $\pm$ 0.04 & 18.88 $\pm$ 0.12 & 4.3 $\pm$ 0.1 & ETG, Elliptical\\ 
19 & 342.17703 & -44.53694 & 19.13 $\pm$ 0.04 & 2.23 $\pm$ 0.15 & 18.11 $\pm$ 0.19 & 2.3 $\pm$ 0.2 & \\ 
20 & 342.17726 & -44.52501 & 23.26 $\pm$ 0.03 & 1.48 $\pm$ 0.05 & 21.35 $\pm$ 0.10 & 1.5 $\pm$ 0.1 & \\ 
21 & 342.17793 & -44.53239 & 20.98 $\pm$ 0.02 & 1.02 $\pm$ 0.03 & 18.25 $\pm$ 0.10 & 2.6 $\pm$ 0.2 & ETG, Elliptical, Red (3$\sigma$), Red (1$\sigma$), Passive\\ 
22 & 342.17795 & -44.52406 & 21.19 $\pm$ 0.01 & 1.49 $\pm$ 0.04 & 19.29 $\pm$ 0.07 & 1.5 $\pm$ 0.1 & Red (3$\sigma$), Passive\\ 
23 & 342.17863 & -44.54655 & 22.20 $\pm$ 0.03 & 0.80 $\pm$ 0.04 & 18.96 $\pm$ 0.14 & 2.2 $\pm$ 0.1 & Red (3$\sigma$), Red (1$\sigma$)\\ 
24 & 342.17891 & -44.52472 & 21.81 $\pm$ 0.02 & 1.07 $\pm$ 0.03 & 19.20 $\pm$ 0.08 & 1.9 $\pm$ 0.1 & Red (3$\sigma$), Red (1$\sigma$), Passive\\ 
25 & 342.17903 & -44.53278 & 20.32 $\pm$ 0.02 & 0.78 $\pm$ 0.05 & 17.01 $\pm$ 0.15 & 3.0 $\pm$ 0.1 & ETG, Elliptical, Red (3$\sigma$), Red (1$\sigma$), Passive\\ 
26 & 342.17961 & -44.52304 & 21.22 $\pm$ 0.02 & 0.85 $\pm$ 0.02 & 18.10 $\pm$ 0.08 & 1.0 $\pm$ 0.1 & \\ 
27 & 342.17967 & -44.52853 & 23.07 $\pm$ 0.02 & 2.43 $\pm$ 0.07 & 22.23 $\pm$ 0.09 & 1.5 $\pm$ 0.1 & Red (3$\sigma$)\\ 
28 & 342.17999 & -44.53568 & 23.42 $\pm$ 0.03 & 1.44 $\pm$ 0.07 & 21.44 $\pm$ 0.14 & 1.5 $\pm$ 0.1 & \\ 
29 & 342.18109 & -44.52920 & 22.61 $\pm$ 0.04 & 0.74 $\pm$ 0.03 & 19.20 $\pm$ 0.14 & 2.1 $\pm$ 0.1 & Passive\\ 
30 & 342.18200 & -44.53857 & 22.81 $\pm$ 0.06 & 1.09 $\pm$ 0.08 & 20.23 $\pm$ 0.22 & 2.1 $\pm$ 0.2 & Red (3$\sigma$)\\ 
31 & 342.18205 & -44.54034 & 21.40 $\pm$ 0.01 & 1.14 $\pm$ 0.06 & 18.92 $\pm$ 0.12 & 3.3 $\pm$ 0.1 & ETG, Elliptical, Passive\\ 
32 & 342.18254 & -44.52687 & 21.78 $\pm$ 0.01 & 0.72 $\pm$ 0.04 & 18.31 $\pm$ 0.12 & 3.0 $\pm$ 0.1 & ETG, Elliptical, Red (3$\sigma$), Red (1$\sigma$), Passive\\ 
33 & 342.18293 & -44.53046 & 20.31 $\pm$ 0.04 & 0.80 $\pm$ 0.03 & 17.05 $\pm$ 0.12 & 3.4 $\pm$ 0.2 & ETG, Elliptical, Passive\\ 
34 & 342.18294 & -44.52492 & 22.82 $\pm$ 0.06 & 1.32 $\pm$ 0.17 & 20.66 $\pm$ 0.34 & 3.2 $\pm$ 0.4 & ETG, Elliptical\\ 
35 & 342.18307 & -44.53310 & 22.18 $\pm$ 0.04& 0.43 $\pm$ 0.01 & 17.57 $\pm$ 0.08 & 2.6 $\pm$ 0.2 & ETG, Elliptical, Passive\\ 
36 & 342.18320 & -44.54385 & 22.58 $\pm$ 0.04& 1.75 $\pm$ 0.09 & 21.03 $\pm$ 0.16 & 1.8 $\pm$ 0.1 & \\ 
37 & 342.18328 & -44.53083 & 17.26 $\pm$ 0.49 & 17.00 $\pm$ 1.60 & 20.65 $\pm$ 0.69 & 1.4 $\pm$ 0.4 & BCG\\ 
38 & 342.18353 & -44.53007 & 20.03 $\pm$ 0.02 & 1.76 $\pm$ 0.11 & 18.50 $\pm$ 0.16 & 3.2 $\pm$ 0.2 & ETG, Elliptical, Passive\\ 
39 & 342.18406 & -44.52692 & 20.45 $\pm$ 0.02 & 1.57 $\pm$ 0.13 & 18.67 $\pm$ 0.20 & 3.3 $\pm$ 0.2 & ETG, Elliptical, Red (3$\sigma$), Red (1$\sigma$), Passive\\ 
40 & 342.18419 & -44.52028 & 22.18 $\pm$ 0.04 & 2.08 $\pm$ 0.14 & 21.01 $\pm$ 0.19 & 1.4 $\pm$ 0.1 & Red (3$\sigma$), Red (1$\sigma$)\\ 
41 & 342.18438 & -44.53619 & 20.12 $\pm$ 0.01 & 2.16 $\pm$ 0.15 & 19.03 $\pm$ 0.16 & 3.3 $\pm$ 0.2 & ETG, Red (3$\sigma$), Red (1$\sigma$), Passive\\ 
42 & 342.18449 & -44.54318 & 20.81 $\pm$ 0.01 & 2.32 $\pm$ 0.14 & 19.87 $\pm$ 0.14 & 4.6 $\pm$ 0.1 & ETG, Elliptical, Red (3$\sigma$), Red (1$\sigma$), Passive\\ 
43 & 342.18453 & -44.52930 & 24.46 $\pm$ 0.07 & 0.41 $\pm$ 0.02 & 19.75 $\pm$ 0.17 & 1.2 $\pm$ 0.1 & Passive\\ 
44 & 342.18487 & -44.54063 & 23.17 $\pm$ 0.05 & 1.11 $\pm$ 0.08 & 20.63 $\pm$ 0.21 & 2.3 $\pm$ 0.2 & Red (3$\sigma$), Red (1$\sigma$)\\ 
45 & 342.18542 & -44.51863 & 19.48 $\pm$ 0.02 & 1.52 $\pm$ 0.16 & 17.62 $\pm$ 0.25 & 5.5 $\pm$ 0.3 & ETG, Red (3$\sigma$), Red (1$\sigma$), Passive\\ 
46 & 342.18555 & -44.53319 & 22.50 $\pm$ 0.02 & 0.51 $\pm$ 0.02 & 18.28 $\pm$ 0.12 & 3.8 $\pm$ 0.4 & ETG, Elliptical, Red (3$\sigma$), Red (1$\sigma$)\\ 
47 & 342.18565 & -44.53125 & 23.36 $\pm$ 0.08 & 0.93 $\pm$ 0.09 & 20.44 $\pm$ 0.30 & 1.8 $\pm$ 0.2 & Passive\\ 
48 & 342.18568 & -44.53050 & 22.90 $\pm$ 0.01 & 1.15 $\pm$ 0.06 & 20.44 $\pm$ 0.12 & 2.9 $\pm$ 0.3 & ETG, Elliptical, Passive\\ 
49 & 342.18570 & -44.52601 & 22.25 $\pm$ 0.01 & 1.35 $\pm$ 0.06 & 20.14 $\pm$ 0.10 & 3.6 $\pm$ 0.1 & ETG, Elliptical, Red (3$\sigma$), Red (1$\sigma$)\\ 
50 & 342.18663 & -44.53108 & 22.78 $\pm$ 0.03 & 0.60 $\pm$ 0.01 & 18.90 $\pm$ 0.08 & 1.7 $\pm$ 0.1 & Passive\\ 
51 & 342.18665 & -44.52247 & 20.91 $\pm$ 0.02& 0.83 $\pm$ 0.04 & 17.75 $\pm$ 0.13 & 3.9 $\pm$ 0.1 & ETG, Red (3$\sigma$), Red (1$\sigma$), Passive\\ 
52 & 342.18666 & -44.54076 & 21.06 $\pm$ 0.01 & 0.86 $\pm$ 0.04 & 17.96 $\pm$ 0.12 & 3.2 $\pm$ 0.1 & ETG, Elliptical, Red (3$\sigma$), Red (1$\sigma$), Passive\\ 
53 & 342.18667 & -44.52058 & 23.85 $\pm$ 0.15 & 1.53 $\pm$ 0.24 & 22.01 $\pm$ 0.50 & 1.2 $\pm$ 0.2 & Red (3$\sigma$), Red (1$\sigma$)\\ 
54 & 342.18678 & -44.52781 & 20.71 $\pm$ 0.01& 1.53 $\pm$ 0.05 & 18.88 $\pm$ 0.07 & 5.3 $\pm$ 0.1 & ETG, Elliptical, Red (3$\sigma$), Red (1$\sigma$), Passive\\ 
55 & 342.18686 & -44.53390 & 22.09 $\pm$ 0.01 & 2.31 $\pm$ 0.11 & 21.15 $\pm$ 0.11 & 2.6 $\pm$ 0.1 & ETG, Elliptical\\ 
56 & 342.18693 & -44.53537 & 20.93 $\pm$ 0.01 & 1.43 $\pm$ 0.05 & 18.94 $\pm$ 0.09 & 2.7 $\pm$ 0.1 & ETG, Elliptical, Red (3$\sigma$), Red (1$\sigma$), Passive\\ 
57 & 342.18811 & -44.53280 & 22.37 $\pm$ 0.03& 1.52 $\pm$ 0.07 & 20.51 $\pm$ 0.13 & 1.7 $\pm$ 0.1 & Elliptical, Red (3$\sigma$), Red (1$\sigma$)\\ 
58 & 342.18813 & -44.52595 & 19.94 $\pm$ 0.03 & 0.90 $\pm$ 0.07 & 16.95 $\pm$ 0.20 & 4.1 $\pm$ 0.3 & ETG, Elliptical, Red (3$\sigma$), Red (1$\sigma$), Passive\\ 
59 & 342.18814 & -44.52972 & 20.10 $\pm$ 0.02& 1.07 $\pm$ 0.05 & 17.48 $\pm$ 0.13 & 3.0 $\pm$ 0.2 & ETG, Red (3$\sigma$), Red (1$\sigma$), Passive\\ 
60 & 342.18844 & -44.51774 & 23.06 $\pm$ 0.10 & 1.11 $\pm$ 0.13 & 20.53 $\pm$ 0.36 & 1.2 $\pm$ 0.1 & \\ 
\hline
\end{tabular}
\caption{Parameters for the spectroscopically confirmed cluster members of AS1063: ID in serial order (col. 1), Right Ascension in degrees (col. 2), Declination in degrees (col. 3), total magnitude in the \textit{F814W} band and its error (col. 4), circularized effective radius in kpc and its error (col. 5), averaged surface brightness at the effective radius in mag arcsec$^{-2}$ and its error (col. 6), S\'ersic index and its error (col. 7) and classification of the galaxy as described in Sect. 4 (col. 8).}
\label{appendixa1}
\end{table*}

\setcounter{table}{0}
\begin{table*}
\centering
\begin{tabular}{l|c|c|c|c|c|c|c|c}
\hline
\hline
ID & RA & DEC & m$_{\mathrm{tot}}$& R$_{\mathrm{e}}$ (kpc) & $\left \langle \mu \right \rangle_{\mathrm{e}}$ (mag arcsec$^{-2}$) & n & Classification \\
\hline
61 & 342.18856 & -44.52670 & 25.05 $\pm$ 0.05 & 0.40  $\pm$  0.07 & 20.30  $\pm$  0.42 & 0.9 $\pm$ 0.2 & \\ 
62 & 342.18893 & -44.54039 & 20.27 $\pm$ 0.07 & 3.01  $\pm$  0.30 & 19.90  $\pm$  0.29 & 3.2 $\pm$ 0.3 & ETG, Elliptical, Red (3$\sigma$)\\ 
63 & 342.18901 & -44.51923 & 22.52 $\pm$ 0.09 & 1.32  $\pm$  0.13 & 20.36  $\pm$  0.30 & 1.4 $\pm$ 0.1 & \\ 
64 & 342.18901 & -44.52469 & 21.31 $\pm$ 0.12 & 4.51  $\pm$  1.40 & 21.82  $\pm$  0.79 & 11.0 $\pm$ 1.6 & ETG, Passive\\
65 & 342.18907 & -44.52643 & 24.46 $\pm$ 0.06 & 0.72  $\pm$  0.06 & 20.99  $\pm$  0.23 & 1.3 $\pm$ 0.1 & \\ 
66 & 342.18916 & -44.52367 & 23.59 $\pm$ 0.02 & 0.61  $\pm$  0.01 & 19.74  $\pm$  0.05 & 2.2 $\pm$ 0.1 & \\ 
67 & 342.18916 & -44.52953 & 20.40 $\pm$ 0.01 & 1.20  $\pm$  0.07 & 18.03  $\pm$  0.13 & 4.1 $\pm$ 0.2 & ETG, Elliptical, Red (3$\sigma$), Red (1$\sigma$), Passive\\ 
68 & 342.18934 & -44.53692 & 21.91 $\pm$ 0.01 & 1.47  $\pm$  0.05 & 19.99  $\pm$  0.08 & 1.8 $\pm$ 0.1 & Red (3$\sigma$), Red (1$\sigma$), Passive\\ 
69 & 342.19002 & -44.52411 & 22.59 $\pm$ 0.02 & 0.76  $\pm$  0.04 & 19.23  $\pm$  0.13 & 4.5 $\pm$ 0.2 & ETG, Elliptical, Red (3$\sigma$), Red (1$\sigma$)\\ 
70 & 342.19019 & -44.51651 & 24.84 $\pm$ 0.04 & 0.55  $\pm$  0.02 & 20.77  $\pm$  0.13 & 0.7 $\pm$ 0.1 & \\ 
71 & 342.19134 & -44.53432 & 21.18 $\pm$ 0.02 & 0.53  $\pm$  0.03 & 17.05  $\pm$  0.15 & 3.0 $\pm$ 0.2 & ETG, Elliptical, Red (3$\sigma$), Red (1$\sigma$), Passive\\ 
72 & 342.19156 & -44.53334 & 21.72 $\pm$ 0.02 & 0.68  $\pm$  0.02 & 18.13  $\pm$  0.08 & 2.9 $\pm$ 0.1 & ETG, Elliptical, Red (3$\sigma$), Red (1$\sigma$), Passive\\ 
73 & 342.19186 & -44.52965 & 22.11 $\pm$ 0.01 & 0.84  $\pm$  0.04 & 18.98  $\pm$  0.11 & 2.5 $\pm$ 0.2 & ETG, Elliptical, Passive\\ 
74 & 342.19269 & -44.51977 & 22.97 $\pm$ 0.11 & 0.76  $\pm$  0.12 & 19.62  $\pm$  0.44 & 2.2 $\pm$ 0.4 & \\ 
75 & 342.19281 & -44.51495 & 21.72 $\pm$ 0.01 & 2.73  $\pm$  0.08 & 21.14  $\pm$  0.06 & 2.3 $\pm$ 0.1 & Red (3$\sigma$)\\ 
76 & 342.19293 & -44.52205 & 22.22 $\pm$ 0.02 & 1.64  $\pm$  0.01 & 20.53  $\pm$  0.03 & 0.7 $\pm$ 0.1 & \\ 
77 & 342.19326 & -44.52412 & 24.37 $\pm$ 0.03 & 0.82  $\pm$  0.02 & 21.18  $\pm$  0.09 & 1.5 $\pm$ 0.1 & \\ 
78 & 342.19330 & -44.51782 & 19.77 $\pm$ 0.04 & 0.93  $\pm$  0.07 & 16.84  $\pm$  0.20 & 4.2 $\pm$ 0.2 & ETG, Elliptical, Red (3$\sigma$), Red (1$\sigma$), Passive\\ 
79 & 342.19330 & -44.52643 & 20.43 $\pm$ 0.03 & 0.96  $\pm$  0.06 & 17.57  $\pm$  0.16 & 2.8 $\pm$ 0.2 & ETG, Red (3$\sigma$), Red (1$\sigma$), Passive\\ 
80 & 342.19368 & -44.51851 & 22.24 $\pm$ 0.08 & 1.08  $\pm$  0.13 & 19.66  $\pm$  0.34 & 2.3 $\pm$ 0.2 & Red (3$\sigma$), Red (1$\sigma$)\\ 
81 & 342.19420 & -44.52210 & 22.46 $\pm$ 0.02 & 1.59  $\pm$  0.03 & 20.71  $\pm$  0.06 & 0.7 $\pm$ 0.1 & \\ 
82 & 342.19461 & -44.53227 & 23.38 $\pm$ 0.03 & 0.42  $\pm$  0.01 & 18.72  $\pm$  0.06 & 1.2 $\pm$ 0.1 & Red (3$\sigma$), Red (1$\sigma$)\\ 
83 & 342.19530 & -44.53490 & 22.72 $\pm$ 0.01 & 1.68  $\pm$  0.07 & 21.09  $\pm$  0.10 & 2.6 $\pm$ 0.1 & ETG, Elliptical, Red (3$\sigma$), Red (1$\sigma$)\\ 
84 & 342.19543 & -44.53245 & 22.53 $\pm$ 0.02 & 1.52  $\pm$  0.07 & 20.68  $\pm$  0.13 & 2.1 $\pm$ 0.1 & Red (3$\sigma$), Red (1$\sigma$)\\ 
85 & 342.19552 & -44.52599 & 19.01 $\pm$ 0.03 & 3.27  $\pm$  0.38 & 18.82  $\pm$  0.29 & 6.3 $\pm$ 0.4 & ETG, Elliptical, Red (3$\sigma$), Red (1$\sigma$), Passive\\ 
86 & 342.19669 & -44.52909 & 22.85 $\pm$ 0.02 & 1.36  $\pm$  0.06 & 20.76  $\pm$  0.11 & 2.0 $\pm$ 0.1 & \\ 
87 & 342.19727 & -44.52327 & 20.39 $\pm$ 0.04 & 1.15  $\pm$  0.07 & 17.93  $\pm$  0.17 & 3.9 $\pm$ 0.3 & ETG, Elliptical, Red (3$\sigma$), Red (1$\sigma$), Passive\\ 
88 & 342.19821 & -44.52740 & 22.50 $\pm$ 0.03 & 1.57  $\pm$  0.07 & 20.72  $\pm$  0.12 & 2.0 $\pm$ 0.1 & Elliptical, Red (3$\sigma$), Red (1$\sigma$)\\ 
89 & 342.20017 & -44.52722 & 22.34 $\pm$ 0.04 & 1.58  $\pm$  0.09 & 20.58  $\pm$  0.17 & 2.3 $\pm$ 0.2 & Red (3$\sigma$), Red (1$\sigma$)\\ 
90 & 342.20029 & -44.52520 & 20.38 $\pm$ 0.05 & 1.96  $\pm$  0.18 & 19.08  $\pm$  0.25 & 3.0 $\pm$ 0.3 & ETG, Elliptical, Red (3$\sigma$), Red (1$\sigma$), Passive\\ 
91 & 342.20084 & -44.52722 & 23.72 $\pm$ 0.02 & 1.01  $\pm$  0.01 & 20.98  $\pm$  0.04 & 1.1 $\pm$ 0.1 & \\ 
92 & 342.20123 & -44.52405 & 23.35 $\pm$ 0.04 & 1.01  $\pm$  0.03 & 20.60  $\pm$  0.09 & 1.9 $\pm$ 0.1 & Red (3$\sigma$), Red (1$\sigma$)\\ 
93 & 342.20137 & -44.52076 & 21.96 $\pm$ 0.02 & 0.78  $\pm$  0.02 & 18.65  $\pm$  0.07 & 2.3 $\pm$ 0.1 & Elliptical, Red (3$\sigma$), Red (1$\sigma$)\\ 
94 & 342.20418 & -44.52524 & 20.55 $\pm$ 0.03 & 1.32  $\pm$  0.09 & 18.39  $\pm$  0.17 & 3.0 $\pm$ 0.2 & ETG, Elliptical, Red (3$\sigma$), Red (1$\sigma$), Passive\\ 
95 & 342.20490 & -44.52587 & 23.46 $\pm$ 0.02 & 1.34  $\pm$  0.04 & 21.33  $\pm$  0.08 & 1.9 $\pm$ 0.2 & \\
\hline
\hline
\end{tabular}
\caption{Continue.}
\end{table*}

\begin{table*}
\centering
\begin{tabular}{l|c|c|c|c|c|c|c|c}
\hline
\hline
ID & RA & DEC & m$_{\mathrm{tot}}$& R$_{\mathrm{e}}$ (kpc) & $\left \langle \mu \right \rangle_{\mathrm{e}}$ (mag arcsec$^{-2}$) & n & Classification \\
\hline
1 & 177.38945 & 22.39391 & 19.68 $\pm$ 0.08 & 3.15 $\pm$ 0.46 & 18.32 $\pm$ 0.40 & 4.8 $\pm$ 0.4& ETG, Elliptical, Red (3$\sigma$), Red (1$\sigma$)\\
2 & 177.38970 & 22.39271 & 20.35 $\pm$ 0.02 & 3.24 $\pm$ 0.30 & 19.05 $\pm$ 0.22 & 3.7 $\pm$ 0.2& ETG, Elliptical, Red (3$\sigma$), Red (1$\sigma$)\\
3 & 177.39015 & 22.40389 & 21.81 $\pm$ 0.02 & 0.98 $\pm$ 0.09 & 17.91 $\pm$ 0.23 & 4.4 $\pm$ 0.3& ETG, Elliptical, Red (3$\sigma$), Red (1$\sigma$)\\
4 & 177.39097 & 22.40168 & 20.94 $\pm$ 0.03 & 1.49 $\pm$ 0.17 & 17.96 $\pm$ 0.28 & 4.8 $\pm$ 0.4& ETG, Elliptical, Red (3$\sigma$), Red (1$\sigma$), Passive\\
5 & 177.39110 & 22.40491 & 20.42 $\pm$ 0.04 & 2.26 $\pm$ 0.33 & 18.33 $\pm$ 0.36 & 6.5 $\pm$ 0.5& ETG, Elliptical, Red (3$\sigma$), Red (1$\sigma$)\\
6 & 177.39120 & 22.39271 & 20.82 $\pm$ 0.12 & 5.96 $\pm$ 0.83 & 20.84 $\pm$ 0.43 & 1.3 $\pm$ 0.3&\\
7 & 177.39138 & 22.40107 & 22.49 $\pm$ 0.06 & 0.67 $\pm$ 0.08 & 17.75 $\pm$ 0.33 & 2.4 $\pm$ 0.3& Red (3$\sigma$), Red (1$\sigma$)\\
8 & 177.39168 & 22.39062 & 20.54 $\pm$ 0.03 & 3.04 $\pm$ 0.25 & 19.10 $\pm$ 0.21 & 2.8 $\pm$ 0.2& ETG, Red (3$\sigma$), Red (1$\sigma$), Passive\\
9 & 177.39183 & 22.40529 & 21.19 $\pm$ 0.01 & 1.01 $\pm$ 0.08 & 17.35 $\pm$ 0.19 & 3.7 $\pm$ 0.4& ETG, Elliptical, Red (3$\sigma$), Red (1$\sigma$), Passive\\
10 & 177.39217 & 22.40128 & 22.80 $\pm$ 0.02 & 0.54 $\pm$ 0.04 & 17.60 $\pm$ 0.20 & 2.6 $\pm$ 0.5& ETG, Elliptical, Red (3$\sigma$), Red (1$\sigma$)\\
11 & 177.39266 & 22.39273 & 20.26 $\pm$ 0.03 & 1.99 $\pm$ 0.36 & 17.90 $\pm$ 0.42 & 2.9 $\pm$ 0.3& ETG, Red (3$\sigma$), Red (1$\sigma$), Passive\\
12 & 177.39269 & 22.39436 & 21.61 $\pm$ 0.05 & 1.37 $\pm$ 0.13 & 18.43 $\pm$ 0.26 & 3.4 $\pm$ 0.2& ETG, Elliptical, Red (3$\sigma$), Red (1$\sigma$), Passive\\
13 & 177.39278 & 22.39810 & 20.52 $\pm$ 0.04 & 1.92 $\pm$ 0.17 & 18.09 $\pm$ 0.23 & 2.3 $\pm$ 0.2& Red (3$\sigma$), Red (1$\sigma$), Passive\\
14 & 177.39288 & 22.39710 & 21.13 $\pm$ 0.01 & 1.79 $\pm$ 0.13 & 18.54 $\pm$ 0.17 & 3.7 $\pm$ 0.2& ETG, Elliptical, Red (3$\sigma$), Red (1$\sigma$)\\
15 & 177.39381 & 22.40231 & 22.66 $\pm$ 0.06 & 1.27 $\pm$ 0.12 & 19.33 $\pm$ 0.25 & 4.0 $\pm$ 0.3& ETG, Elliptical, Red (3$\sigma$), Red (1$\sigma$)\\
16 & 177.39416 & 22.39503 & 23.61 $\pm$ 0.18 & 1.39 $\pm$ 0.34 & 20.46 $\pm$ 0.72 & 2.3 $\pm$ 0.6& Red (3$\sigma$), Red (1$\sigma$)\\
17 & 177.39453 & 22.40063 & 22.29 $\pm$ 0.04 & 2.76 $\pm$ 0.16 & 20.64 $\pm$ 0.17 & 3.1 $\pm$ 0.1& ETG, Elliptical\\
18 & 177.39484 & 22.39292 & 23.06 $\pm$ 0.06 & 1.22 $\pm$ 0.16 & 19.63 $\pm$ 0.34 & 3.6 $\pm$ 0.5& ETG, Elliptical, Red (3$\sigma$), Red (1$\sigma$)\\
19 & 177.39503 & 22.39602 & 22.62 $\pm$ 0.09 & 1.10 $\pm$ 0.15 & 18.97 $\pm$ 0.39 & 2.0 $\pm$ 0.1& Red (3$\sigma$), Red (1$\sigma$)\\
20 & 177.39507 & 22.38985 & 22.77 $\pm$ 0.29 & 0.77 $\pm$ 0.48 & 18.36 $\pm$ 1.60 & 6.0 $\pm$ 3.8& ETG, Elliptical, Red (3$\sigma$), Red (1$\sigma$)\\
21 & 177.39527 & 22.40105 & 22.07 $\pm$ 0.03 & 1.19 $\pm$ 0.08 & 18.58 $\pm$ 0.18 & 3.6 $\pm$ 0.2&ETG, Elliptical, Red (3$\sigma$), Red (1$\sigma$), Passive\\
22 & 177.39584 & 22.39350 & 21.31 $\pm$ 0.01 & 2.38 $\pm$ 0.08 & 19.34 $\pm$ 0.08 & 7.0 $\pm$ 0.1&ETG, Elliptical, Red (3$\sigma$), Red (1$\sigma$)\\
23 & 177.39688 & 22.39231 & 22.53 $\pm$ 0.12 & 3.14 $\pm$ 0.30 & 21.16 $\pm$ 0.33 & 4.5 $\pm$ 0.4&ETG, Elliptical\\
24 & 177.39694 & 22.39297 & 20.55 $\pm$ 0.01 & 2.89 $\pm$ 0.07 & 19.00 $\pm$ 0.06 & 4.2 $\pm$ 0.1& ETG, Elliptical\\
25 & 177.39726 & 22.40029 & 22.71 $\pm$ 0.45 & 3.90 $\pm$ 1.70 & 21.81 $\pm$ 1.40 & 2.5 $\pm$ 0.8&ETG, Elliptical\\
26 & 177.39752 & 22.39953 & 20.65 $\pm$ 0.10 & 7.70 $\pm$ 1.20 & 21.23 $\pm$ 0.44 & 6.8 $\pm$ 0.7& ETG, Elliptical, Red (3$\sigma$), Passive\\
27 & 177.39762 & 22.40288 & 22.78 $\pm$ 0.07 & 1.55 $\pm$ 0.18 & 19.87 $\pm$ 0.32 & 2.2 $\pm$ 0.3&Red (3$\sigma$)\\
28 & 177.39779 & 22.39545 & 20.53 $\pm$ 0.06 & 2.40 $\pm$ 0.51 & 18.57 $\pm$ 0.53 & 3.5 $\pm$ 0.4& ETG, Elliptical, Red (3$\sigma$), Red (1$\sigma$), Passive\\
29 & 177.39792 & 22.40394 & 23.07 $\pm$ 0.07 & 1.48 $\pm$ 0.19 & 20.07 $\pm$ 0.35 & 1.6 $\pm$ 0.2&Red (3$\sigma$)\\
30 & 177.39795 & 22.40105 & 20.23 $\pm$ 0.06 & 2.51 $\pm$ 0.37 & 18.38 $\pm$ 0.38 & 4.4 $\pm$ 0.4& ETG, Elliptical, Red (3$\sigma$), Red (1$\sigma$), Passive\\
31 & 177.39846 & 22.40536 & 22.50 $\pm$ 0.06 & 1.37 $\pm$ 0.15 & 19.32 $\pm$ 0.30 & 2.7 $\pm$ 0.4& ETG, Elliptical, Red (3$\sigma$), Red (1$\sigma$)\\
32 & 177.39855 & 22.38979 & 22.10 $\pm$ 0.03 & 0.86 $\pm$ 0.06 & 17.91 $\pm$ 0.19 & 1.6 $\pm$ 0.2& Red (3$\sigma$), Passive\\
33 & 177.39860 & 22.39808 & 21.54 $\pm$ 0.08 & 4.06 $\pm$ 0.53 & 20.73 $\pm$ 0.37 & 7.1 $\pm$ 0.6& ETG, Elliptical, Red (3$\sigma$), Passive\\
34 & 177.39869 & 22.39230 & 21.17 $\pm$ 0.03 & 1.40 $\pm$ 0.15 & 18.05 $\pm$ 0.25 & 3.1 $\pm$ 0.3& ETG, Elliptical, Red (3$\sigma$)\\
35 & 177.39874 & 22.39853 & 17.83 $\pm$ 0.10 & 35.60 $\pm$ 5.30 & 21.73 $\pm$ 0.42 & 3.4 $\pm$ 0.2&BCG\\
36 & 177.39886 & 22.40181 & 21.62 $\pm$ 0.02 & 1.24 $\pm$ 0.08 & 18.24 $\pm$ 0.16 & 3.6 $\pm$ 0.1& ETG, Elliptical, Red (3$\sigma$), Red (1$\sigma$), Passive\\
37 & 177.39919 & 22.40090 & 23.77 $\pm$ 0.10 & 1.84 $\pm$ 0.26 & 21.24 $\pm$ 0.41 & 3.4 $\pm$ 0.2& ETG, Elliptical, Red (3$\sigma$)\\
38 & 177.39965 & 22.39961 & 22.09 $\pm$ 0.02 & 1.16 $\pm$ 0.06 & 18.55 $\pm$ 0.13 & 2.7 $\pm$ 0.2& ETG, Elliptical, Red (3$\sigma$), Red (1$\sigma$), Passive\\
39 & 177.39982 & 22.39726 & 20.26 $\pm$ 0.09 & 4.41 $\pm$ 1.00 & 19.63 $\pm$ 0.58 & 5.8 $\pm$ 0.6& ETG, Elliptical, Red (3$\sigma$), Passive\\
40 & 177.40020 & 22.39441 & 23.11 $\pm$ 0.17 & 1.83 $\pm$ 0.50 & 20.56 $\pm$ 0.76 & 2.9 $\pm$ 0.8&ETG, Elliptical\\
41 & 177.40040 & 22.39821 & 22.63 $\pm$ 0.10 & 0.46 $\pm$ 0.09 & 17.09 $\pm$ 0.50 & 3.8 $\pm$ 1.5& ETG, Elliptical, Red (3$\sigma$), Red (1$\sigma$)\\
42 & 177.40078 & 22.39625 & 22.75 $\pm$ 0.08 & 0.68 $\pm$ 0.09 & 18.06 $\pm$ 0.37 & 2.6 $\pm$ 0.6& ETG, Elliptical, Red (3$\sigma$), Red (1$\sigma$)\\
43 & 177.40080 & 22.39377 & 23.55 $\pm$ 0.08 & 0.72 $\pm$ 0.09 & 18.97 $\pm$ 0.35 & 1.9 $\pm$ 0.5& Red (3$\sigma$), Red (1$\sigma$)\\
44 & 177.40103 & 22.39788 & 20.45 $\pm$ 0.07 & 1.67 $\pm$ 0.26 & 17.71 $\pm$ 0.41 & 5.0 $\pm$ 0.7& ETG, Elliptical, Red (3$\sigma$), Red (1$\sigma$), Passive\\
45 & 177.40121 & 22.40033 & 20.55 $\pm$ 0.02 & 1.32 $\pm$ 0.11 & 17.30 $\pm$ 0.21 & 4.2 $\pm$ 0.4& ETG, Elliptical, Red (3$\sigma$), Red (1$\sigma$), Passive\\
46 & 177.40173 & 22.39880 & 22.72 $\pm$ 0.06 & 0.60 $\pm$ 0.07 & 17.76 $\pm$ 0.30 & 3.1 $\pm$ 0.6& ETG, Elliptical, Red (3$\sigma$), Red (1$\sigma$)\\
47 & 177.40225 & 22.39975 & 21.27 $\pm$ 0.02 & 1.42 $\pm$ 0.09 & 18.17 $\pm$ 0.16 & 2.7 $\pm$ 0.2& ETG, Elliptical, Red (3$\sigma$), Red (1$\sigma$), Passive\\
48 & 177.40239 & 22.39803 & 23.36 $\pm$ 0.06 & 1.25 $\pm$ 0.09 & 19.99 $\pm$ 0.22 & 1.5 $\pm$ 0.2&\\
49 & 177.40262 & 22.39618 & 20.90 $\pm$ 0.10 & 5.83 $\pm$ 1.00 & 20.87 $\pm$ 0.49 & 5.9 $\pm$ 0.9& ETG, Elliptical, Red (3$\sigma$), Passive\\
50 & 177.40288 & 22.40201 & 22.64 $\pm$ 0.05 & 2.46 $\pm$ 0.28 & 20.74 $\pm$ 0.30 & 4.1 $\pm$ 0.3& ETG, Elliptical, Red (3$\sigma$)\\
51 & 177.40306 & 22.40439 & 21.49 $\pm$ 0.02 & 1.15 $\pm$ 0.08 & 17.93 $\pm$ 0.17 & 3.8 $\pm$ 0.3& ETG, Elliptical, Red (3$\sigma$), Red (1$\sigma$), Passive\\
52 & 177.40358 & 22.39638 & 19.86 $\pm$ 0.02 & 2.85 $\pm$ 0.28 & 18.28 $\pm$ 0.24 & 3.0 $\pm$ 0.2& ETG, Red (3$\sigma$), Red (1$\sigma$), Passive\\
53 & 177.40366 & 22.39194 & 21.26 $\pm$ 0.04 & 1.03 $\pm$ 0.09 & 17.48 $\pm$ 0.23 & 3.8 $\pm$ 0.5& ETG, Elliptical, Red (3$\sigma$), Red (1$\sigma$), Passive\\
54 & 177.40369 & 22.38911 & 20.97 $\pm$ 0.03 & 2.39 $\pm$ 0.22 & 19.01 $\pm$ 0.24 & 3.9 $\pm$ 0.3& ETG, Elliptical, Red (3$\sigma$), Red (1$\sigma$)\\
55 & 177.40372 & 22.40458 & 20.66 $\pm$ 0.03 & 2.16 $\pm$ 0.21 & 18.48 $\pm$ 0.24 & 3.9 $\pm$ 0.3& ETG, Red (3$\sigma$), Red (1$\sigma$), Passive\\
56 & 177.40401 & 22.40213 & 22.80 $\pm$ 0.05 & 1.93 $\pm$ 0.18 & 20.38 $\pm$ 0.26 & 4.2 $\pm$ 0.3& ETG, Elliptical, Red (3$\sigma$)\\
57 & 177.40403 & 22.40302 & 21.84 $\pm$ 0.02 & 2.83 $\pm$ 0.08 & 20.25 $\pm$ 0.08 & 0.9 $\pm$ 0.1&\\
58 & 177.40516 & 22.39977 & 22.76 $\pm$ 0.03 & 1.00 $\pm$ 0.11 & 18.91 $\pm$ 0.27 & 2.9 $\pm$ 0.4& ETG, Elliptical, Red (3$\sigma$), Red (1$\sigma$)\\
59 & 177.40536 & 22.39165 & 23.17 $\pm$ 0.05 & 1.46 $\pm$ 0.13 & 20.14 $\pm$ 0.24 & 2.8 $\pm$ 0.3& ETG, Elliptical, Red (3$\sigma$), Red (1$\sigma$)\\
60 & 177.40544 & 22.39787 & 20.57 $\pm$ 0.01 & 2.13 $\pm$ 0.18 & 18.36 $\pm$ 0.20 & 4.4 $\pm$ 0.2& ETG, Elliptical, Red (3$\sigma$), Red (1$\sigma$), Passive\\
\hline
\hline
\end{tabular}
\caption{Parameters for the spectroscopically confirmed cluster members of M1149: ID in serial order (col. 1), Right Ascension in degrees (col. 2), Declination in degrees (col. 3), total magnitude in the \textit{F814W} band and its error (col. 4), circularized effective radius in kpc and its error (col. 5), averaged surface brightness at the effective radius in mag arcsec$^{-2}$ and its error (col. 6), S\'ersic index and its error (col. 7) and classification of the galaxy as described in Sect. 4 (col. 8).}
\label{appendixa2}
\end{table*}

\setcounter{table}{1}
\begin{table*}
\centering
\begin{tabular}{l|c|c|c|c|c|c|c|c}
\hline
\hline
ID & RA & DEC & m$_{\mathrm{tot}}$& R$_{\mathrm{e}}$ (kpc) & $\left \langle \mu \right \rangle_{\mathrm{e}}$ (mag arcsec$^{-2}$) & n & Classification \\
\hline
61 & 177.40629 & 22.40540& 22.52 $\pm$ 0.02 & 0.89 $\pm$ 0.04 & 18.41 $\pm$ 0.14 & 0.6 $\pm$ 0.1\\
62 & 177.40646 & 22.38958 & 19.02 $\pm$ 0.02 & 6.59 $\pm$ 0.34 & 19.26 $\pm$ 0.13 & 3.4 $\pm$ 0.1& ETG, Elliptical, Red (3$\sigma$), Red (1$\sigma$), Passive\\
63 & 177.40663 & 22.39555 & 22.26 $\pm$ 0.05 & 0.97 $\pm$ 0.12 & 18.33 $\pm$ 0.32 & 1.8 $\pm$ 0.3& Red (3$\sigma$), Passive\\
64 & 177.40690 & 22.39582 & 21.62 $\pm$ 0.05 & 1.65 $\pm$ 0.24 & 18.85 $\pm$ 0.37 & 2.9 $\pm$ 0.4& ETG, Elliptical, Red (3$\sigma$), Red (1$\sigma$), Passive\\
65 & 177.40726 & 22.39144 & 23.27 $\pm$ 0.11 & 1.41 $\pm$ 0.21 & 20.16 $\pm$ 0.44 & 2.4 $\pm$ 0.5& Red (3$\sigma$), Red (1$\sigma$)\\
66 & 177.40738 & 22.39479 & 23.71 $\pm$ 0.14 & 1.40 $\pm$ 0.19 & 20.58 $\pm$ 0.43 & 0.9 $\pm$ 0.5\\
67 & 177.40747 & 22.39912 & 23.00 $\pm$ 0.03 & 1.19 $\pm$ 0.13 & 19.53 $\pm$ 0.26 & 1.9 $\pm$ 0.3\\
68 & 177.40752 & 22.40305 & 19.77 $\pm$ 0.03 & 2.51 $\pm$ 0.22 & 17.92 $\pm$ 0.23 & 4.4 $\pm$ 0.2& ETG, Red (3$\sigma$), Red (1$\sigma$), Passive\\
\hline
\hline
\end{tabular}
\caption{Continue.}
\end{table*}


\bsp	
\label{lastpage}
\end{document}